\documentclass[twocolumn]{aastex7}
\usepackage{booktabs}
\usepackage[flushleft]{threeparttable}
\DeclareRobustCommand{\ion}[2]{%
\relax\ifmmode
\ifx\testbx\f@series
{\mathbf{#1\,\mathsc{#2}}}\else
{\mathrm{#1\,\mathsc{#2}}}\fi
\else\textup{#1\,{\mdseries\textsc{#2}}}%
\fi}

\begin{document}

\title{PyEMILI: A New Generation Computer-aided Spectral Line Identifier\\
-- II. Emission-line Identification and Plasma Diagnostics of A Sample of Gaseous Nebulae}

\author[0009-0000-7976-7383]{Zhijun Tu}
\affiliation{CAS Key Laboratory of Optical Astronomy, National Astronomical Observatories, Chinese Academy of Sciences (NAOC), Beijing 100101, P.~R.\ China}
\email{zjtu@nao.cas.cn}

\author[0000-0003-1286-2743]{Xuan Fang}
\affiliation{CAS Key Laboratory of Optical Astronomy, National Astronomical Observatories, Chinese Academy of Sciences (NAOC), Beijing 100101, P.~R.\ China}
\affiliation{School of Astronomy and Space Sciences, University of Chinese Academy of Sciences, Beijing 100049, P.~R.\ China}
\affiliation{Xinjiang Astronomical Observatory, Chinese Academy of Sciences, 150 Science 1-Street, Urumqi, Xinjiang, 830011, P.~R.\ China}
\affiliation{Laboratory for Space Research, Faculty of Science, The University of Hong Kong, Pokfulam Road, Hong Kong, P.~R.\ China}
\email[show]{fangx@nao.cas.cn}

\author[0000-0002-6138-1869]{Jorge Garc\'{i}a-Rojas}
\affiliation{Instituto de Astrof\'{i}sica de Canarias, E-38205 La Laguna, Tenerife, Spain}
\affiliation{Departamento de Astrof\'{i}sica, Universidad de La Laguna, E-38206 La Laguna, Tenerife, Spain}
\email{jogarcia@iac.es}

\author[0000-0002-3742-8460]{Robert Williams}
\affiliation{Space Telescope Science Institute, 3700 San Martin Drive, Baltimore, MD 21218, USA}
\affiliation{Department of Astronomy \& Astrophysics, University of California, Santa Cruz, 1156 High Street, Santa Cruz, CA 95064, USA}
\email{wms@stsci.edu}

\author{Jifeng Liu}
\affiliation{CAS Key Laboratory of Optical Astronomy, National Astronomical Observatories, Chinese Academy of Sciences (NAOC), Beijing 100101, P.~R.\ China}
\affiliation{School of Astronomy and Space Sciences, University of Chinese Academy of Sciences, Beijing 100049, P.~R.\ China}
\affiliation{Institute for Frontiers in Astronomy and Astrophysics, Beijing Normal University, Beijing 102206, P.~R.\ China}
\affiliation{New Cornerstone Science Laboratory, National Astronomical Observatories, Chinese Academy of Sciences, Beijing 100012, P.~R.\ China}
\email{jfliu@nao.cas.cn}

\correspondingauthor{Xuan Fang}

\begin{abstract}
In order to test the robustness and reliability of the new generation spectral-line identifier PyEMILI, as initially introduced in Paper~I, in line identification and establish a reference/benchmark dataset for future spectroscopic studies, we run the code on the line lists of a selected sample of emission-line nebulae, including planetary nebulae (PNe), \ion{H}{ii} regions, and Herbig-Haro (HH) objects with deep high-dispersion spectroscopic observations published over the past two decades.  The automated line identifications by PyEMILI demonstrate significant improvements in both completeness and accuracy compared to the previous manual identifications in the literature.  Since our last report of PyEMILI, the atomic transition database used by the code has been further expanded by cross-matching the Kurucz Line Lists.  Moreover, to aid the PyEMILI identification of numerous faint optical recombination lines (ORLs) of \ion{C}{ii}, \ion{N}{ii}, \ion{O}{ii} and \ion{Ne}{ii}, we compiled a new dataset of effective recombination coefficients for these nebular lines, and created a new subroutine in the code to generate theoretical spectra of heavy-element ORLs at various electron temperature and density cases; these theoretical spectra can be used to fit the observed recombination spectrum of a PN to obtain the electron temperature, density and ionic abundances using the Markov-Chain Monte Carlo (MCMC) method.  We present MCMC-derived parameters for a sample of PNe.  This work establishes PyEMILI as a robust and versatile tool for both line identification and plasma diagnostics in deep spectroscopy of gaseous nebulae. 
\end{abstract}


\section{Introduction}
\label{sec:intro}

Emission-line nebulae such as planetary nebulae (PNe) and \ion{H}{ii} regions are among the most important probes of chemical composition and physical conditions of the interstellar medium (ISM).  PNe, a short-lived evolutionary phase of the low- and intermediate-mass ($\sim$1--8\,$M_{\odot}$) stars, preserve the chemical signatures of their progenitors and trace stellar populations across different galactic environments, from bulges to halos.  In contrast, \ion{H}{ii} regions, ionized by young massive stars, represent the present-day chemical composition of galaxies.  Together, they provide complementary insights into the past and current chemical evolution of the Milky Way as well as other galaxies \citep[e.g.][]{Sanders_2012,2018ApJ...853...50F,2018ApJ...862...45S}. 

The nebular spectra of PNe and \ion{H}{ii} regions are dominated by strong collisionally excited lines (CELs) and hydrogen and helium recombination lines \citep[e.g.][]{Kwok_2000,2006agna.book.....O}.  The deepest spectroscopy also reveal faint optical recombination lines (ORLs) of heavy elements (mainly the second-row elements C, N, O and Ne).  This ensemble of emission lines enables measurements of elemental abundances, plasma conditions, and stellar nucleosynthesis yields.  Deep, high-dispersion spectroscopic observations of PNe using large-aperture ground-based telescopes now reveal hundreds of emission lines in the optical-NIR wavelength region (from $\sim$3600\,{\AA} to 1\,$\mu$m), including numerous extremely faint ORLs of the C, N, O, and Ne ions (e.g.\ \citealt{2016MNRAS.461.2818M, 2018MNRAS.473.4476G, 2022MNRAS.510.5444G, Gomez_Llanos_2024}; see also a review of \citealt{2020rfma.book...89G} and the references therein), which are crucial for abundance determinations.  These advances highlight the importance of systematic and reliable line identification with high efficiency as a foundation of nebular spectroscopy. 

To address this need, we have developed PyEMILI \citep{https://doi.org/10.5281/zenodo.14054096}, a new Python-based tool for computer-aided spectral-line identification \citep[][Paper~I]{Tu_2025}, which was developed from the previous Fortran-based emission-line identification code EMILI \citep{Sharpee_2003}.  PyEMILI refines the logic of traditional line-identification methods, incorporates updated atomic transition data \citep{2018vanhoof}, and includes a dedicated dataset of effective recombination coefficients for nebular lines to better identify the faint ORLs. Initial tests in Paper~I demonstrated significant improvements in the accuracy and efficiency of line identification compared to manual identifications, providing a robust framework for analyzing deep, high-dispersion spectra of emission-line nebulae.

In the landscape of nebular analysis softwares, PyEMILI is unique.  In functionality, PyEMILI plays a complementary role to the widely used packages such as {\sc pyneb} \citep{2015A&A...573A..42L}, {\sc cloudy} \citep{1998PASP..110..761F,2017RMxAA..53..385F,2023RNAAS...7..246G}, and {\sc alfa} \citep{2016MNRAS.456.3774W}.  While {\sc cloudy} is primarily designed for photoionization modeling of nebulae -- to predict the strengths of the most prominent CELs (of heavy elements) and ORLs (mainly of H\,{\sc i}, He\,{\sc i} and He\,{\sc ii}) under the physical conditions that models converge to, PyEMILI instead focuses on the identification of all observed emission lines, particularly the numerous faint ORLs of heavy elements that have been overlooked in various modeling approaches.  The main purpose of PyEMILI is to assign the most reliable ID to every detected feature in an observed spectrum, enabling robust analysis of the weak and previously unidentified lines. 

The {\sc alfa} code is dedicated to emission line fitting and measurement, constructing synthetic spectra that reproduce the observed profiles.  PyEMILI has also implemented a line-measurement module (see Paper~I, Section\,2.7 therein), which, although slightly slower than the optimized fitting routine in {\sc alfa}, offers a \texttt{matplotlib}-based interactive interface for manual refinement and supports the fitting of both emission and absorption features in a spectrum.  For constraining the key parameters/properties, such as the central-star temperature, chemical composition, or the ionization structures of PNe, tools like {\sc cloudy} and {\sc pyneb} remain indispensable.  PyEMILI thus occupies a unique niche in the field of nebular spectroscopy -- serving as a bridge between the spectral data and subsequent physical modeling. 

A typical workflow of nebular analysis can be summarized as follows:  (1) emission-line measurement using {\sc alfa}, PyEMILI, or manual fitting; (2) emission-line identification with PyEMILI; (3) plasma diagnostics and ionic abundance determination using {\sc pyneb}, based on both CELs and ORLs (note that PyEMILI also implements its own Bayesian module for the ORL-based $T_{\rm e}$, $N_{\rm e}$, and ionic abundance fitting using an expanded dataset of the heavy-element ORLs; see Section\,\ref{MCMC}); and (4) comprehensive photoionization modeling using {\sc cloudy} to constrain the global nebular properties such as elemental abundances, ionization and thermal structures, as well as the effective temperature and luminosity of the PN central star. 

In this paper, the second in the series, we present extensive PyEMILI identification runs on the emission-line lists of a large sample of PNe and \ion{H}{ii} regions with deep, high-dispersion spectroscopic observations published over the past two decades.  Our goals are to establish a reference database of optical line identifications, to revise and supplement line identifications in the literature (particularly for the lines previously unidentified or likely misidentified), and to introduce the new functionality of recombination-line fitting using effective recombination coefficients of ORLs.  These results provide both a benchmark for future high-dispersion spectroscopic studies and a demonstration of PyEMILI’s reliability and versatility. 

The structure of this paper is as follows:  In Section\,2, we describe further improvements in PyEMILI since the publication of Paper~I.  In Section\,3, we report our test runs of PyEMILI on the line lists of a sample of nebulae, as well as our line-identification results compared with the previously published manual identifications.  In Section\,4, we present plasma diagnostics for the PNe in our sample using the newly compiled database of the effective recombination coefficients for the \ion{O}{ii} and \ion{N}{ii} nebular lines, and discuss the results.  Finally, in Section\,5, we summarize the work, draw our conclusions, and provide perspectives on future developments.

\section{New Features in PyEMILI}

In this section, we first briefly introduce the atomic transition database used by PyEMILI, and then describe a newly added subroutine in the code that can generate theoretical recombination spectra for various electron temperature and density cases, based on the dataset of effective recombination coefficients for the heavy-element ORLs.  These theoretical recombination spectra are used to fit the observed recombination spectrum of a PN (or an \ion{H}{ii} region) so that optimal values of $T_{\rm e}$, $N_{\rm e}$, and ionic abundances can be obtained for the nebula.

\subsection{Atomic Transition Database} 

The atomic transition database used by PyEMILI was developed based on the Atomic Line List v3.00b4 \citep[hereafter AtLL;][]{2018vanhoof}, which has been supplemented with huge volumes of transition probabilities retrieved from the literature, including the Kurucz Line Lists\footnote{\url{http://kurucz.harvard.edu/linelists/}} (see description in Section\,2.3 of Paper~I).  However, after numerous tests of PyEMILI for line identification and result verification, we found it evident that certain atomic transitions, such as the permitted transitions of \ion{O}{ii} from the 3d--4f array, in particular those from upper levels of the G[3]$^{\rm o}$ spectral term (e.g.\ $\lambda$4291.25 3d\,$^{4}$P$_{5/2}$--4f\,G[3]$^{\rm o}_{7/2}$ of the M55 multiplet, $\lambda$4344.42 3d\,$^{4}$D$_{5/2}$--4f\,G[3]$^{\rm o}_{7/2}$ of the M65c multiplet, and $\lambda$4477.90 3d\,$^{2}$P$_{3/2}$--4f\,G[3]$^{\rm o}_{5/2}$ of the M88 multiplet), were still missing from the atomic transition database. 

In order to construct a complete atomic transition database, which is critical for more comprehensive identifications of spectral lines, we incorporated additional transition data from the Kurucz Line Lists \citep{2017CaJPh..95..825K,2018ASPC..515...47K}.  This database contains a relatively more extensive collection of atomic transition probabilities, including those with very low transition probabilities.  However, the authenticity of many transitions is yet to be proved, and thus not all transitions are appropriate for PyEMILI.  The criteria for retrieving the atomic transition data from the Kurucz Line Lists are standardized as follows: 
\begin{itemize} 
    \item The atomic numbers span from 1 to 36 (from H to Kr). 
    \item Wavelength coverage is from 1000\,{\AA} to 2\,$\mu$m, consistent with the wavelength range of the atomic transition database currently used by PyEMILI (this is also the wavelength region covered by the majority of the high-dispersion spectrographs on the ground-based large telescopes as well as the Space Telescope Imaging Spectrograph (STIS) on board the \emph{Hubble Space Telescope}). 
    \item The ionization energy of the previous lower ionization stage, X$^{(i-1)+}$, of an ion X$^{i+}$ is less than 100\,eV. 
    \item For the permitted transitions (i.e.\ electric-dipole allowed transitions, mostly are recombination lines) and the intercombination transitions (i.e.\ semi-forbidden transitions), only those with transition probabilities $>$10$^5$\,s$^{-1}$ are considered. 
    \item For the forbidden transitions, Equation\,10 in Paper~I is used to calculate the theoretical line fluxes; here $T_{\rm e}$ = 10,000\,K and $N_{\rm e}$ = 10$^{4}$\,cm$^{-3}$ are assumed.  The dilution factor ($D$ in Equation\,10 of Paper~I) is set to be a typical value of 0.02, and the ionic abundance is the solar value \citep{2009ARA&A..47..481A}.  As a result, the transitions with theoretical line fluxes exceeding 10$^{-5}$ H$\beta$ are included\footnote{At current detection limits of the high-dispersion spectrographs on the 8--10\,m large telescopes, the faintest emission lines in an optical spectrum can be detected to levels of $\sim$10$^{-5}\times\,I$(H$\beta$).  An even lower limit of $\lesssim$10$^{-6}\times\,I$(H$\beta$) might be achieved in the future, given the expected increase in spectrophotometric depths for the instruments on the next-generation giant telescopes (such as the 25.4\,m GMT, 39\,m ELT and 30\,m TMT)}. 
\end{itemize}

\begin{figure*}[t]
\centering 
\epsscale{1.0}
\plotone{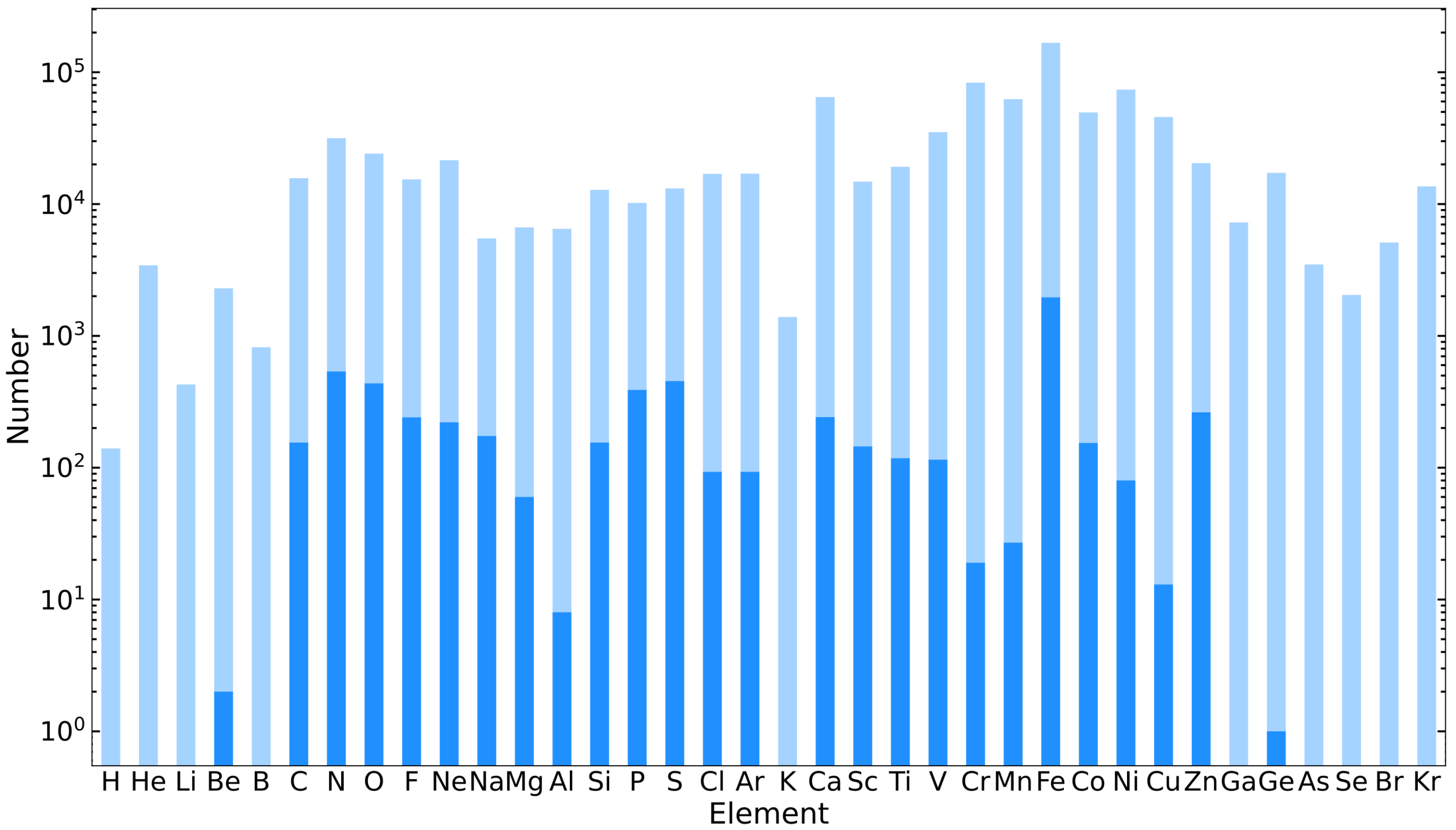}
\caption{Number distribution of atomic transitions, both the newly added (6154 transitions, in dark blue) and the entire atomic transition database ($\sim$891,000 transitions, in light blue) used by PyEMILI, classified by elements.} 
\label{fig1} 
\end{figure*}

Here, we set the ionization energy threshold at 100\,eV to ensure that the atomic transitions added to the atomic transition database are well matched with the ionization structure parameters of PyEMILI (see Section\,2.2.1 in Paper~I), where the lower limit of the last energy Bin (Bin~5, Section\,2.2 in Paper~I, see also Table\,1 therein) is 100\,eV.  In addition, among the numerous emission lines detected in the deep high-dispersion spectrum of a PN, only very few manually identified emission lines can be attributed to Bin 5; thus the ionization energy threshold of 100\,eV is adequate for PyEMILI's line identification. 

The transition probability threshold of 10$^{5}$\,s$^{-1}$ is used in retrieving the transition data of the permitted and intercombination transitions from the Kurucz Line Lists; this threshold value is reasonably chosen given the parameter setting in  PyEMILI, where the permitted transitions lacking transition probabilities were assigned a specified value of 10$^{4}$\,s$^{-1}$ (see the description in Section\,2.4.2 of Paper~I).  Hence the transition probability threshold for permitted transitions is established to be higher than 10$^{5}$\,s$^{-1}$, i.e., at least one order of magnitude higher than the specified value in PyEMILI.  Regarding the threshold of theoretical line fluxes for the forbidden transitions, we adopt 10$^{-5}$ of H$\beta$ flux, as determined by the typical detection limits of very faint nebular emission lines in the current deep high-dispersion spectroscopic observations of PNe and \ion{H}{ii} regions with large ground-based telescopes.


A total of 6154 atomic transitions were added to PyEMILI's atomic transition database, including 4951 permitted transitions and 1203 forbidden transitions.  The number distribution of atomic transitions as classified by elements is visually demonstrated in Figure\,\ref{fig1}, where both the newly added 6154 atomic transitions and the entire atomic transition database ($\sim$891,000 transitions) of PyEMILI are presented.  It should be noted that the treatment of each fine-structure transition within a multiplet as an item in the database results in huge numbers of transitions for some elements, particularly those with numerous electrons and complicated atomic structures, such as Fe. 

\subsection{PyEMILI Fitting of the Recombination Spectra of Nebulae} 
\label{MCMC}

\subsubsection{Dataset of Effective Recombination Coefficients} \label{MCMC:part1}

In order to help identify numerous optical recombination lines, the majority of which are very faint ($\lesssim$10$^{-4}$--10$^{-3}$ of H$\beta$ flux), in deep high-dispersion spectra of PNe and \ion{H}{ii} regions, we compiled for PyEMILI a dataset of state-of-the-art effective recombination coefficients, denoted as $\alpha_{\rm eff}$($\lambda$), for the nebular lines of the relatively abundant heavy-element ions (\ion{C}{ii}, \ion{N}{ii}, \ion{O}{ii} and \ion{Ne}{ii}) in PNe as well as the \ion{H}{i}, \ion{He}{i} and \ion{He}{ii} recombination lines (see Section\,2.5 in Paper~I).  These coefficients were collected from the literature \citep{HIcoe,NeIIcoe,CIIcoe,NIIcoe,NIIcoe_corr,CIIdicoe,OIIcoe,HeIcoe}.

\begin{figure*}
\begin{center}
\includegraphics[width=16.5cm,angle=0]{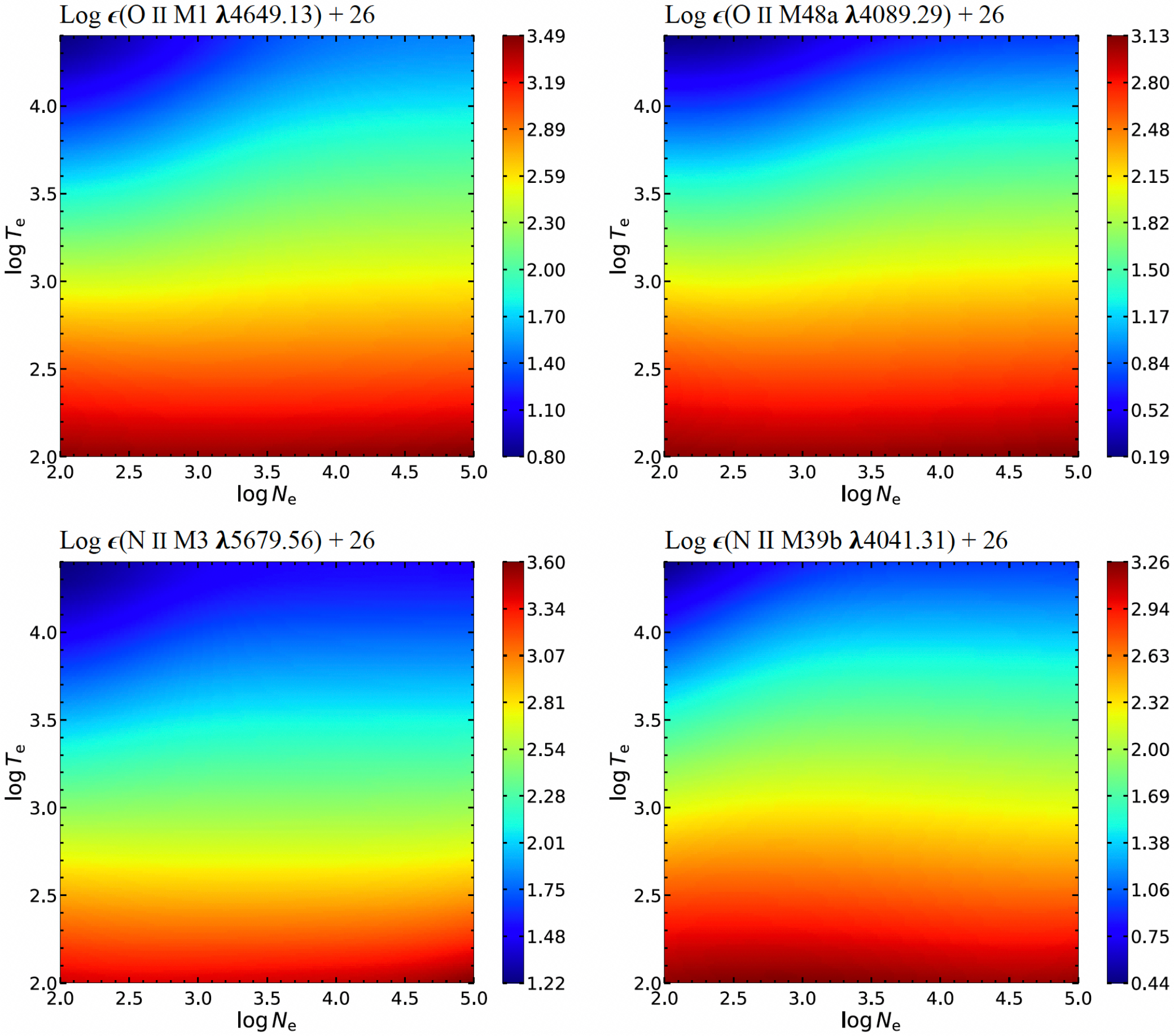}
\caption{Emissivities of the \ion{O}{ii} $\lambda$4649.13 M1 2p$^{2}$3s\,$^{4}$P$_{5/2}$ -- 2p$^{2}$3p\,$^{4}$D$^{\rm o}_{7/2}$ (\emph{top-left}) and $\lambda$4089.29 M48a 2p$^{2}$3d\,$^{4}$F$_{9/2}$ -- 2p$^{2}$4f\,G[5]$^{\rm o}_{11/2}$ (\emph{top-right}), and \ion{N}{ii} $\lambda$5679.56 M3 2p3s\,$^{3}$P$^{\rm o}_{2}$ -- 2p3p\,$^{3}$D$_{3}$ (\emph{bottom-left}) and $\lambda$4041.31 M39b 2p3d\,$^{3}$F$^{\rm o}_{4}$ -- 2p4f\,GG[9/2]$_{5}$ (\emph{bottom-right}) nebular lines, as a function of electron temperature and density; these four lines are the strongest transitions of the 3s--3p and 3d--4f arrays of the two ions.  Emissivity $\epsilon$ is defined by Eq\,1 using the effective recombination coefficients calculated by \citet[][for \ion{N}{ii} ORLs]{NIIcoe,NIIcoe_corr} and \citet[][for \ion{O}{ii} ORLs]{OIIcoe}.  The color bar of each panel indicates the emissivity value in logarithm, $\log{\epsilon(\lambda)}$+26.} 
\label{fig2}
\end{center}
\end{figure*}

The dataset of these effective recombination coefficients for nebular recombination lines is stored as a three-dimensional data cube in the form of emissivity $\epsilon$($\lambda$,\,$T_{\rm e}$,\,$N_{\rm e}$), in units of erg\,cm$^{3}$\,s$^{-1}$.  The emissivity of a transition with wavelength $\lambda$ is defined as 
\begin{equation} \label{eq1}
    \epsilon(\lambda, T_{\rm e}, N_{\rm e}) = \alpha_{\rm eff}(\lambda, T_{\rm e}, N_{\rm e})\times h\frac{c}{\lambda}\,,
\end{equation}
and is a function of two free parameters, electron temperature $T_{\rm e}$, and electron density $N_{\rm e}$; in Eq\,\ref{eq1}, $h$ is the Plank constant, and $c$ is the speed of light.  Variation in $T_{\rm e}$ and/or $N_{\rm e}$ directly influences the value of $\alpha_{\rm eff}$($\lambda$) and consequently the theoretical intensity of the nebular recombination line (at a given ionic abundance).  As an example, to show that $\alpha_{\rm eff}$($\lambda$) is a function of electron temperature and electron density, Figure\,\ref{fig2} shows the variation of the emissivities of \ion{N}{ii} $\lambda$5679.56 (2p3s\,$^{3}$P$^{\rm o}_{2}$ -- 2p3p\,$^{3}$D$_{3}$) and \ion{O}{ii} $\lambda$4649.13 (2p$^{2}$3s\,$^{4}$P$_{5/2}$ -- 2p$^{2}$3p\,$^{4}$D$^{\rm o}_{7/2}$), the strongest transitions of the two ions, in the $\log{T_{\rm e}}$--$\log{N_{\rm e}}$ diagram.

\subsubsection{A New Subroutine in PyEMILI}
To further leverage the emissivity, we have developed a new subroutine, \texttt{recomb}\footnote{Named after ``recombination'', one of the basic atomic processes in ionized gaseous nebulae.}, for PyEMILI, which can be used to constrain $T_{\rm e}$, $N_{\rm e}$, and ionic abundances in a photoionized gaseous nebula, based on the observed intensities of a number of recombination lines (emitted by the same ion) as measured from deep-dispersion spectra.  The methodology employed in the subroutine \texttt{recomb} relies on Bayesian inference facilitated by the Markov-Chain Monte Carlo (MCMC) technique, and is implemented by using the Python package \texttt{emcee}\footnote{\url{https://github.com/dfm/emcee};\\ \url{https://pypi.org/project/emcee/}},  
which is a Python implementation of the affine-invariant ensemble sampler for MCMC first proposed by \citet{2010CAMCS...5...65G}, and later used by others in astrophysical research \citep[e.g.][]{Foreman_Mackey_2013}. 

Within the PyEMILI subroutine \texttt{recomb}, three independent parameters can be configured: electron temperature, electron density, and ionic abundance.  Users have the flexibility to specify value(s) for one or two of these parameters, allowing the remaining parameter(s) to be sampled. 

Our approach assumes a flat prior distribution over the three parameters.  We considered the effective recombination coefficients for nebular lines of individual ions calculated at various electron temperature and density cases from the literature.  Furthermore, the ionic abundance (C$^{2+}$, N$^{2+}$, O$^{2+}$ and Ne$^{2+}$) used to calculate the predicted/theoretical line intensity ranges from 0.01 to 100 times the solar abundance of the corresponding element (given that under the physical conditions of PNe, the majority of C, N, O and Ne are in the doubly ionized stage, and thus X/H is close to X$^{2+}$/H$^{+}$).  To ensure reliable analysis, we performed a bilinear interpolation of $\epsilon$($\lambda$) in the two-dimensional grid of ($T_{\rm e}$, $N_{\rm e}$). 

In order to compare the predicted line intensities with those measured from a nebular spectrum to obtain the properties of a nebula, we use the likelihood function (similar to the method utilized in \citealt{2013MNRAS.428.3443M} and \citealt{Mendez_Delgado_2021}), which is defined as 
\begin{equation}
    \ln \mathcal{L} = -\frac{1}{2}~\sum_{l}\left(\frac{I_{\mathrm{obs}}(\lambda_l)-I_{\rm pred}(\lambda_l)}{\sigma(I_{\rm obs})}\right)^{2}\,,
\end{equation}
where $l$ represents individual recombination lines emitted by the same ion X$^{\rm i+}$.  $I_{\mathrm{obs}}(\lambda_l)$ is the observed, extinction-corrected line intensity relative to H$\beta$, i.e.\ $I_{\rm obs}$(H$\beta$)=1, and $\sigma$($I_{\rm obs}$) refers to the uncertainty in $I_{\rm obs}$($\lambda_{l}$).  $I_{\rm pred}$($\lambda_l$) is the predicted/theoretical line intensity of the recombination line, which is defined as 
\begin{equation}
    I_{\rm pred}(\lambda_l) = N({\rm X}^{\rm  i+}) \times \frac{\epsilon(\lambda_{l}, T_{\rm e}, N_{\rm e})}{\epsilon_{{\rm H}\beta}(T_{\rm e}, N_{\rm e})}\,, 
\end{equation}
where $N$(X$^{\rm i+}$) represents the ionic abundance X$^{\rm i+}$/H$^{+}$, and $\epsilon$($\lambda_l$,$T_{\rm e}$,$N_{\rm e}$) and $\epsilon_{{\rm H}\beta}$($T_{\rm e}$,$N_{\rm e}$) correspond to the emissivities of the recombination line $\lambda_{l}$ and H$\beta$, respectively. 

\subsubsection{Line Blending Issue}
The actual spectral-line measurements are always more or less affected by spectral resolution ($R$), which may cause the blending of emission lines with very close wavelengths.  This line-blending issue is common among the \ion{O}{ii} and \ion{N}{ii} ORLs, the majority of which are very faint ($\lesssim$10$^{-4}$--10$^{-3}$ H$\beta$).  Many \ion{O}{ii} and \ion{N}{ii} transitions are very close in wavelength, some even almost identical, and to distinguish these faint lines is impossible even at high spectral resolution.  Line blending mostly occurs in the \ion{O}{ii} and \ion{N}{ii} ORLs of the 3d--4f transition array, given the small energy difference between the fine-structure levels of the spectral terms of the 4f electron configuration. 

For example, \ion{N}{ii} $\lambda$4041.31 (M39b 3d\,$^{3}$F$^{\rm o}_{4}$ -- 4f\,G[9/2]$_{5}$) is the strongest transition in the 3d--4f array, but it is blended with \ion{O}{ii} $\lambda$4041.28 (M50c 3d\,$^{4}$F$_{5/2}$ -- 4f\,F[2]$^{\rm o}_{5/2}$).  Consequently, the observed flux of \ion{N}{ii} $\lambda$4041.31 is a blend of at least two components.  Another example is the blend of fine-structure lines within the same multiplet; e.g., \ion{O}{ii} $\lambda$4069.89 (M10 3p\,$^{4}$D$^{\rm o}_{3/2}$ -- 3d\,$^{4}$F$_{5/2}$) is blended with 4069.62\footnote{Under typical physical conditions of PNe ($T_{\rm e}\sim$10,000\,K and $N_{\rm e}\sim$10$^{2}$--10$^{4}$\,cm$^{-3}$), the theoretical fluxes of the two \ion{O}{ii} lines, $\lambda$4069.62 and $\lambda$4069.89, are comparable to each other, because their effective recombination coefficients only differ by $\sim$33\% \citep{OIIcoe}.} (M10 3p\,$^{4}$D$^{\rm o}_{1/2}$ -- 3d\,$^{4}$F$_{3/2}$).  Even in the case where these emission lines, which are close in wavelength, can be resolved using high-dispersion spectroscopy ($R\gtrsim$20,000), the nebular expansion of PNe (with a typical expansion velocity $\sim$20--30 km\,s$^{-1}$) will cause double-peak in emission line profile, further complicating the flux measurements \citep[e.g.,][]{2015MNRAS.452.2606G,2016MNRAS.461.2818M}. 

To consider the line-blending issue in spectral fits, we introduce a parameter of velocity difference $\Delta{v}$, which is defined based on the spectral resolution.  We sum the emissivities $\epsilon$($\lambda$,$T_{\rm e}$,$N_{\rm e}$) of all the recombination lines emitted by the same ion within a range of the observed wavelength, $\lambda_{\rm obs} \pm$ ($\lambda_{\rm obs}\times\Delta{v}$/$c$), where $\lambda_{\rm obs}$ is the observed wavelength corrected for radial velocity (determined through measurements of the \ion{H}{i} Balmer lines) and $\lambda_{\rm lab}$ is the laboratory wavelength of the corresponding transition.  This cumulative emissivity represents the final total theoretical intensity of the observed lines, expressed as 
\begin{equation} 
    I_{\rm pred}(\lambda_{l}) = N({\rm X}^{\rm i+}) \times \frac{\sum_{m} {\epsilon(\lambda_{m}, T_{\rm e}, N_{\rm e})}}{\epsilon_{{\rm H}\beta}(T_{\rm e}, N_{\rm e})}\,,
\end{equation}
where $\lambda_{m}$ signifies the individual recombination lines with wavelengths falling within the wavelength difference $\Delta\lambda$ (= $\lambda_{\rm obs}\times\Delta{v}$/$c$) of an observed emission line.

\subsubsection{Applicability and Parameter Settings of the Subroutine}


To robustly evaluate the model parameters (i.e., \ electron temperature, electron density, and ionic abundance) and ensure reliable posterior distributions, PyEMILI pre-sets specific configurations in \texttt{emcee} before sampling the parameter space, aiming to optimize the performance and convergence of the MCMC chains.  The MCMC sampling employs 16 ``walkers'', each performing up to 30,000 steps, with the autocorrelation time $\tau$ evaluated every 100 steps using the function \texttt{get\_autocorr\_time()} from the Python package \texttt{emcee}.  Convergence is considered to be achieved when the total length of the chain exceeds 100$\times\tau$ and $\tau$ changes by less than 1\%, i.e.\ $|\tau-\tau_{\mathrm{old}}|/\tau < 0.01$.  After convergence is detected, the sampler continues for an additional 1000 steps to ensure stability.  The final `burn-in' period is defined as two times the maximum $\tau$ value (2$\times\tau_{\rm max}$), while the thinning factor is set to be half of the minimum $\tau$ (0.5$\times\tau_{\rm min}$) to reduce sample autocorrelation. 

The subroutine \texttt{recomb} of PyEMILI can create the plots of posterior distributions of the model parameters by utilizing the scatterplot-matrices generating package \texttt{corner.py}\footnote{\url{https://github.com/dfm/corner.py};\\ \url{https://pypi.org/project/corner/}}, 
proposed by \citet{Mackey_2016}, a Python module that uses \texttt{matplotlib} \citep{2007CSE.....9...90H} to visualize multi-dimensional samples using a scatterplot matrix; in these visualizations, each one- and two-dimensional projection of the sample is plotted to reveal covariances.

\begin{deluxetable*}{lccclclr}

\setlength{\tabcolsep}{0.14cm}
\tabletypesize{\small}
\tablecaption{Test Sample for PyEMILI \label{testsample}}
\tablehead{\colhead{Object}  & \colhead{EC$^{a}$} & \colhead{Instrument} & \colhead{$R$} & \colhead{Wavelength} & \colhead{$N_{\rm line}\,^{b}$} & \colhead{Agreement$^{c}$} & \colhead{Ref.} \\ 
\colhead{} & \colhead{} & \colhead{(Telescope)} & \colhead{} &\colhead{(\AA)} & \colhead{} & \colhead{(\%)} & \colhead{} } 
\startdata
\multicolumn{8}{c}{Planetary Nebulae}\\
\smallskip
NGC\,3918 & 9 & UVES (VLT) &40000& 3100--10420 & 659 & 88.9\% &(1) \\
NGC\,6153 & 8 & B\&C (ESO) &3000& 3040--7400 & 340 & 92.6\% &(2) \\
Abell\,46 & 11 & ISIS (WHT) &3000& 3630--9230 & 132 & 99.2\%$^{d}$ &(5) \\
Cn\,1--5 & 3 & MIKE (Magellan Clay) &27000& 3440--9130 & 246 & 97.1\% &(6) \\
Hb\,4 & 9 & MIKE (Magellan Clay) &27000& 3340--9130 & 251 & 98.8\% &(6) \\
He\,2--86 & 3 & MIKE (Magellan Clay) &27000& 3710--9230 & 338 & 97.6\% &(6) \\
M\,1--25 & 2 & MIKE (Magellan Clay) &27000& 3340--9130 & 269 & 98.1\% &(6) \\
M\,1--30 & 1 & MIKE (Magellan Clay) &27000& 3700--9310 & 325 & 92.6\% &(6) \\
M\,1--32 & 2 & MIKE (Magellan Clay) &27000& 3700--9270 & 210 & 92.8\% &(6) \\
M\,1--61 & 3 & MIKE (Magellan Clay) &27000& 3510--9130 & 295 & 98.3\% &(6) \\
M\,3--15 & 3 & MIKE (Magellan Clay) &27000& 3490--9130 & 131 & 96.2\% &(6) \\
NGC\,5189 & 10 & MIKE (Magellan Clay) &27000& 3340--9130 & 264 & 97.7\% &(6) \\
NGC\,6369 & 3 & MIKE (Magellan Clay) &27000& 3700--9230 & 185 & 98.9\% &(6) \\
PC\,14 & 5 & MIKE (Magellan Clay) &27000& 3340--9130 & 266 & 97.7\% &(6) \\
Pe\,1--1 & 3 & MIKE (Magellan Clay) &27000& 3610--9130 & 220 & 95.8\% &(6) \\
NGC\,6543 &*$^{f}$& ISIS (WHT) &3000& 3630--7820 & 187 & 92.0\% &(7) \\
IC\,4776 & 3 & UVES (VLT) &30000& 3310--9970 & 337 & 94.3\%$^{d}$ &(9) \\
H\,1--40 & 3 & UVES (VLT) &15000& 3720--10410 & 196 & 98.5\% &(10) \\
H\,1--50 & 6 & UVES (VLT) &15000& 3120--10340 & 288 & 97.9\% &(10) \\
He\,2--73 & 8 & UVES (VLT) &15000& 3130--10410 & 317 & 97.2\% &(10) \\
He\,2--96 & 3 & UVES (VLT) &15000& 3690--10410 & 202 & 99.0\% &(10) \\
He\,2--158 & 2 & UVES (VLT) &15000& 3180--10340 & 137 & 98.5\% &(10) \\
M\,1--31  & 3 & UVES (VLT) &15000& 3690--10410 & 219 & 98.6\% &(10) \\
M\,1--33 & 3 & UVES (VLT) &15000& 3630--10410 & 322 & 96.8\% &(10) \\
M\,1--60 & 3 & UVES (VLT) &15000& 3630--10410 & 275 & 99.3\% &(10) \\
M\,2--31 & 3 & UVES (VLT) &15000& 3180--10400 & 206 & 99.5\% &(10) \\
NGC\,2022 & 12 & B\&C (ESO) &3000& 3750--7330 & 129 & 83.9\% &(11) \\
NGC\,5315 & 3 & UVES (VLT) &40000& 3100--10410 & 448 & 93.0\% &(12) \\
NGC\,5315 (NIR) &3& FIRE (Magellan Baade) &4800& 8310--19550$^{e}$ & 145 & 92.0\% &(12) \\
\midrule
\multicolumn{8}{c}{\ion{H}{ii} Regions}\\
30\,Doradus & * & UVES (VLT) &8800& 3180--10340 & 364 & 87.9\% &(8) \\
Orion Nebula (HH\,514) &*& UVES (VLT) &40000& 3180--10400 & 555 & 91.3\% &(4) \\
\midrule
\multicolumn{8}{c}{Herbig-Haro Objects}\\
HH\,204 (combined) &*& UVES (VLT) &40000& 3180--10340 & 272 & 97.4\%$^{d}$ &(3) \\
HH\,204 (cut 1) &*& UVES (VLT) &40000& 3180--10400 & 430 & 90.2\%   &(3) \\
HH\,514 (jet) &*& UVES (VLT) &40000& 3180--10340 & 88 & 99.0\% &(4) \\
\enddata
\tablecomments{
References to the deep optical spectroscopy: (1) \citet{2015MNRAS.452.2606G}; (2) \citet{2000MNRAS.312..585L}; (3) \citet{2021ApJ...918...27M}; (4) \citet{2022MNRAS.514..744M}; (5) \citet{2015ApJ...803...99C}; (6) \citet{2012AA...538A..54G}; (7) \citet{2004MNRAS.351.1026W}; (8) \citet{2003ApJ...584..735P}; (9) \citet{2017MNRAS.471.3529S}; (10) \citet{2018MNRAS.473.4476G}; (11) \citet{2003MNRAS.345..186T}; (12) \citet{2017MNRAS.471.1341M}. \\
\vspace{-1.0mm}
\tablenotetext{a}{Excitation class (EC) of PNe as defined by \citet{1991ApSS.181...73G}.} 
\vspace{-1.5mm}
\tablenotetext{b}{Number of emission lines detected in the spectrum.}
\vspace{-1.5mm}
\tablenotetext{c}{The percentage of agreement (or agreement rate) is defined with the number of lines whose manually assigned IDs are also ranked as ``A'' by PyEMILI.}
\vspace{-1.5mm}
\tablenotetext{d}{The wavelengths input to PyEMILI in these samples are laboratory wavelengths.}
\vspace{-1.5mm}
\tablenotetext{e}{Only the emission lines with wavelengths less than 20000\,\AA\ are included due to the wavelength range limitation of PyEMILI's transition database.}
\vspace{-1.5mm}
\tablenotetext{f}{Measurement of the [\ion{O}{iii}] $\lambda5007$ emission line was not available for this PN due to wavelength coverage; thus its excitation class could not be calculated.}}
\end{deluxetable*}

Running of the subroutine \texttt{recomb} requires an input line list which comprises (1) velocity-corrected observed wavelengths, (2) extinction-corrected line intensities, 
(3) uncertainties in line intensity, and (4) the proper spectral notations (e.g.\ \ion{C}{ii}, \ion{N}{ii}, \ion{O}{ii}, etc.) of atomic transitions.  Currently, only the \ion{O}{ii} and \ion{N}{ii} ORLs are used in the PyEMILI/\texttt{recomb} fitting of the nebular optical recombination spectra, because so far only the effective recombination coefficients for the nebular lines of these two ions have been reliably calculated \citep{NIIcoe,NIIcoe_corr,OIIcoe}.  These recombination calculations of the \ion{O}{ii} and \ion{N}{ii} lines were carried out entirely in the intermediate coupling scheme for all the $J$-resolved fine-structure transitions between the states with $n\leq$11 and $l\leq$4, taking into account the density dependence of the coefficients arising from the relative populations of the fine-structure levels of the ground term of the recombining ion (O$^{2+}$ $^{3}$P$_{0,\,1,\,2}$ in the case of \ion{O}{ii}, and N$^{2+}$ $^{2}$P$^{\rm o}_{1/2,\,3/2}$ in the case of \ion{N}{ii}), and have opened up the possibility of electron temperature and density diagnostics via recombination line analysis \citep[e.g.][]{2013MNRAS.428.3443M}. 

In contrast, no efforts have been attempted so far for nebular plasma diagnostics based on the \ion{Ne}{ii} recombination spectrum, partly due to the lack of reliable effective recombination coefficients.  Since all the recombination calculations of the \ion{Ne}{ii} nebular lines were calculated under the $LS$ coupling scheme \citep{NeIIcoe}, and relative populations of the $^{3}$P$_{2,\,1,\,0}$ fine-structure levels of the recombining ion Ne$^{2+}$ were assumed to be proportional the statistical weights (5\,:\,3\,:\,1), no density diagnostic is possible with the current available atomic data, although the \ion{Ne}{ii} ORLs have been detected in the deep spectra of a number of Galactic PNe. 

For the \ion{C}{ii} recombination spectrum of PNe, at the current detection limit of the large ground-based telescopes, the emission lines detected in the optical region ($\sim$3800--7500\,{\AA}) are much fewer than those of \ion{O}{ii} and \ion{N}{ii}, given the atomic structure of C$^{+}$.  The effective recombination coefficients were calculated by \citet{CIIcoe} for the \ion{C}{ii} transitions between doublet states, only at a single electron density of 10$^{4}$\,cm$^{-3}$.  Moreover, the nebular recombination lines of \ion{C}{ii} were also calculated by \citet{CIIdicoe}, who utilized the $R$-matrix method in intermediate coupling but only considered the dielectronic recombination processes (autoionization and near-threshold resonances); these calculations are therefore not suitable for the \ion{C}{ii} lines that are mainly produced via radiative recombination, e.g.\ the strongest transition of \ion{C}{ii} in the optical, $\lambda$4267 M6 2s$^{2}$3d\,$^{3}$D -- 2s$^{2}$4f\,$^{2}$F$^{\rm o}$. 

Apart from the availability of reliable effective recombination coefficients, our decision to focus on the \ion{O}{ii} and \ion{N}{ii} ORLs in spectral fits is also due to their being generally stronger and more numerous than the recombination lines of \ion{C}{ii} and \ion{Ne}{ii} as well as other heavy-element ions, in the deep optical spectra of Galactic PNe so far obtained.  In Section\,\ref{diagnostic}, we elucidate the diagnostic results derived for a dozen PNe with high-quality spectroscopic data, demonstrating the efficacy of this method applied to the \ion{O}{ii} and \ion{N}{ii} ORLs.

\section{Test of PyEMILI on Samples} 
\label{test sample}

\subsection{Sample Selection}

Although consensus has been reached regarding the identification of strong and prominent emission lines in the spectra of photoionized gaseous nebulae \citep[e.g.][]{2006agna.book.....O}, many faint emission lines are yet to be identified and investigated.  PyEMILI was specifically designed for the identification of faint emission lines, which are detected in large numbers, in the deep high-dispersion spectra of PNe and \ion{H}{ii} regions.  The results of PyEMILI's identification of the emission lines with fluxes $>$0.01 H$\beta$ generally highly agree with those of the manual identifications (IDs\footnote{Hereafter, we use ``manual IDs'' to refer to the IDs (atomic transitions) of the emission lines in a nebular spectrum that were assigned through manual identification.}); this has been demonstrated in the test runs of PyEMILI on the emission-line list of the Galactic PN Hf\,2-2 in Paper~I.  In other words, all candidates with reliable manual IDs had been ranked as ``A'' by PyEMILI, as shown in Table\,7 of Paper~I. 

To comprehensively evaluate the performance of PyEMILI in the identification of emission lines, we have selected a sample of Galactic emission-line objects with deep high-dispersion spectroscopic observations published over the past two decades.  The test sample is summarized in Table\,\ref{testsample}, including 28 PNe\footnote{For Galactic PN NGC\,5315, there are two spectra retrieved from \citet{2017MNRAS.471.1341M}: the optical spectrum (3100--10,420\,{\AA}, $R\sim$40,000) obtained with VLT/UVES, and the near-IR spectrum (0.8--2.5\,$\mu$m, $R\sim$4800) obtained with the FIRE spectrograph on the Magellan Baade Telescope.  Thus there are 29 spectra of PNe in our test sample (see Table\,\ref{testsample}).}, two \ion{H}{ii} regions and three Herbig-Haro (HH) objects.  
As the ionizing radiation fields from the central stars of PNe are harder than the OB stars associated with \ion{H}{ii} regions, they produce different nebular excitations, which can be discerned by the differences in relative strengths of nebular emission lines in the spectra of the two ISM.  HH objects are often referred to those associated with early stages of star formation, and are produced by the fast jets from the young stars in the birth (protostars) colliding with the surrounding ISM; the emission lines (especially those in the UV) from HH objects may be shock excited.  However, the HH objects in our sample (see Table\,\ref{testsample}) belong to the photoionized objects that are immersed in the strong radiation field of the Orion Nebula, where photoionization dominates over shock excitation. 

Deep spectra of the above three types of emission-line nebulae are used to test the PyEMILI code and assess its performance and reliability in line identification.  PNe dominate the size of our sample, given that they represent the bulk of the stellar populations in the universe and, more importantly, they exhibit the most complicated emission-line spectra among all the ISM.  The PNe in our sample have differences in properties (nebular size, evolutionary stage, central star temperature $T_{\rm eff}$, etc.) and encompass a broad range in terms of excitation class (EC), a parameter proposed by \citet{1991ApSS.181...73G}.  EC has a direct relation to the ionization and thermal equilibrium of nebulae, which is partially determined by the $T_{\rm eff}$ of the PN central star; in \citet{1991ApSS.181...73G} it was defined by the intensity ratios of nebular emission lines: $EC$ = $F_{[{\rm O\,III}]}(\lambda4959+\lambda5007)/F({\rm H}\beta)$ for low-excitation ($EC$=1--3) PNe, and $EC$ = $\log{F_{[{\rm O\,III}]}(\lambda4959+\lambda5007)/F_{{\rm He\,II}}(\lambda4686)}$ for the middle-excitation ($EC$=4--8) and high-excitation ($EC$=9--12) PNe.  The $EC$ values of the PNe in our sample are summarized in Table\,\ref{testsample}. 

For the majority of objects in our test sample, deep high-resolution spectroscopic data (in the form of emission-line tables) have been published.  To illustrate the limitation of PyEMILI in identifying the emission lines in the low-resolution spectra, a small number of objects with lower-resolution spectra are also included. 

Many line lists of \ion{H}{ii} regions in the literature are relatively short ($\lesssim100$ lines) and exhibit a high level of agreement in line identifications when using PyEMILI.  Therefore, we selected two of the most representative ones, the Orion Nebula and 30~Doradus.  For the Orion Nebula, we selected the most recent VLT/UVES high-dispersion spectroscopy of the photoionized object HH\,514 therein as reported by \citet{2022MNRAS.514..744M}, who detected more than 500 emission lines.  30~Doradus, also known as the Tarantula Nebula, is a large \ion{H}{ii} region in the Large Magellanic Cloud (LMC), which is a dwarf satellite of the Milky Way and whose average metallicity is lower than that of the Sun, at a distance of 50\,kpc \citep{Freedman_2010}.  We adopted the line list of 30 Doradus reported by \citet{2003ApJ...584..735P}, who detected more than 300 emission lines.

\subsection{The Spectroscopic Data: Emission-line Tables}

Before using PyEMILI to identify the emission lines from the line tables of each object in the test sample, as retrieved from the literature (see Table\,\ref{testsample}), we need to make revisions in the input line data.  For each object, the input observed wavelengths of emission lines are uniformly corrected for the radial velocity (usually the systemic velocity of the object), which can be derived by comparing the observed and the laboratory wavelengths of \ion{H}{i} Balmer lines, mainly the low-level transitions of hydrogen, H$\alpha$, H$\beta$, H$\gamma$, and H$\delta$.  In exceptional cases where the observed wavelengths are absent in the literature, the laboratory wavelengths of the known emission lines are used in replacement for the velocity-corrected observed wavelengths.  For a spectrum with spectral resolution $R\geq$10,000 (e.g.\ obtained with VLT UVES), the wavelength uncertainties of all emission lines are set to be 10--15 km\,s$^{-1}$; for $R<$10,000, the wavelength uncertainty in velocity is set to be 20--30 km\,s$^{-1}$.  The extinction-corrected line intensities are used in line identification. 

The final emission-line tables (with velocity-corrected observed wavelengths, wavelength uncertainties, and dereddened line intensities) of the test sample are then read by the PyEMILI code for line identifications.  The final identified emission-line tables of the test sample, in the simplified format where the most probable ID (i.e.\ with the highest rankings) assigned by PyEMILI is presented for each observed line (see Table\,\ref{Lines_NGC3918} in Appendix\,A as an example), are presented in Tables\,A.1--A.34.  The PyEMILI output files in the original complete format, where multiple IDs with different rankings assigned by PyEMILI are presented for each observed emission line (see Table\,\ref{output_NGC3918} in Appendix\,B as an example), are presented in Tables\,B.1--B.34.  We emphasize that the emission-line tables with PyEMILI's identification results, in the two formats as aforementioned, of the 34 sample spectra in our test sample are all publicly available online via the general-purpose open repository Zenodo\footnote{\url{https://doi.org/10.5281/zenodo.17540949}}\citep{tu_2025_17540949} for free usage and reference in astrophysical spectroscopy. 

Nebular abundances of ions/elements are needed to calculate the predicted template fluxes of nebular lines 
(see definition in Section\,2.4.2 of Paper~I).  For the PNe in our sample, we use the pre-set nebular abundance tables in PyEMILI, which are adopted from \citet{2006agna.book.....O}, to calculate the predicted template fluxes; for \ion{H}{ii} regions and HH objects, the pre-set solar abundance table reported by \citet{2009ARA&A..47..481A} is adopted.  Default values of the electron temperature and electron density, $T_{\rm e}$=10,000\,K and $N_{\rm e}$=10$^{4}$\,cm$^{-3}$, are utilized unless otherwise specified.

\subsection{Amendments and Supplements to the Manual Identifications}

After test runs of PyEMILI on the emission-line list of each object in our sample, we carefully check the outputs and scrutinize the IDs assigned to each observed emission line by PyEMILI, and then compare them with the corresponding manual ID available in the literature.  In Table\,\ref{testsample}, the column with the header ``Agreement'' presents the percentage of agreement in line identifications, which is defined as the proportion of the lines for which the manual IDs also received an ``A'' ranking assigned by PyEMILI.  The ranking of a candidate ID is based on its identification index (\emph{IDI}) value, which is defined to quantify the reliability of this candidate ID (the definition of \emph{IDI} is presented in Paper~I).  All the manual IDs of the sample objects adhere strictly to the identification results documented in the literature. 

Upon comparison, we found that for some emission lines, the IDs assigned by PyEMILI were more reasonable than the corresponding manual IDs given in the literature.  Moreover, for many of the emission lines that were unidentified in the literature, PyEMILI assigned ranked IDs, which helped us to identify their most probable transitions.  Through careful visual check, comparison, and scrutinization of the PyEMILI results (against the identifications published in the literature), we made amendments and supplements to the manual IDs in the literature.  These are summarized in Table~\ref{amendedlinelist}, where we present the amended and supplemented identifications to the manual IDs of the emission lines of the objects in the test sample. 

We used standardized criteria to determine which observed lines should be incorporated into Table\,\ref{amendedlinelist}:  (1) the emission lines whose manual IDs do not receive an ``A'' ranking by PyEMILI are considered; in cases where the manual IDs (given in the literature) are ranked as ``A'' by PyEMILI, we regard these identifications as more credible and do not subject them to further scrutinization.  (2) For the manual IDs falling into the case described in (1), we examine their rankings given by PyEMILI, the predicted line fluxes, the multiplet check results, and the wavelength differences.  For the case that a manual ID is a forbidden transition, we also assess its upper-level energy, as the line intensity of a forbidden transition can be proportional to ${\rm e}^{-E/kT}$, where $-E$ represents the excitation energy of the upper level of the transition. 

For the same observed line being checked, in addition to its manual ID, we also scrutinize the high-ranking candidate IDs assigned by PyEMILI, primarily those with rankings ``A'' and ``B''.  If the predicted line flux of the manual ID appears excessively weak or if the velocity difference of the manual ID systemically deviates from other fine-structure components/members in the same multiplet or from H$\beta$, it is deemed to be an unreliable manual ID.  Subsequently, we evaluate whether the high-ranking candidate IDs can serve as plausible identifications for the line.  In the case of faint lines, the results of the multiplet check are for reference only, given the difficulty in detecting other fine-structure members of the same multiplet for this faint line.  When the predicted line fluxes of the manual IDs and those of the ``A''-ranking candidates (assigned by PyEMILI) are not easily distinguishable, this case is considered as line-blending, indicating both identifications are plausible. 

A comprehensive compilation of the detailed amendments and supplements to the manual IDs we finally made for the objects in our test sample is presented in Table\,\ref{amendedlinelist}, which is segmented into contents, with each section corresponding to the line identification results of an object.  
It is essential to emphasize that all identifications presented in Table\,\ref{amendedlinelist} adhere strictly to the criteria established above, with no consideration given to the line profiles (e.g., NGC\,3918, see Section~\ref{NGC3918}).

\begin{deluxetable*}{ccccccccccc}
\tablecaption{Amended and Supplemental Identifications to the Manual IDs 
\label{amendedlinelist}}
\tablehead{\colhead{$\lambda_{\rm obs}$} & \colhead{$I_{\lambda}$} & \colhead{$\lambda_{\rm lab\,1}$} & \colhead{$\mathrm{ID_1}$} & \colhead{$\mathrm{Transition\,1}$} & \colhead{$\mathrm{Electron\,Configuration\,1}$} & \colhead{$\lambda_{\rm lab\,2}$} & \colhead{$\mathrm{ID_2}$} & \colhead{$\mathrm{Transition\,2}$} & \colhead{$\mathrm{Electron\,Configuration\,2}$} & \colhead{$\mathrm{Note}$} \\ 
\colhead{(\AA)} & \colhead{(H$\beta$=1)} & \colhead{(\AA)} & \colhead{} & \colhead{} & \colhead{} & \colhead{(\AA)} & \colhead{} & \colhead{} & \colhead{} & \colhead{} \\
\colhead{(1)} & \colhead{(2)} & \colhead{(3)} & \colhead{(4)} & \colhead{(5)} & \colhead{(6)} & \colhead{(7)} & \colhead{(8)} & \colhead{(9)} & \colhead{(10)} & \colhead{(11)}} 
\startdata
\multicolumn{11}{l}{Object Name: NGC\,3918}\\
\multicolumn{11}{l}{$V_\mathrm{H\beta}=-23.43\,\mathrm{km\,s^{-1}}$}\\
\multicolumn{11}{l}{}\\
3230.12&\multicolumn{1}{l|}{7.51E-04}&3230.54&\ion{N}{ii}&$\mathrm{^3D-{^3P^o}}$&\multicolumn{1}{l|}{$\mathrm{2s^2.2p.({^2P^o}).4p-2s^2.2p.({^2P^o}).10s}$}&3230.06&\ion{Ne}{ii}&$\mathrm{^2D-{^2D^o}}$&\multicolumn{1}{l|}{$\mathrm{2s^2.2p^4.({^1D}).3s-2s^2.2p^4.({^1D}).3p}$}&ERC, Wav\\
3571.36&\multicolumn{1}{l|}{5.31E-04}&…&…&…&\multicolumn{1}{l|}{…}&3571.231&\ion{Ne}{ii}]&$\mathrm{^4S^o-{^4F}}$&\multicolumn{1}{l|}{$\mathrm{2s^2.2p^4.({^3P}).3p-2s^2.2p^4.({^3P}).3d}$}&TP\\
3756.14&\multicolumn{1}{l|}{3.23E-04}&3756.1&\ion{He}{i}&$\mathrm{^1P^o-{^1D}}$&\multicolumn{1}{l|}{$\mathrm{1s.2p-1s.14d}$}&3755.7&[\ion{Fe}{v}]&$\mathrm{^5D-{^3F^4}}$&\multicolumn{1}{l|}{$\mathrm{3d^4-3d^4}$}&Blend, Mult, TP\\
4003.89&\multicolumn{1}{l|}{2.60E-04}&4003.58&\ion{N}{iii}&$\mathrm{^2D-{^2F^o}}$&\multicolumn{1}{l|}{$\mathrm{2s^2.4d-2s^2.5f}$}&4003.21&[\ion{Fe}{v}]&$\mathrm{^5D-{^3P^4}}$&\multicolumn{1}{l|}{$\mathrm{3d^4-3d^4}$}&Blend, Mult, TP\\
4376.58&\multicolumn{1}{l|}{3.38E-04}&4376.53&\ion{O}{iii}&$\mathrm{F[7/2]-G[9/2]^o}$&\multicolumn{1}{l|}{$\mathrm{2s^2.2p.({^2P^o}).4f.F-2s^2.2p.({^2P^o}).5g.G^o}$}&4376.582&\ion{C}{ii}&$\mathrm{^4P^o-{^4D}}$&\multicolumn{1}{l|}{$\mathrm{2s.2p.({^3P^o}).3d-2s.2p.({^3P^o}).4f}$}&Blend, Mult, TP\\
4428.52&\multicolumn{1}{l|}{1.09E-04}&4428.52&\ion{Ne}{ii}&$\mathrm{^2D-{^2[3]^o}}$&\multicolumn{1}{l|}{$\mathrm{2s^2.2p^4.({^3P}).3d-2s^2.2p^4.({^3P\left \langle2\right \rangle}).4f}$}&4428.34&[\ion{Mn}{iv}]&$\mathrm{a{^5D}-a{^3F}}$&\multicolumn{1}{l|}{$\mathrm{3d^4-3d^4}$}&Blend, Mult, TP\\
4452.88&\multicolumn{1}{l|}{5.35E-04}&4452.38&\ion{O}{ii}&$\mathrm{^2P-{^2D^o}}$&\multicolumn{1}{l|}{$\mathrm{2s^2.2p^2.({^3P}).3s-2s^2.2p^2.({^3P}).3p}$}&4453&\ion{Ne}{iv}&$\mathrm{^2S^o-{^2P}}$&\multicolumn{1}{l|}{$\mathrm{2s^2.2p^2.({^3P}).4p-2s^2.2p^2.({^3P}).4d}$}&Wav, TP\\
4481.67&\multicolumn{1}{l|}{4.41E-04}&4481.21&\ion{Mg}{ii}&$\mathrm{^2D-{^2F^o}}$&\multicolumn{1}{l|}{$\mathrm{2p^6.3d-2p^6.4f}$}&4481.38&[\ion{Mn}{v}]&$\mathrm{^2G-{^2F}}$&\multicolumn{1}{l|}{$\mathrm{3d^3-3d^3}$}&Mult, Wav, TP\\
4554.99&\multicolumn{1}{l|}{3.80E-05}&…&…&…&\multicolumn{1}{l|}{…}&4555&[\ion{Fe}{ii}]&$\mathrm{a{^6D}-b{^4F}}$&\multicolumn{1}{l|}{$\mathrm{3d^6.({^5D}).4s-3d^6.({^3F2}).4s}$}&TP, Wav\\
4620.53&\multicolumn{1}{l|}{1.58E-04}&4621.39&\ion{N}{ii}&$\mathrm{{^3P^o}-{^3P}}$&\multicolumn{1}{l|}{$\mathrm{2s^2.2p.({^2P^o}).3s-2s^2.2p.({^2P^o}).3p}$}&4620.19&\ion{C}{ii}&$\mathrm{^2D-{^2F^o}}$&\multicolumn{1}{l|}{$\mathrm{2s^2.4d-2s^2.8f}$}&ERC, Wav\\
4643.4&\multicolumn{1}{l|}{2.83E-04}&4643.3&[\ion{Mn}{iv}]&$\mathrm{b{^3G}-d{^3F}}$&\multicolumn{1}{l|}{$\mathrm{3d^3.({^2G}).4s-3d^3.({^2F}).4s}$}&4643.086&\ion{N}{ii}&$\mathrm{{^3P^o}-{^3P}}$&\multicolumn{1}{l|}{$\mathrm{2s^2.2p.({^2P^o}).3s-2s^2.2p.({^2P^o}).3p}$}&ERC\\
4673.75&\multicolumn{1}{l|}{1.98E-04}&4673.48&\ion{O}{iii}&$\mathrm{^3P-{^3P^o}}$&\multicolumn{1}{l|}{$\mathrm{2s^2.2p.({^2P^o}).4p-2s^2.2p.({^2P^o}).5s}$}&4673.73&\ion{O}{ii}&$\mathrm{^4P-{^4D^o}}$&\multicolumn{1}{l|}{$\mathrm{2s^2.2p^2.({^3P}).3s-2s^2.2p^2.({^3P}).3p}$}&ERC, Mult\\
4707.22&\multicolumn{1}{l|}{1.47E-04}&4707.22&[\ion{Mn}{iv}]&$\mathrm{b{^3G}-d{^3F}}$&\multicolumn{1}{l|}{$\mathrm{3d^3.({^2G}).4s-3d^3.({^2F}).4s}$}&4707.31&\ion{N}{iv}&$\mathrm{^3F^o-{^3G}}$&\multicolumn{1}{l|}{$\mathrm{2s.5f-2s.6g}$}&TP\\
\\
\enddata
\tablecomments{
The contents of this table are arranged in sections corresponding to the objects in the order they appear in Table\,\ref{testsample}.  In each section, the first two rows are the object name and the radial velocity used to correct the observed wavelengths of emission lines.  The radial velocity was determined by comparing the observed and laboratory wavelengths of H$\beta$; $V_{{\rm H}\beta}$=0, if the velocity correction has already been made in the literature. \\
\smallskip\\
Columns (1) and (2):  The observed wavelength and the line intensity (relative to H$\beta$).  The wavelengths have been corrected for the radial velocity as derived through measurements of H$\beta$. \\
Columns (3), (4), (5), and (6):  The laboratory wavelength, emitting ion, transition (lower and upper spectral terms), and electron configurations (lower and upper) of the corresponding manual ID given in the literature.  Tridot ``...'' means no identification. \\
Columns (7), (8), (9), and (10):  New transition information was selected from the candidate IDs assigned by PyEMILI.  The most probable ID is adopted. \\
Column (11):  The reason why the new ID assigned by PyEMILI is more robust.  \textbf{ERC}: the predicted line flux of the new ID is calculated using the effective recombination coefficient in our atomic transition database (see the description in Section\,2.5 of Paper~I), and is more consistent with the observed intensity of this line.  \textbf{Mult}: the new ID has other multiplet members (i.e.\ fine-structure components of the same multiplet as this ID) identified in this object.  \textbf{TP}: for the permitted transitions, this ID has a higher transition probability; for the forbidden transitions, this ID also refers to a relatively lower energy of the upper level.  Both cases lead to a higher value of the predicted line flux.  \textbf{Wav}: the laboratory wavelength of the new ID agrees better with the velocity-corrected observed wavelength (while the laboratory wavelength of the manual ID deviates too much from the observed wavelength).  \textbf{Blend}: the primitive ID is also possible but probably has a line-blending issue.  \textbf{Abun}: a higher ionic abundance for the new ID (usually for hydrogen and helium). \\
\smallskip\\
(Only a portion of this table is shown here to demonstrate its form and content.  This table is published in its entirety online only in the machine-readable format.)} 
\end{deluxetable*}

\subsection{Comments on Individual Objects}

The results of PyEMILI's line identification for the test sample are summarized in Table\,\ref{testsample}, where for each object, the number percentage of the emission lines whose manual IDs are also assigned the ``A'' ranking by PyEMILI is presented.  For the absolute majority of the objects, in particular the PNe in the sample, the line identification results of PyEMILI in overall have agreement rates exceeding 90\% with the manual IDs.  For each PN or \ion{H}{ii} region, we believe that the actual reliability of PyEMILI's identification is probably even higher, given that manual identification of huge numbers of spectral lines is more prone to be incomplete than the systematic identification by machines, and may inevitably bring uncertainties \citep{Sharpee_2003,Fang_2011}.  In this section, we comment on specific objects in our test sample, and carefully examine the manual IDs of the emission lines detected in them.  All the observed wavelengths of emission lines have been corrected for radial velocities as derived through measurements of the observed wavelengths of H$\beta$.

\subsubsection{NGC\,3918} \label{NGC3918}

NGC\,3918 is a highly excited PN whose optical spectrum exhibits numerous CELs across multiple ionization stages.  Additionally, \citet{2015MNRAS.452.2606G} identified emission lines from various refractory elements (Ca, K, Cr, Mn, Fe, Co, Ni, and Cu) and several neutron-capture elements (Se, Kr, Rb, and Xe).  The utilization of the UVES echelle spectrograph on the 8.2\,m VLT produces very deep spectra of NGC\,3918 and enables the detection of more than 750 emission lines, intensifying the intricacy of manual line identification for this PN.  Moreover, the expansion velocity field of the PN, accompanied by such high spectral resolution, is reflected in the line profile pattern, transitioning from a double-peaked profile in the lowest ionization species to a simpler profile in the highest ionization species.

Nevertheless, our test of PyEMILI using default parameters demonstrated a high level of agreement with the manual IDs assigned by \citet{2015MNRAS.452.2606G}.  This agreement rate has considered the inclusion of transitions not present in PyEMILI's atomic transition database, such as the \ion{He}{ii} transitions with the upper-level principal quantum number $n>50$ and neutron-capture elements with atomic number $Z>36$ (Rb, Xe).  If these transitions (not included in PyEMILI) are not taken into account, the agreement rate can be higher.  Notably, in the manual identifications of \citet{2015MNRAS.452.2606G}, the faintest \ion{He}{ii} lines are from $n\,=67$ to 5. Hence 17 \ion{He}{ii} lines in NGC\,3918 remain unidentified by PyEMILI.

Certain manual IDs encompass very weak transitions, which we deem unreliable. For instance, the manual ID of \ion{N}{ii} $\mathrm{\lambda3230.54\ 4p\,{^3D}\text{--}10s\,{^3P^o}}$, received a ranking of `D' by PyEMILI for the observed line at 3230.12\,\AA. Conversely, the candidate \ion{Ne}{ii} $\mathrm{\lambda3230.07\ 3s\,{^2D}\text{--}3p\,{^2D^o}}$, was ranked as ``A'' for the same observed line, attributed to its more substantial predicted line intensity and smaller wavelength difference. In such instances, \ion{Ne}{ii} $\mathrm{\lambda3230.07}$ is considered more reliable and is incorporated into Table~\ref{amendedlinelist} with reasons denoted as `ERC' and `Wav' (see Table~\ref{amendedlinelist} for more details).

Other notable weak transitions include electric quadrupole transitions [\ion{Mn}{iv}] $\mathrm{\lambda\lambda4643.3,4707.22}$, which should not be deemed possible identifications for the observed lines due to their high upper-level energies. The upper-level energies for both [\ion{Mn}{iv}] transitions surpass 150,000 $\mathrm{cm^{-1}}$, equivalent to approximately 19 eV. Such forbidden lines, characterized by such high upper-level energies, are very faint under nebular electron temperatures (for NGC\,3918, the electron temperature $T_\mathrm{e}$ of the forbidden lines is approximately 15,000 K \citep{2015MNRAS.452.2606G}). 

Within some of the reliable manual IDs, PyEMILI found an additional potential blending transition within the observed line. For instance, the manual ID \ion{He}{i} $\mathrm{\lambda3756.1\ 2p\,{^1P^o}\text{--}14d\,{^1D}}$, for the observed line at 3756.14\,{\AA}, is considered reliable as other \ion{He}{i} lines with the transitions $\mathrm{2p}\text{--}n\mathrm{d}\ (n>14)$, were also detected. However, PyEMILI assigned the highest rank to [\ion{Fe}{v}] $\mathrm{\lambda3755.7}$, based on its strong predicted line intensity and the detection of five additional multiplet members, all assigned the same ranking of ``A''.  Consequently, [\ion{Fe}{v}] $\mathrm{\lambda3755.7}$ is also a potential identification for this observed line.  Considering the observed line intensities of its multiplet members, the line intensity of [\ion{Fe}{v}] may be comparable to or even stronger than the \ion{He}{i} line. 

However, it should be noted that those amended and supplemental identifications in Table\,\ref{amendedlinelist} do not take into account emission-line profiles, which can be associated with the ionization energy of the line-emitting ions.  In the VLT/UVES spectrum of NGC\,3918, the observed line at 3756.14\,{\AA} has a double-peak line profile, indicating a lower ionization species (i.e.\ our identification of [Fe\,{\sc v}] might be questionable, because this high-ionization species exists in the central region of the PN and is expected to exhibit single-peak emission).

\subsubsection{NGC\,6153}

The limited spectral resolution of NGC\,6153 introduces considerable wavelength uncertainty in line identification, and contributes to instances of line blending, resulting in a predominant source of discrepancies between the manual IDs and those assigned by PyEMILI.  For example, the manual ID for the observed line at 3045.8\,{\AA}, is \ion{O}{iii} $\mathrm{\lambda3047.1 3s\,{^3P^o}\text{--}3p\,{^3P}}$, featuring a velocity difference of up to 120 $\mathrm{km\,s^{-1}}$. In this case, identification by PyEMILI for this manual ID is hindered since the wavelength difference exceeds 5$\sigma$ of the input wavelength uncertainty (20 $\mathrm{km\,s^{-1}}$ set as 1$\sigma$ for this object). A larger input wavelength uncertainty is avoided to prevent false positives, 
ensuring the overall reliability of line identifications. Nevertheless, PyEMILI also calculates the predicted line intensity of candidates within the wavelength uncertainty range of 5--10$\, \sigma$, with the strongest candidate incorporated into the final row of candidates for this observed line in the PyEMILI's output, serving as a reference. The manual ID of \ion{O}{iii} $\mathrm{\lambda3047.1}$, serves as a typical example of this case (refer to the online file \textsc{NGC6153.out} for additional details).

\begin{figure*}
\begin{center}
\includegraphics[width=13.5cm,angle=0]{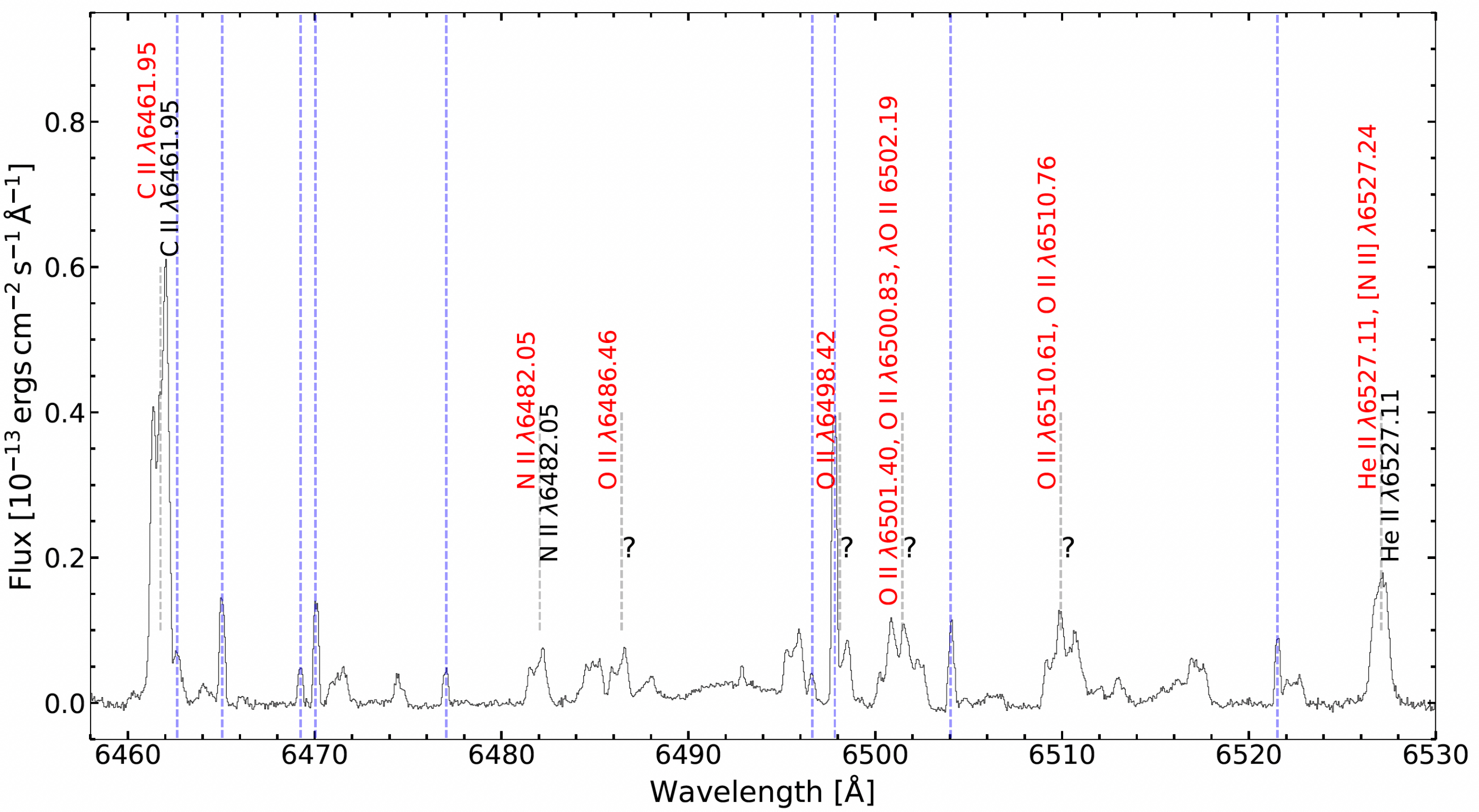}
\caption{VLT/UVES echelle spectrum of NGC\,6153 showing the faint nebular emission lines identified by PyEMILI (this work, labeled with red IDs) and also detected and identified in the deep spectroscopy of \citet[][labeled with black IDs]{2000MNRAS.312..585L}.  The black ``?'' indicates the line was unidentified in \citet{2000MNRAS.312..585L}.  The spectrum has been velocity-corrected through the measurements of the \ion{H}{i} Balmer lines.  Sky emission lines are marked by blue vertical-dashed lines.} 
\label{fig3}
\end{center}
\end{figure*}

Given that the manual identifications were conducted two decades ago, some observed emission lines remained unidentified due to the absence of comprehensive atomic transition data (see e.g.\ Figure\,\ref{fig3}).  The advancement in the atomic transition database has facilitated the identification of previously unidentified lines. Notably, an intriguing discovery emerged, wherein the PyEMILI analysis proposed that certain unidentified observed lines in NGC\,6153 are likely attributed to \ion{O}{ii} transitions from the 4f--6g electron configuration, owing to the remarkably high ionic abundance of $\mathrm{O^{2+}}/\mathrm{H^{+}}$ in this object.

The observed emission lines at wavelengths 6486.43\,\AA, 6497.64\,\AA, 6501.46\,\AA, and 6510.89\,\AA, have been identified by PyEMILI as \ion{O}{ii} candidate transitions with ``A'' ranking, corresponding to \ion{O}{ii} $\mathrm{\lambda6486.46}$, $\mathrm{\lambda6498.42}$, $\mathrm{\lambda6501.40}$, and $\mathrm{\lambda6510.61}$, respectively. Specifically, the line at 6501.46\,\AA\ may also exhibit a blend with \ion{O}{ii} $\mathrm{\lambda6500.83}$ and $\mathrm{\lambda6502.19}$, while the line at 6510.89\,\AA\ may be associated with \ion{O}{ii} $\mathrm{\lambda6510.76}$. All identified \ion{O}{ii} transitions mentioned above originate from the 4f--6g electron configuration, except for \ion{O}{ii} $\mathrm{\lambda6510.76\ 3d\,{^2F}\text{--}4p\,{^2F^o}}$. 

To further validate the reliability of these \ion{O}{ii} candidate transitions, we determined the $\mathrm{O^{2+}}/\mathrm{H^{+}}$ abundances using other manually identified \ion{O}{ii} ORLs. Subsequently, we utilized the emissivities of these \ion{O}{ii} candidate transitions to correspond to the four observed lines in question. By multiplying by the obtained ionic abundances, we calculated the theoretical line intensities of these four observed lines, revealing a general agreement with the observed line fluxes. As a result, we assert that these four observed lines are highly likely associated with these \ion{O}{ii} candidate transitions.

However, due to limitations in the wavelength coverage of the spectrograph used in the literature (3000 to 7400\,\AA), the detection of \ion{O}{ii} 5f--6g or 4f--5g transitions (with wavelength typically greater than 9500\,\AA), which are relative to the 4f--6g transitions, is not feasible. In the event of high-resolution observations in the near-infrared (NIR) band for this object, we anticipate the potential detection of \ion{O}{ii} 5f--6g or 4f--5g transitions.

\subsubsection{\ion{H}{ii} Regions and HH Objects}

Upon examination, the line list of 30 Doradus, processed through PyEMILI, shows a relatively lower agreement rate with those candidate IDs proposed by PyEMILI. PyEMILI encounters challenges in furnishing reasonable identifications for several emission lines marked by uncertain manual IDs and many unidentified lines. Upon critical scrutiny, we cast doubt on the reliability of some manual IDs within this line list. We are skeptical about the authenticity of many weak lines in the 30 Doradus line list, accompanied by concerns regarding potentially significant flux errors associated with these weak lines. Notably, 11 emission lines in the line list are manually identified as \ion{Ne}{i}, with the strongest transition being \ion{Ne}{i} $\mathrm{\lambda5326.40\ 3p\,{^{3/2}[1/2]}\text{--}4d\,{^{3/2}{[3/2]}^o}}$. Intriguingly, the observed flux for this strongest transition is merely 0.019 (relative to $I_\mathrm{H\beta}=100$), contrasting sharply with other theoretically weaker \ion{Ne}{i} transitions that exhibit the same or even higher observed fluxes in the line list. This inconsistency points to a possible issue with the manual IDs and fluxes assigned to these emission lines.

In contrast, the situation with the Orion Nebula is more favorable, as evidenced by a relatively satisfactory agreement rate with the extensive line list. The total number of input emission lines is 555, with 13 lines still lacking identifications in the literature. Of the emission lines with manual IDs, 46 were not ranked as ``A'' by PyEMILI. However, upon careful examination, we reassessed 43 of these lines, including both initially unidentified lines and those without an ``A'' ranking in PyEMILI, resulting in the provision of more plausible results. Further details are presented in Table~\ref{amendedlinelist}.

In the context of HH objects, we adopted two distinct line lists associated with HH\,204\footnote{The HH objects in our sample belong to the photoionized HH objects in the Orion nebula, where the strong radiation field dominates over shock excitation.} as the test samples. These include the combined spectrum and the cut 1 spectrum. The cut 1 spectrum contains emission lines originating from three different regions---HH\,204, the Orion Nebula, and the diffuse blue layer, as detailed by \citet{2021ApJ...918...27M}. For our analysis, we specifically extracted the lines corresponding to HH\,204 from this spectrum. The input wavelengths for the combined spectrum align with the laboratory wavelengths of manual IDs, given that observed wavelengths were not available. In instances where laboratory wavelengths are provided, we assigned a wavelength uncertainty of 10 $\mathrm{km\,s^{-1}}$ as a standard parameter in PyEMILI.

The overall agreement between manual IDs of HH\,204 and those ``A'' ranking candidates provided by PyEMILI is satisfactory. PyEMILI exhibits effective identification capabilities, particularly for various forbidden lines such as [\ion{Fe}{ii}], [\ion{Fe}{iii}], and [\ion{Ni}{ii}].  The literature table designates uncertain manual IDs with a question mark ``?'', mainly consisting of intercombination lines (e.g., \ion{Ca}{i}], \ion{Fe}{ii}], and \ion{Cr}{ii}]), whose authenticity we have also questioned. According to the output of PyEMILI, several of these uncertain lines can be amended with other high-confidence candidates. For instance, the dubious manual ID of \ion{Ca}{i}] $\mathrm{\lambda9260.94\ 46d\,{^1D}\text{--}8p\,{^3D^o}}$, is replaced by \ion{O}{i} $\mathrm{\lambda9260.85\ 3p\,{^5P}\text{--}3d\,{^5D^o}}$.

Furthermore, PyEMILI identified some \ion{Fe}{ii} permitted transitions in HH\,204 that were previously seldom identified in HH objects.  In particular, transitions such as \ion{Fe}{ii} $\mathrm{\lambda9095.12,\lambda9204.10,\lambda9417.32}$ are all originating from the 5s--5p transition within the same multiplet and attained an ``A'' ranking by PyEMILI.  We speculate that these transitions may be excited through resonance fluorescence, although confirmation needs consideration of electron temperature, electron density, and the ambient radiation field.  Past studies, such as that by \citet{hh-fe1}, have identified a dozen \ion{Fe}{ii} permitted transitions in the UV spectra of HH\,47A, by examining observed intensities and corresponding transition probabilities within the same multiplet for identifications.

Regarding another sample of HH object, namely the jet component of HH\,514, the predominant emission lines within the jet base consist of the iron forbidden lines, e.g., [\ion{Fe}{ii}] and [\ion{Fe}{iii}].  The agreement rate for the HH\,514 jet base approaches nearly 100\%, with the exception of the observed line at 6813.91\,\AA, annotated in the literature as [\ion{Ni}{ii}] $\mathrm{\lambda6813.57}$.  This manual ID is designated with a ``B'' ranking due to none of the four multiplet members associated with this manual ID being detected.

\subsubsection{PNe with [WR]-type Central Stars}

In their study, \citet{2012AA...538A..54G} provided twelve high-resolution spectra of PNe featuring central stars with [WR] characteristics. Each object has an individual line list in the literature, and PyEMILI was employed to identify all line lists within these objects. The input wavelengths were derived from the observed wavelengths corrected for the velocity difference of the $\mathrm{H\beta}$ line of the respective objects. A uniform wavelength uncertainty of 15 $\mathrm{km,s^{-1}}$ was assigned across the samples, considering the high-resolution nature of the spectra.

The line identifications generated by PyEMILI for the twelve [WR] PNe are deemed satisfactory, with all agreement rates objectively exceeding 90\%. However, a notable issue arises with the majority of unidentified observed lines, indicating the difficulty in obtaining reasonable identifications through PyEMILI.  We attribute this issue to the substantial uncertainties in measured fluxes of faint lines and the potential presence of spurious emission line features in the spectra. It is also possible that these lines come from molecular transitions, and further identifications are needed.

For instance, the observed line at 3693.49\,\AA\ in M1-25, with an observed line flux of 0.03 (relative to $I_\mathrm{H\beta}=100$), remains unidentified in the literature. PyEMILI designates the ranking ``A'' and ``B'' candidates for this line as \ion{He}{i} $\mathrm{\lambda3693.26\ 2p\,{^1P^o}\text{--}32d\,{^1D}}$ and \ion{Ne}{ii} $\mathrm{\lambda3693.39\ 3d\,{^4F}\text{--}5p\,{^4D^o}}$, respectively. The conditions of these two transitions are not sufficient to be reliable candidates. For \ion{He}{i} transitions from 2p--$n$d, the highest $n$ with manually identified transitions in this object is only $n=15$ (\ion{He}{i} $\mathrm{\lambda3478.97\ 2p\,{^3P^o}\text{--}15d\,{^3D}}$), making the likelihood of the transition with 2p--32d very low. Similarly, \ion{Ne}{ii} is improbable due to its low predicted line intensity. Even when considering the potential blending of these two transitions, the combined line intensity does not match the observed magnitude.

Another illustrative example is the unidentified line at 4032.14\,\AA\ in M1--61. Although PyEMILI proposes a highly probable identification of \ion{O}{ii} $\mathrm{\lambda3693.39\ 3d\,{^4F}\text{--}4f\,{F[4]^o}}$, the predicted line intensity of this \ion{O}{ii} transition relative to the corresponding observed line flux is notably lower compared to other \ion{O}{ii} transitions in this object with the same 3d--4f electron configuration. Consequently, such observed lines lacking robust candidates are not included in Table~\ref{amendedlinelist}.

Regarding the observed lines with manual IDs in the literature, there is also an issue associated with the velocity differences for certain manual IDs.  Within the line list of M3-15, the manual IDs for the two observed lines at 3703.87\,{\AA} and 3712.08\,{\AA} were \ion{He}{i} $\mathrm{\lambda3705.00\ 2p\,{^3P^o}\text{--}7d\,{^3D}}$ and \ion{Ne}{ii} $\mathrm{\lambda3713.08\ 3s\,{^2P}\text{--}3p\,{^2D^o}}$, respectively. Despite the line intensities of these transitions, their laboratory wavelengths deviate significantly from the observed values. More plausible identifications for these two observed lines would be \ion{H}{i} $\mathrm{\lambda3703.85\ 2\text{--}16}$, and \ion{H}{i} $\mathrm{\lambda3711.97\ 2\text{--}15}$.

In the line list for the same object, two manual IDs of \ion{Ne}{ii} $\mathrm{\lambda\lambda4391.94,4409.30}$ exhibit a velocity difference of approximately -40 $\mathrm{km\,s^{-1}}$ relative to their corresponding observed lines. In contrast, the velocity difference of $\mathrm{H\beta}$ to its observed line is 96.2 $\mathrm{km\,s^{-1}}$. Such substantial velocity differences for these two \ion{Ne}{ii} lines compared to $\mathrm{H\beta}$ imply their origin from a distinct component of the object. Apparently, identifying such observed lines by correcting the velocity difference of $\mathrm{H\beta}$ is unreliable.

In certain line lists in the literature, writing errors exist. For example, in the line lists of M1--25 and M1--30, the observed lines 4294.89 and 4294.8 should be \ion{O}{ii} $\mathrm{\lambda4294.78\ 3d\,{^4P}\text{--}4f\,{D[2]^o}}$, instead of \ion{S}{ii} $\mathrm{\lambda4294.78}$ as stated in the literature. Furthermore, an erroneous transition is present in the line list of NGC\,5189, where the observed line at 5677.54\,\AA, was incorrectly attributed to \ion{N}{ii} $\mathrm{\lambda5677.66}$ of multiplet 3, which does not contain a transition at the wavelength 5677.66\,\AA. Nevertheless, PyEMILI could not offer a reasonable candidate for this observed line, leading us to question its authenticity.

\subsubsection{NGC\,6543 and NGC\,2022}

The spectrum of NGC\,6543, as presented by \citet{2004MNRAS.351.1026W}, encounters analogous issues to other low-resolution spectra, characterized by substantial uncertainties in the observed wavelengths. Striking a suitable balance for the wavelength uncertainty is a critical point in line identification, as setting it too high introduces numerous false positive terms while setting it too low compromises the accuracy of the results. This constitutes a primary drawback when utilizing PyEMILI for line lists derived from low-resolution spectra.

As an illustration, the observed lines at 6713.04\,\AA\ and 6727.41\,\AA\ were manually identified as \ion{S}{ii} $\mathrm{\lambda\lambda6716.44,6730.82}$. Despite the considerable wavelength differences, both lines exhibit predicted line intensities exceeding one per cent of $\mathrm{H\beta}$. Moreover, there are no other prominent transitions in proximity to these wavelengths. Consequently, we confidently affirm that they are indeed \ion{S}{ii} lines. However, such substantial differences pose challenges for PyEMILI's fully automated processing. Within the NGC\,6543 line list, seven lines presenting similar issues contribute to discrepancies between line identifications of PyEMILI and the manual counterparts.

In a prior study, \citet{2003MNRAS.345..186T} analyzed 15 PNe. However, after testing the samples with PyEMILI, we found many issues with the manual line identifications. Thus we present results solely for NGC\,2022 as an illustrative example. The utilization of low-resolution spectra with substantial wavelength uncertainties renders the identification of emission lines more intricate.

To illustrate the manual IDs for NGC\,2022 is not self-consistent, we offer two examples. Firstly, consider the observed line at 4025.32\,\AA, with the manual ID \ion{He}{i} $\mathrm{\lambda4026.21\ 2p\text{--}5d}$. Two compelling reasons led us to question the appropriateness of the \ion{He}{i} identification: (1) The wavelength difference between the observed line and \ion{He}{i} $\mathrm{\lambda4026.21}$ exceeds 60 $\mathrm{km\,s^{-1}}$, whereas the wavelength differences for all other \ion{He}{i} lines in this line list are generally less than 10 $\mathrm{km\,s^{-1}}$. (2) The observed line flux of the theoretically stronger line \ion{He}{i} $\mathrm{\lambda4471.50\ 2p\text{--}4d}$ in this line list is 0.687 (relative to $I_\mathrm{H\beta}=100$), whereas the observed line flux of the line at 4025.32\,\AA\ is 1.59, even exceeding that of \ion{He}{i} $\mathrm{\lambda4471.50}$ by twofold. The correct line identification for this observed line 4025.32\,\AA, should be \ion{He}{ii} $\mathrm{\lambda4025.6\ 4\text{--}13}$, consistent with other observed \ion{He}{ii} lines in terms of wavelength differences and observed line fluxes.

Similarly, consider the observed line at 4714.42\,\AA, with the manual ID \ion{He}{i} $\mathrm{\lambda4713.22\ 2p\text{--}4s}$. Due to the larger wavelength difference and inconsistently observed line flux relative to other \ion{He}{i} 2p--$n$s transitions in the line list, we propose the correct line identification as [\ion{Ne}{iv}] $\mathrm{\lambda4714.25}$, which is self-consistent with the two multiplet members (\ion{Ne}{iv}] $\mathrm{\lambda\lambda4724.15,4725.57}$) identified with similarly observed fluxes.

\subsubsection{Nine Galactic PNe}

The utilization of PyEMILI with the nine Galactic PNe spanning from H\,1--40 to M\,2--31 reported by \citet{2018MNRAS.473.4476G}, yielded exemplary test results, with all samples achieving an agreement rate surpassing 95\%. A promising result is the capacity of PyEMILI to provide appropriate identifications for many previously unidentified observed lines in M\,1--31, M\,1--33, and M\,1--60. The majority of these identifications pertain to \ion{N}{ii} and \ion{O}{ii} from the 4f--5g transitions within the NIR band. For instance, the transition \ion{N}{ii} $\mathrm{\lambda10035.44\ 4f\,{G[9/2]}\text{--}5g\,{H[11/2]^o}}$, 
is suggested by PyEMILI to be the most likely candidate of the corresponding observed line, which was unidentified in all these three PNe. Further details can be found in Table~\ref{amendedlinelist}.

\subsubsection{NGC\,5315}

The NGC\,5315 is the one sample with wavelength coverage extending up to 25000\,\AA. Nevertheless, due to the wavelength limitations of the atomic transition database in PyEMILI, we have only tested the line list with wavelengths within 20,000\,\AA. To facilitate distinct identifications, we segregated the two line lists extracted from the literature, which cover the optical and NIR bands, into separate samples. This division was necessary due to the different spectrographs used in these two bands.

Regarding the sample in the optical band, it comprises numerous very faint lines with observed line fluxes $I_\mathrm{obs}<0.01$, relative to $I_\mathrm{H\beta}=100$. The reliability of manual IDs for these faint lines, as well as the corresponding high-ranking candidates suggested by PyEMILI, cannot be assured. Consequently, only a limited number of emission lines, for which we deemed PyEMILI's candidates more appropriate, have been amended and supplemented. Despite these challenges, the overall agreement is considered satisfactory, and PyEMILI successfully identifies several n-capture element ions such as [\ion{Kr}{iii}] $\mathrm{\lambda6826.7}$, [\ion{Kr}{iv}] $\mathrm{\lambda\lambda5346.02,5867.74}$, and [\ion{Se}{iii}] $\mathrm{\lambda8854.0}$. Notably, in the case of the observed line at 6555.99\,\AA, for which \citet{2017MNRAS.471.1341M} expressed doubt regarding the authenticity of the [\ion{Br}{iii}] $\mathrm{\lambda6555.56}$ identification, PyEMILI proposed more plausible IDs---several \ion{O}{ii} 4f--6g transitions (refer to Table~\ref{amendedlinelist}).

For the line list of NIR band samples, no adjustments were made to the observed wavelengths. The originally observed wavelengths were employed by PyEMILI for line identifications, and the outcomes illustrate that this also allows the determination of velocity differences between numerous emission lines and their respective laboratory wavelengths of the real IDs, yielding favorable line identification results. The velocity differences between the observed wavelengths and the laboratory wavelengths of the real IDs amount to approximately 50 $\mathrm{km\,s^{-1}}$. This discrepancy is not markedly distinct from the initially input 1$\sigma$ wavelength uncertainty (specified as 30 $\mathrm{km\,s^{-1}}$), thus making the automatic correction of this velocity difference feasible.  The emission lines within the NIR band are mainly emitted by \ion{H}{i}, \ion{He}{i}, and [\ion{Fe}{ii}], all of which can be effectively identified, except for certain lines with excessively large wavelength differences from their real IDs.

\section{Plasma Diagnostics with Optical Recombination Lines} 
\label{diagnostic}

We apply the MCMC methodology, as outlined in Section\,\ref{MCMC}, in the plasma diagnostic analyses of PNe in our sample based on the \ion{O}{ii} and \ion{N}{ii} ORLs.  12 PNe from our test sample (as introduced in Section\,\ref{test sample}) are analyzed in this diagnostic effort, supplemented with one additional object, Hf\,2-2, an archetypal PN renowned for high abundance discrepancy factor (ADF) values \citep{2006MNRAS.368.1959L}; these objects are among the Galactic PNe with the best measurements of the \ion{N}{ii} and \ion{O}{ii} ORLs.  Our selection of PNe gives preference to the objects that have a substantial number of \ion{O}{ii} and/or \ion{N}{ii} lines with high-quality measurements (reliable line fluxes with moderate or small errors) in the published emission-line lists. 

In conjunction with the prescribed selection criteria applied to the sample, additional criteria are imposed on the \ion{O}{ii} and \ion{N}{ii} ORLs within each sample. Firstly, ORLs with flux errors exceeding 40 per cent of observed line fluxes are excluded. Secondly, rigorous filtering is applied to eliminate ORLs with possible line blending and those presenting significant contributions from fluorescence excitation (refer to Section\,\ref{ORL select}).

In our MCMC sampling, all three parameters---electron temperature, electron density, and ionic abundance are treated as free parameters. The sampling ranges for electron temperature and electron density are derived from the literature values. The ionic abundance is sampled within the range of 0.01 to 100 times the corresponding solar elemental abundance. The velocity parameter $\Delta v$ (see section \ref{MCMC}) is uniformly set at 20 km\,s$^{-1}$ for each ion of every object, unless explicitly specified otherwise.

\subsection{Recombination Lines Selection} 
\label{ORL select}

\subsubsection{Fluorescence Excitation} 
\label{Fluorescence excitation}

The observed line fluxes associated with the permitted transitions of \ion{O}{ii} and \ion{N}{ii} may include contributions not only from the recombination process but also from fluorescence excitation \citep[e.g.][]{1976ApJ...206..658G,2012MNRAS.426.2318E} and charge exchange \citep[e.g.][]{1986MNRAS.221P..61C,1993MNRAS.261..465L,1993MNRAS.262..699L}. The computations of effective recombination coefficients, however, exclusively account for the recombination process. Consequently, it becomes imperative to exclude those ORLs with substantial fluorescence contributions. 

As highlighted by \citet{2012MNRAS.426.2318E}, the PN IC\,418, characterized by low-excitation conditions, exhibits a typical scenario. In this instance, \ion{N}{ii} ORLs originating from the low-lying 3s--3p and 3p--3d states are predominantly excited by fluorescence mechanism, constituting approximately 80 per cent of the observed line fluxes. It is noteworthy that the fluorescence mechanism has a limited influence on populating higher angular momentum states, such as f and beyond g, h, etc., within the ion $\mathrm{N^+}$. Consequently, the 4f and other higher states are primarily populated through the recombination process, rendering them appropriate for inclusion in our test samples.

\citet{2013A&A...558A.122G} calculated the ionic abundances of $\mathrm{N^{2+}/H^+}$ for several PNe with different excitation degree, utilizing \ion{N}{ii} ORLs from the M3 (of the 3s--3p transition array), M39 and M48 (of the 3d--4f transition array) multiplets.  The results clearly show that in low-excitation PNe, the ionic abundances of $\mathrm{N^{2+}/H^+}$ derived from the M3 lines systemically exceed those obtained from the M39 and M48 lines.  However, this relation is not evident in high-excitation PNe, such as Hb\,4 and NGC\,5189 \citep{2013A&A...558A.122G}.  Conversely, for another high-excitation PN, NGC\,3918, as reported by \citet{2015MNRAS.452.2606G}, the ionic abundance values of $\mathrm{N^{2+}/H^+}$ derived from the M39 and M48 nebular lines are even higher than those determined with the M3 lines. 

Furthermore, \citet{2006MNRAS.368.1959L} computed $\mathrm{N^{2+}/H^+}$ for Hf\,2-2 with high-excitation class, utilizing \ion{N}{ii} ORLs from the M3 multiplet and those from the 3d--4f transitions.  Analogously, the ionic abundances derived from the 3s--3p and 3d--4f transitions do not exhibit systematic differences.  Consequently, the fluorescence excitation effects for \ion{N}{ii} in high-excitation class PNe are considered insignificant.  Thus all available \ion{N}{ii} ORLs will be adopted under such nebular conditions, i.e.\ $EC\geq$8.

Concerning the \ion{O}{ii} ORLs, \citet{2012MNRAS.426.2318E} stated that fluorescence excitation does not have a dominant influence on their observed line fluxes in low-excitation PNe.  On average, only about 15 per cent of the observed line flux is contributed by fluorescence excitation.  Therefore, we consider the impact of fluorescence excitation on the \ion{O}{ii} ORLs to be negligible and include all available \ion{O}{ii} ORLs in our test samples.

\subsubsection{Line Blending}

The condition of line blending is an inevitable occurrence, even in a high-resolution spectrum. In the situation of unresolved blended lines, the separation of these blended lines proves to be an elusive task unless precise theoretical calculations of the individual components are carried out. However, in comparison to the existing atomic transition database, only a limited number of recombination lines have undergone the computation of effective recombination coefficients in prior works. Consequently, several recombination lines, identified as being strongly blended with other lines through PyEMILI's line identifications and pertinent literature, have been excluded from our samples.

The eliminated recombination lines are presented in Table~\ref{blendion}. Given that \ion{O}{ii} ORLs exhibit greater prevalence and line fluxes in the optical spectra of most PNe in contrast to \ion{N}{ii} ORLs, the identification of \ion{O}{ii} ORLs would be easier, even in the presence of line blending. 

Concerning the influence of fluorescence excitation in low-excitation PNe, results in the unavailability of numerous \ion{N}{ii} ORLs from the strongest multiplets, thereby complicating the plasma diagnostics with \ion{N}{ii} ORLs. Furthermore, \ion{N}{ii} ORLs originating from 4f and higher states exhibit even fainter line intensities than those from 3d or 3p in spectra of PNe. Hence, We endeavor to retain a maximal number of usable \ion{N}{ii} ORLs, arguing that even in the presence of line blending, the impact on the line fluxes is not significant.

\begin{deluxetable}{cccccc}[h]
\label{blendion}
\tablecaption{Optical Recombination Lines Excluded from Plasma Diagnostics due to Possible Blended Components} 
\tablehead{\colhead{Ion} & \colhead{$\lambda_0$} & \colhead{Mult.} & \colhead{Blended Ion} & \colhead{$\lambda_0$} & \colhead{Mult.} \\ 
\colhead{} & \colhead{({\AA})} & \colhead{} & \colhead{} & \colhead{({\AA})} & \colhead{} } 
\startdata
\ion{O}{ii} & 3856.13 & M12 & \ion{Si}{ii} & 3856.02 & M1 \\
\ion{O}{ii} & 4097.26 & M48b & \ion{N}{iii} & 4097.33 & M1 \\
\ion{O}{ii} & 4120.54 & M20 & \ion{He}{i} & 4120.81 & M16 \\
\ion{O}{ii} & 4156.53 & M19 & \ion{C}{iii} & 4156.50 & M21 \\
\ion{O}{ii} & 4169.22 & M19 & \ion{He}{i} & 4168.97 & M52 \\
\ion{O}{ii} & 4638.86 & M1 & \ion{C}{ii} & 4638.92 & M12.01 \\
\ion{O}{ii} & 4641.81 & M1 & \ion{N}{iii} & 4641.85 & M2 \\
\enddata 
\end{deluxetable}

\startlongtable
\begin{deluxetable*}{llcccrc}
\tablecaption{Plasma Diagnostics and Ionic Abundance Determinations for PNe Using the \ion{O}{ii} and \ion{N}{ii} Optical Recombination Lines (ORLs)
\label{DIAGNOSTICS}} 
\setlength{\tabcolsep}{0.5cm}
\tablehead{\colhead{Object} & \colhead{ORLs} & \colhead{$\log{T_{\rm e}}$} \hspace{0.5cm} & \colhead{$\log{N_{\rm e}}$} \hspace{0.5cm} & \colhead{12+$\log{({\rm X}^{2+}/{\rm H}^{+})}$} & \colhead{\hspace{0.5cm}Ref.} & ADF$^{a}$\\ 
\colhead{} & \colhead{} & \colhead{(K)} & \colhead{(cm$^{-3}$)} & \colhead{(X=O,\,N)} & \colhead{} & \colhead{}} 
\startdata
IC\,4776	 & \ion{O}{ii} & $3.33_{-0.02}^{+0.02}$ & $4.12_{-0.14}^{+0.26}$ & $8.81_{-0.01}^{+0.01}$ & (1) & 1.75\\
 &  & $3.55_{-0.35}^{+0.31}$ & $3.45_{-0.24}^{+0.18}$ & $8.78_{-0.03}^{+0.02}$ &  & \\
 & \ion{N}{ii} & $2.44_{-0.16}^{+0.23}$ & $2.74_{-0.35}^{+0.51}$ & $7.30_{-0.18}^{+0.26}$ &  & \\
 &  &  &  & $8.24_{-0.03}^{+0.03}$ &  & \\
Hf\,2-2 (2 arcsec)& \ion{O}{ii} & $2.90_{-0.21}^{+0.20}$ & $3.08_{-0.09}^{+0.10}$ & $9.78_{-0.06}^{+0.08}$ & (2) & 83\\
 &  & $2.80$ &$3.69$  & $9.87$ &  & \\
 & \ion{N}{ii} & $2.37_{-0.15}^{+0.20}$ & $2.63_{-0.22}^{+0.38}$ & $9.00_{-0.15}^{+0.20}$ &  & \\
 &  &  &  & $9.43$ &  & \\
Hf\,2-2 (4 arcsec)& \ion{O}{ii} & $2.44_{-0.09}^{+0.18}$ & $3.08_{-0.24}^{+0.51}$ & $9.46_{-0.10}^{+0.16}$ & (2) & \\
 &  & $2.80$ &$3.60$  & $9.85$ &  & \\
Hf\,2-2 (8 arcsec)& \ion{O}{ii} & $2.46_{-0.06}^{+0.13}$ & $>3.0$ & $9.48_{-0.06}^{+0.11}$ & (2) & \\
 &  & $2.80$ &$4.13$  & $9.84$ &  & \\
 & \ion{N}{ii} & $2.41_{-0.16}^{+0.26}$ & $3.03_{-0.42}^{+0.68}$ & $9.08_{-0.20}^{+0.27}$ &  & \\
 &  &  &  & $9.42$ &  & \\
NGC\,6153 (minor) & \ion{O}{ii} & $3.08_{-0.01}^{+0.01}$ & $>4.2$ & $9.54_{-0.01}^{+0.01}$ & (3) & 10\\
 &  &  &  & $9.61_{-0.03}^{+0.03}$ &  & \\
 & \ion{N}{ii}$^{a}$ & $3.51_{-0.09}^{+0.09}$ & $>3.9$ & $9.14_{-0.01}^{+0.01}$ &  & \\
 &  &  &  & $9.24$ &  & \\
NGC\,6153 (entire)& \ion{O}{ii} & $3.03_{-0.01}^{+0.01}$ & $>3.9$ & $9.49_{-0.01}^{+0.01}$ & (3) & \\
 &  &  &  & $9.61_{-0.03}^{+0.03}$ &  & \\
 & \ion{N}{ii}$^{a}$ & $3.25_{-0.17}^{+0.12}$ & $>3.2$ & $9.09_{-0.03}^{+0.02}$ &  & \\
 &  &  &  & $9.23$ &  & \\
NGC\,3918 & \ion{O}{ii} & $3.30_{-0.02}^{+0.02}$ & $>3.8$ & $8.69_{-0.01}^{+0.01}$ & (4) & 1.85\\
 &  &  &  & $8.72_{-0.07}^{+0.05}$ &  & \\
 & \ion{N}{ii}$^{b}$ & $3.41_{-0.37}^{+0.28}$ & $>3.0$ & $8.04_{-0.04}^{+0.03}$ &  & \\
 &  &  &  & $8.00_{-0.05}^{+0.04}$ &  & \\
M\,1-30 & \ion{O}{ii} & $3.20_{-0.05}^{+0.05}$ & $3.68_{-0.20}^{+0.37}$ & $8.91_{-0.02}^{+0.02}$ & (5) & 2.4\\
 &  & $>3.78$ &  & $8.91_{-0.02}^{+0.02}$ &  & \\
PC\,14 & \ion{O}{ii} & $3.50_{-0.15}^{+0.16}$ & $3.51_{-0.15}^{+0.19}$ & $9.05_{-0.02}^{+0.02}$ & (5) & 1.9\\
 &  & $3.46_{-0.51}^{+0.54}$ &  & $9.03_{-0.04}^{+0.03}$ &  & \\
He\,2-86 & \ion{O}{ii} & $3.25_{-0.02}^{+0.02}$ & $>3.9$ & $9.07_{-0.01}^{+0.01}$ & (5) & 1.9\\
 &  & $3.79_{-0.25}^{+0.22}$ &  & $9.02_{-0.02}^{+0.02}$ &  & \\
H\,1-50 & \ion{O}{ii} & $3.61_{-0.13}^{+0.13}$ & $3.38_{-0.15}^{+0.17}$ & $9.08_{-0.03}^{+0.03}$ & (6) & 2.4\\
 &  &  &  & $9.00_{-0.11}^{+0.09}$ &  & \\
M\,1-33 & \ion{O}{ii} & $3.37_{-0.05}^{+0.06}$ & $>3.6$ & $9.12_{-0.02}^{+0.02}$ & (6) & 2.77\\
 &  &  &  & $9.17_{-0.06}^{+0.06}$ &  & \\
M\,1-60 & \ion{O}{ii} & $4.15_{-0.11}^{+0.10}$ & $4.05_{-0.20}^{+0.33}$ & $9.14_{-0.02}^{+0.02}$ & (6) & 2.75\\
 &  &  &  & $9.15_{-0.07}^{+0.06}$ &  & \\
M\,2-31 & \ion{O}{ii} & $3.85_{-0.16}^{+0.14}$ & $3.55_{-0.21}^{+0.24}$ & $8.99_{-0.04}^{+0.04}$ & (6) & 2.42\\
 &  &  &  & $8.95_{-0.10}^{+0.09}$ &  & \\
NGC\,5315 & \ion{O}{ii} & $3.33_{-0.04}^{+0.05}$ & $3.76_{-0.20}^{+0.33}$ & $8.84_{-0.02}^{+0.02}$ & (7) & 1.58\\
 &  &  &  & $8.91_{-0.07}^{+0.06}$ &  & \\
Abell\,46 & \ion{O}{ii} & $2.41_{-0.07}^{+0.14}$ & $>2.5$ & $9.56_{-0.08}^{+0.11}$ & (8) & 120\\
 & &$2.90_{-2.90}^{+0.66}$ & 3.47  & $9.93_{-0.04}^{+0.04}$ &  & \\
\enddata
\tablecomments{ 
Plasma diagnostics and ionic abundance determinations based on the ORLs were carried out using the MCMC approach.  Results were obtained from the median value of the marginalized posterior distribution and 68\% probability interval, i.e., 16th, 50th, and 84th quantiles.  For each ion (\ion{O}{ii} or \ion{N}{ii}), the data in the first row were derived in the current work, while those in the second row are from the literature as indicated in the final column ``Ref.''.  Only the ORLs with flux uncertainties below 40\% are used here.  Several ORLs of \ion{O}{ii} that are possibly blended with other emission lines were excluded in the analysis.  Due to possible contribution by fluorescence in low-excitation PNe, only the \ion{N}{ii} ORLs from the 3d--4f transitions array are used in the analysis unless otherwise stated. 
\tablenotetext{a}{ADF values are adopted from the compilation of \citep{2018MNRAS.480.4589W}; see also \url{https://nebulousresearch.org/adfs/}.}
\tablenotetext{b}{All the \ion{N}{ii} transitions are used due to the high-excitation PN.}}
\tablerefs{(1) \citet{2017MNRAS.471.3529S}; (2) \citet{2006MNRAS.368.1959L}; (3) \citet{2000MNRAS.312..585L}; (4) \citet{2015MNRAS.452.2606G}; (5) \citet{2012AA...538A..54G}; (6) \citet{2018MNRAS.473.4476G}; (7) \citet{2017MNRAS.471.1341M}; (8) \citet{2015ApJ...803...99C}.}
\end{deluxetable*}

\begin{deluxetable}{ccccc}[h] 
\label{reclinelist}
\setlength{\tabcolsep}{0.3cm}
\tablecaption{Optical Recombination Lines Used for Plasma Diagnostics}
\tablehead{\colhead{$\lambda_{\mathrm{obs}}$} & \colhead{$I_\lambda/I_{\mathrm{H\beta}}$} & \colhead{$I_\mathrm{err}/I_{\mathrm{H\beta}}$} & \colhead{Ion} & \colhead{$\lambda_0$} \\ 
\colhead{} & \colhead{} & \colhead{} & \colhead{} & \colhead{} } 
\startdata
\multicolumn{5}{l}{Object Name: IC\,4776}\\
3882.19&1.70E-04&5.00E-05&\ion{O}{II}&3882.19\\
3907.46&1.00E-04&3.00E-05&\ion{O}{II}&3907.46\\
4072.16&1.25E-03&9.00E-05&\ion{O}{II}&4072.15\\
4078.84&4.50E-04&9.00E-05&\ion{O}{II}&4078.84\\
4089.29&8.80E-04&1.40E-04&\ion{O}{II}&4089.29\\
4092.93&1.50E-04&4.00E-05&\ion{O}{II}&4092.93\\
4110.78&2.90E-04&3.00E-05&\ion{O}{II}&4110.79\\
4119.22&6.60E-04&3.00E-05&\ion{O}{II}&4119.22\\
4121.46&2.70E-04&3.00E-05&\ion{O}{II}&4121.46\\
4132.80&4.60E-04&4.00E-05&\ion{O}{II}&4132.80\\
4153.30&5.30E-04&2.00E-05&\ion{O}{II}&4153.30\\
\enddata
\tablecomments{This table presents the recombination lines used for MCMC sampling. The Col. (Ion) and ($\lambda_0$) are the primary components of the observed lines (fine-structure lines of other \ion{O}{ii} are possibly blended in.)\\
(This table is published in its entirety online only in the machine-readable format.)}
\end{deluxetable}

\subsection{Results and Discussion}

Table~\ref{DIAGNOSTICS} presents the results of MCMC sampling with respect to the electron temperature, electron density, and ionic abundance for each object. These outcomes are derived from the median values of the marginalized posterior distribution, accompanied by the 68\% probability interval, delineated by the 16th, 50th, and 84th quantiles. The catalog of \ion{O}{ii} and \ion{N}{ii} ORLs employed in the MCMC sampling for each object is detailed in Table~\ref{reclinelist}. A comprehensive test was conducted on a total of 13 objects, including \ion{O}{ii} ORLs in all 13 objects and \ion{N}{ii} ORLs in 4 objects.

The $T_\mathrm{e}$(\ion{O}{ii} ORLs) in this study exhibit robust constraints with corresponding reference values in the literature. The average $\log{T_\mathrm{e}}$(\ion{O}{ii} ORLs) across the 12 objects is approximately $3.3\pm0.4$.  The precision of $\log{N_\mathrm{e}}$(\ion{O}{ii} ORLs) is not universally well constrained due to the relatively weak dependence of emissivity ($\epsilon$) on electron density.  Achieving a more refined constraint on electron density requires more accurate line flux measurements to trace such a weak correlation and derive a more precise electron density limit.

For the fainter \ion{N}{ii} ORLs, which are typically less frequently detected in deep spectra than \ion{O}{ii} ORLs, we have derived broader ranges for the posterior distributions of $\log{T_\mathrm{e}}$(\ion{N}{ii} ORLs) and $\log{N_\mathrm{e}}$(\ion{N}{ii} ORLs). 

On average, the ionic abundances of $\mathrm{O^{2+}/H^+}$ and $\mathrm{N^{2+}/H^+}$ agree with the values given in the literature within a deviation of 0.05\,dex.  Nevertheless, huge discrepancies have been observed in a few samples with respect to the literature, which can be attributed to (1) the utilization of different electron temperatures and electron densities for determining ionic abundances, and (2) more \ion{O}{ii} and \ion{N}{ii} ORLs used in MCMC sampling, resulting in a different set of recombination lines utilized in abundance determination.  Detailed discussion on these deviations will be presented in the following section.

\subsubsection{2D Distribution of Electron Temperature and Ionic Abundance}

Examining the two-dimensional (2D) posterior distributions for all samples reveals that the electron temperature and ionic abundance $\log(\mathrm{X^{i+}/H^+})$ often exhibit distinct correlations.  Beyond the well constrained cases of Gaussian distribution, three representative morphologies are identified in the 2D distributions: (1) a positive correlation between electron temperature and ionic abundance (e.g., the right panel of Figure\,\ref{MCMC_IC4776}); (2) a negative correlation (e.g., the left panel of Figure\,\ref{MCMC_IC4776}); and (3) a combination of both, forming a ``heart-shaped'' distribution (e.g., the right panel of Figure\,\ref{MCMC_NGC6153_minor}). 

These correlations probably originate from the temperature dependence of the effective recombination coefficients of \ion{O}{ii} and \ion{N}{ii} ORLs and H$\beta$.  Taking \ion{O}{ii} as an example, the theoretical intensity ratio of an \ion{O}{ii} recombination line (with wavelength $\lambda$) to H$\beta$ can be expressed as
\begin{equation}
    \frac{I_{\mathrm{O\,II}}}{I_{\mathrm{H}\beta}}
    \propto 
    \frac{\mathrm{O^{2+}}}{\mathrm{H^+}} \times
    \frac{\alpha_{\rm eff}(\lambda)}%
         {\alpha_{\rm eff}({\rm H}\beta)} ,
\end{equation}
where $\alpha_{\rm eff}$($\lambda$) and $\alpha_{\rm eff}$(H$\beta$) are the effective recombination coefficients of the \ion{O}{ii} nebular line and H$\beta$, respectively.  The ionic abundance is thus 
\begin{equation}
    \frac{\mathrm{O^{2+}}}{\mathrm{H^+}}
    \propto
    \frac{I_{\mathrm{O\,II}}}{I_{\mathrm{H}\beta}} \times 
    \frac{\alpha_{\rm eff}({\rm H}\beta)}%
         {\alpha_{\rm eff}(\lambda)} .
\end{equation}
Assuming $\alpha_{\rm eff} \propto T_{\rm e}^{-\beta}$, where the power-law index $\beta\sim1$ for radiative recombination \citep{2006agna.book.....O}, one obtains a derivative 
\begin{equation}\label{eq7}
    \frac{d\log(\mathrm{O^{2+}/H^+})}{d\log T_{\mathrm{e}}}
    = \beta_{\mathrm{O\,II}} - \beta_{\mathrm{H}\beta} ,
\end{equation}
where $\beta_{\rm O\,II}$ and $\beta_{{\rm H}\beta}$ are the power-law indices of the \ion{O}{ii} line and H$\beta$, respectively.  For different recombination lines (in particular, recombination lines emitted by different ions), the $\beta$ values are different; this results in a possible correlation/anticorrelation between ionic abundance and electron temperature (derived from ORLs). 

In the 2D posterior distribution diagrams, a positive correlation between ionic abundance and $T_{\rm e}$ corresponds to $\beta_{\mathrm{O\,II}} > \beta_{{\rm H}\beta}$, while a negative correlation corresponds to $\beta_{\mathrm{O\,II}} < \beta_{{\rm H}\beta}$.  Because multiple O\,{\sc ii} recombination lines are jointly fitted in the MCMC sampling, the effective $\beta_{\rm O\,II}$ used here represents a weighted average of the $\beta_{\rm O\,II}$ values from all individual \ion{O}{ii} lines involved in the fit.  In Eq.~\ref{eq7}, we have assumed that $\beta$ remains approximately constant over a limited temperature range, i.e., $\frac{d\beta}{d\log T_e} \approx 0$. However, practical data indicate that $\beta$ may exhibit significant variations within specific temperature intervals. This temperature dependence of $\beta$ consequently produces the diverse morphological characteristics observed in the posterior distributions of $\log(\mathrm{O}^{2+}/\mathrm{H}^{+})$ versus $\log T_e$. At very low electron temperatures ($\log{T_{\rm e}}<2.7$), the apparent strong degeneracy mainly arises from the code implementation, where $\alpha_{\rm eff}$(H$\beta$) is fixed to its lower-bound value at $\log{T_{\rm e}}=2.7$ due to the tabulated range limit \citep{HIcoe}, which leads to slightly underestimated ionic abundances when the fitted electron temperature falls below this limit. This constitutes one of the reasons why the fitted ionic abundances for \ion{O}{ii} and \ion{N}{ii} ORLs with fitted $\log{T_{\rm e}}<2.7$ exhibit significant discrepancies from the values reported in the literature.

\subsubsection{IC\,4776}

In a study reported by \citet[hereafter PS17]{2017MNRAS.471.3529S}, the electron temperature of \ion{O}{ii} ORLs was determined to be $3560^{+3710}_{-1960}$ K, while the corresponding electron density was estimated as $2850^{+1430}_{-1240} \mathrm{cm^{-3}}$ using the ratio \ion{O}{ii} $\lambda4649/\lambda4089$. Utilizing our methodology with the \ion{O}{ii} ORLs measured by PS17, we obtained an electron temperature of $2140^{+100}_{-100}$ K and an electron density of $13180^{+10810}_{-3630} \mathrm{cm^{-3}}$, as illustrated in Figure\,\ref{MCMC_IC4776}.

Our derived $T_\mathrm{e}$ for \ion{O}{ii} ORLs agrees with that of PS17, and the ionic abundance of $\mathrm{O^{2+}/H^+}$ is also in approximate agreement. Utilizing the electron temperature and density of \ion{O}{ii} ORLs determined by PS17 for sampling the ionic abundance with our methodology, yields consistent results with smaller uncertainties compared to those of PS17. Concerning $N_\mathrm{e}$ for \ion{O}{ii} ORLs, the disparity in electron density could potentially be attributed to the fact that PS17 relied on only two \ion{O}{ii} ORLs (specifically, $\lambda4649/\lambda4089$), while our approach incorporates a broader set of observed \ion{O}{ii} ORLs.

We carried out a comparative analysis between the observed line fluxes of \ion{O}{ii} ORLs and the corresponding theoretical line intensities computed with the parameters derived from MCMC sampling. Notably, discrepancies between the observed and theoretical line intensities were identified, reaching approximately a factor of two sigma or even larger of observed flux error for specific \ion{O}{ii} ORLs, such as $\lambda4185.44, \lambda4303.82, \lambda4596.18,$ and $\lambda4610.20$.  Furthermore, these \ion{O}{ii} ORLs showed consistent deviations when utilizing the electron temperature, electron density, and ionic abundance obtained by PS17. We thus infer that some \ion{O}{ii} ORLs may have underestimated observed flux errors, potentially being the secondary factor in the observed differences in electron density.

The electron temperature in IC\,4776, derived from the 3d--4f transitions of \ion{N}{ii}, is exceptionally low, with $T_\mathrm{e}\text{(\ion{N}{ii} ORLs)}=275^{+190}_{-80}$\,K, accompanied by a correspondingly low electron density, $N\mathrm{e}$(\ion{N}{ii} ORLs)$=550^{+1220}_{-300}$\,$\mathrm{cm^{-3}}$.  The selection of only \ion{N}{ii} ORLs from 3d--4f transitions for MCMC sampling was driven by the low-excitation nature of IC\,4776, as discussed in section \ref{Fluorescence excitation}. A discrepancy of approximately 1 dex was observed in the ionic abundances of $\mathrm{N^{2+}/H^+}$ when comparing our results with those obtained by PS17. This difference may be attributed to two factors: firstly, the extremely low electron temperature and density we determined, as opposed to potentially higher values used by PS17 in determining the ionic abundance of $\mathrm{N^{2+}/H^+}$; secondly, PS17 incorporated additional 3s--3p and 3p--3d transitions of \ion{N}{ii} ORLs, which are prone to strong contributions from fluorescence excitation in low-excitation PNe, thus introducing bias in the obtained ionic abundances.

\subsubsection{Hf\,2-2}

Hf\,2-2 is a well-known PN that shows an extremely high ADF ($\sim$80) for $\mathrm{O^{2+}/H^+}$ \citep{2006MNRAS.368.1959L,2016MNRAS.461.2818M,2018MNRAS.480.4589W,2022MNRAS.510.5444G}. The \ion{O}{ii} and \ion{N}{ii} ORLs we used to do the MCMC sampling and the corresponding observed data are from \citet{2006MNRAS.368.1959L}, and we set the velocity parameter $\Delta v$ to 30 $\mathrm{km\,s^{-1}}$ for all the \ion{O}{ii} and \ion{N}{ii} ORLs in Hf\,2-2 we used because the lines were from the median-resolution (R$\sim$3000) spectra.

The \ion{O}{ii} ORLs with the wavelengths from $\lambda4276$ to $\lambda4317$ are not included since the fluxes are the sum of the fluxes of many \ion{O}{ii} fine structure lines and the velocity differences between these fine structure lines are large, which cannot be well corrected by a uniform velocity parameter $\Delta v$. The \ion{O}{ii} ORL $\lambda4676.24$ was also not included because of a problematic flux uncertainty (too low).

We have tested the observed data from the spectra of three different slitwidths, i.e., 2, 4, and 8 arcsec slitwidths. Only the \ion{N}{ii} ORLs from the spectra of 4 arcsec slitwidth were not tested, because there were some \ion{N}{ii} ORLs observed in the spectra of 2 and 8 arcsec slitwidths but not in the spectra of 4 arcsec slitwidth, resulting in too few \ion{N}{ii} ORLs for MCMC sampling.

The electron temperatures extracted from the \ion{O}{ii} optical recombination lines (ORLs) in the spectra obtained with slitwidths of 2, 4, and 8 arcseconds are $790^{+470}_{-300}$, $280^{+140}_{-50}$, and $290^{+100}_{-40}$\,K, respectively. These values exhibit general concordance with the electron temperature, $T_\mathrm{e}\text{(\ion{O}{ii} ORLs)}=630$\,K, determined by \citet{2006MNRAS.368.1959L}, and the electron temperature $T_\mathrm{e}\text{(\ion{O}{ii} ORLs)}<2000$\,K as derived by \citet{2018MNRAS.480.4589W}.

The corresponding electron densities for \ion{O}{ii} ORLs obtained with slitwidths of 2, 4, and 8 arcsec are $1200^{+310}_{-220}$, $1200^{+2690}_{-510}$, and $4270^{+13930}_{-2790}$\,cm$^{-3}$, respectively. Notably, these values are consistently lower than the electron densities $N_\mathrm{e}\text{(\ion{O}{ii} ORLs)}=$ 4850, 4000, and 13500\,cm$^{-3}$ reported by \citet{2006MNRAS.368.1959L} for slitwidths of 2, 4, and 8 arcsec. However, they are generally in agreement with the electron density $N_\mathrm{e}\text{(\ion{O}{ii} ORLs)}=1250^{+1500}_{-1250}$\,cm$^{-3}$ determined by \citet{2018MNRAS.480.4589W}.

It is noteworthy that \citet{2006MNRAS.368.1959L} derived the electron density $N_\mathrm{e}$(\ion{O}{ii} ORLs) based on seven \ion{O}{ii} ORLs of multiplet M1 at an electron temperature $T_\mathrm{e} = 10000$\,K, which may contribute to the discrepancies in electron densities.

As shown in Figures~\ref{MCMC_Hf2-2_2arcsec}--\ref{MCMC_Hf2-2_4arcsec}, the ionic abundance of $\mathrm{O^{2+}/H^+}$ exhibits a positive correlation with $T_\mathrm{e}$ with large degeneracy when $\log{T_\mathrm{e}}\lesssim$2.75.  Consequently, the ionic abundance, expressed as $12+\log (\mathrm{O^{2+}/H^+})$, derived from the spectrum obtained with 2\arcsec\ slit width is $9.78^{+0.06}_{-0.08}$, which is in alignment with the ionic abundance derived by \citet{2006MNRAS.368.1959L}, who determined the ionic abundances for all heavy elements assuming a nebular condition of $T_\mathrm{e} = 900$\,K and $N_\mathrm{e}=1000$\,cm$^{-3}$.  Our result is also in excellent agreement with the abundance value reported by \citet{2022MNRAS.510.5444G}, who utilized the \ion{O}{ii} $\lambda\lambda4649.13+4650.84$ ORLs and adopted a much higher electron temperature of 4000\,K.  However, the ionic abundance determined from data obtained with the 4\arcsec\ and 8\arcsec\ slit widths is diminished to approximately 9.5 due to their $\log{T_\mathrm{e}}<$2.75 [K].

The $T_\mathrm{e}$, $N_\mathrm{e}$ and ionic abundances of $\mathrm{N^{2+}/H^+}$ derived from the spectra of 2 and 8 arcsec slitwidths are in good agreement.  Both show extremely low electron temperatures, $T_\mathrm{e}\lesssim500$\,K, and the electron densities $N_\mathrm{e}\lesssim3000$\,cm$^{-3}$.  A similar relation also exists between the ionic abundance of $\mathrm{N^{2+}/H^+}$ and electron temperature for \ion{N}{ii} ORLs, i.e., positive correlation with large degeneracy, resulting in a difference of about 0.4 dex between the $12+\log (\mathrm{N^{2+}/H^+})$ we obtained and that determined by \citet{2006MNRAS.368.1959L}.

\subsubsection{NGC\,6153}

NGC 6153 is a bright PN located in the southern sky, which shows super-metal-rich characteristics reported by \citet{2000MNRAS.312..585L}.  We use both the spectral data of the `minor axis' and the `entire nebula' from \citet{2000MNRAS.312..585L} to analyze the \ion{O}{ii} and \ion{N}{ii} ORLs.

Many emission lines have been deblended so that the fluxes of individual components were shown. Therefore, We used the laboratory wavelengths corresponding to the manual IDs as the input wavelengths and set the velocity parameter $\Delta v$ to 10 $\mathrm{km\,s^{-1}}$ for all the \ion{O}{ii} and \ion{N}{ii} ORLs.  For emission lines from the same parent ion with very close laboratory wavelengths, we combine them into a single line, and the flux is the sum of the fluxes of the individual emission lines, e.g., the \ion{N}{ii} $\lambda4236.91$ and $\lambda4237.05$.

The line flux errors follow the criteria: (1) assuming 5 per cent for lines with dereddened fluxes $I(\lambda)\geq0.2$ relative to $I_\mathrm{H\beta}=100$; (2) 10 per cent for those lines with $0.1\leq I(\lambda)<0.2$; (3) 20 per cent for $0.05\leq I(\lambda)<0.1$; and (4) 30 per cent for those lines with $ I(\lambda)<0.05$.

\begin{figure*}
\centering 
\includegraphics[width=0.8\textwidth]{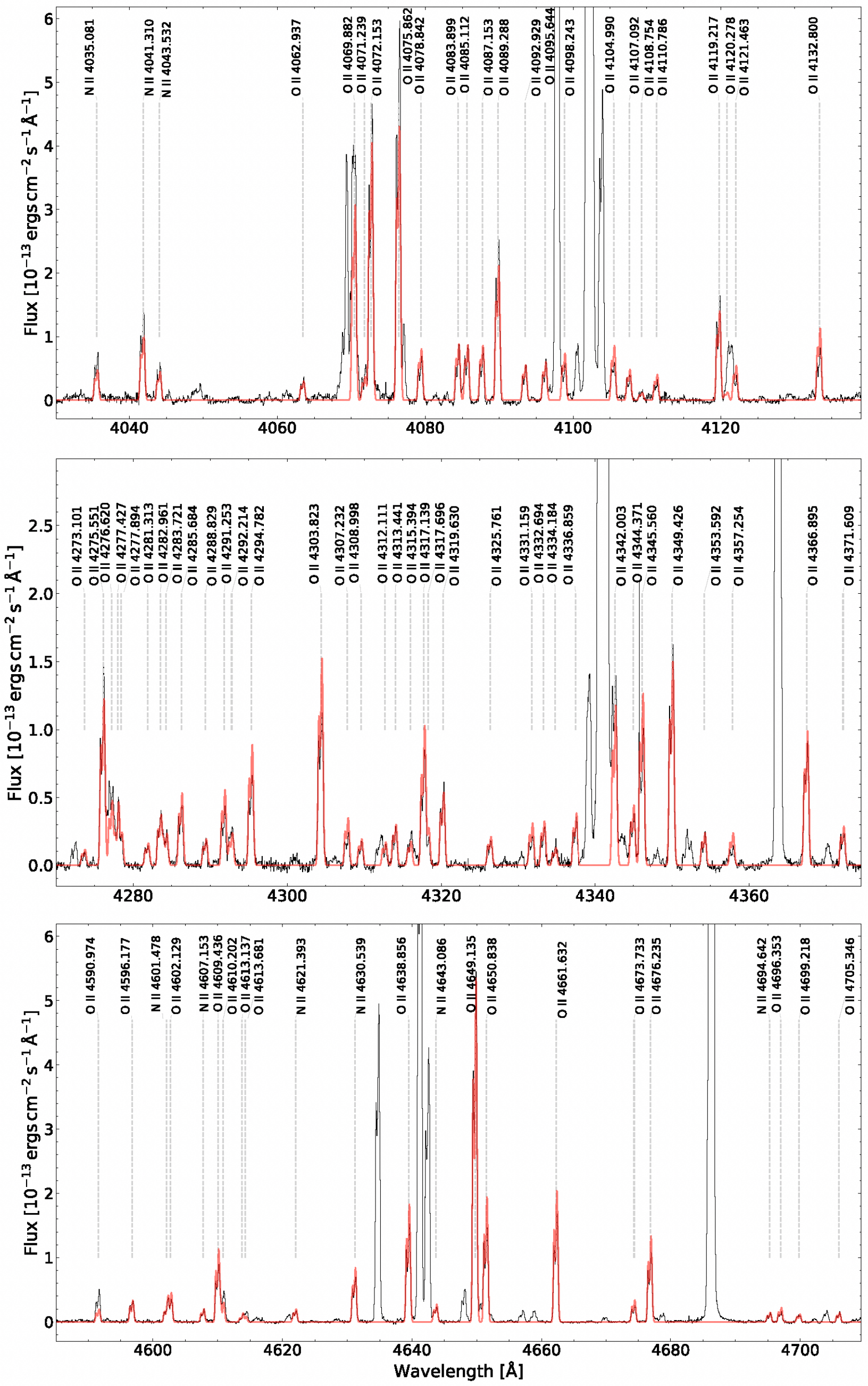}
\caption{Comparison of the observed spectrum (black) of NGC\,6153 with the theoretical spectra (red) of \ion{O}{ii} and \ion{N}{ii} obtained using the specified electron temperature, electron densities, and ionic abundances.} 
\label{fig4}
\end{figure*}

The posterior distribution (Figures\,\ref{MCMC_NGC6153_minor} and \ref{MCMC_NGC6153_entire}) shows that $T_\mathrm{e}$, $N_\mathrm{e}$, and ionic abundances for \ion{O}{ii} and \ion{N}{ii} ORLs in the `minor axis' and the `entire nebula' are in good agreement. The $T_\mathrm{e}$(\ion{O}{ii} ORLs) is about 1000\,K, which is smaller than the $T_\mathrm{e}\text{(\ion{N}{ii} ORLs)}\lesssim4000$\,K.  \citet{2022AJ....164..243R} obtained the $T_\mathrm{e}$(\ion{O}{ii} ORLs) around 1800 to 5000\,K using the line ratio $\lambda4089/\lambda4649$, and derived $T_\mathrm{e}\text{(\ion{N}{ii} ORLs)}>2500$\,K using the line ratio $\lambda4041/\lambda5680$.

Both the $N_\mathrm{e}$(\ion{O}{ii} ORLs) and $N_\mathrm{e}$(\ion{N}{ii} ORLs) we obtained are general greater than 10,000\,cm$^{-3}$, which are in general agreement with those obtained by \citet{2022AJ....164..243R}, who reported the $N_\mathrm{e}\text{(\ion{O}{ii} ORLs)}>5000$\,cm$^{-3}$ and $N_\mathrm{e}\text{(\ion{N}{ii} ORLs)}>10,000$\,cm$^{-3}$.

The differences between ionic abundances of $\mathrm{N^{2+}/H^+}$ and $\mathrm{O^{2+}/H^+}$ we obtained and those determined by \citet{2000MNRAS.312..585L} are around 0.1 dex, which is mainly due to the different electron temperatures and electron densities we used (\citet{2000MNRAS.312..585L} determined the ionic abundances of $\mathrm{N^{2+}/H^+}$ and $\mathrm{O^{2+}/H^+}$ under the conditions that $T_\mathrm{e}=9100$\,K, $N_\mathrm{e}=3500\,\text{cm}^{-3}$).

It is noteworthy that all the $T_\mathrm{e}$, $N_\mathrm{e}$, and ionic abundances we derived are based on the assumption that all the \ion{O}{ii} and \ion{N}{ii} ORLs come from the same nebular region. However, as reported in \citet{2000MNRAS.312..585L}, NGC\,6153 probably have two nebular components with different electron temperatures (a high-$T_{\rm e}$ component with normal value close to $\sim$10,000\,K, and a low-$T_{\rm e}$ component $\lesssim$1000\,K) and densities.  The \ion{H}{i} lines come from the normal region with relatively high $T_\mathrm{e}$, and ORLs of heavy elements mainly come from cold ($\lesssim$1000\,K), metal-rich components, maybe in the form of very small clumps that are unresolvable even with the instruments on the \emph{Hubble Space Telescope} (R.\ Williams, private communivation). 

Both \citet{2011MNRAS.411.1035Y} and \citet{2020MNRAS.497.3363G} used the bi-abundance photoionization models to fit the observed fluxes of heavy-element ORLs and CELs.  It is clear from their analysis that the observed fluxes of the \ion{O}{ii} or \ion{N}{ii} ORLs do not come exclusively from the same region.  Therefore, the $T_\mathrm{e}$ and $N_\mathrm{e}$ we measured from a spectrum are the average values weighted by the line flux contributions from the two different nebular regions (a large fraction of the line flux from the low-temperature region biases the results in favour of the low temperatures).  In order for a more accurate treatment of nebular emission (and consequently, more reliable determination of the ionic abundances from ORLs and CELs), it is also important to consider the weight of \ion{H}{i} emissivity in both nebular components.  Such attempt has been made recently on MUSE spectroscopy of NGC\,6153 by \citet{Gomez_Llanos_2024}, who defined an abundance contrast factor (ACF) between the two plasma components.

Figure \ref{fig4} presents the high-resolution spectra of NGC\,6153 obtained from the VLT/UVES, with fluxes corrected for dust extinction, assuming $c(\mathrm{H\beta}) = 1.3$ \citep{2000MNRAS.312..585L}. The exceptionally high resolution of these spectra enables the detailed observation of the double-peaked features of the emission lines.  

The theoretical spectra of \ion{O}{ii} and \ion{N}{ii} ORLs are also displayed for comparison. To generate the double-peak features of each emission line, we assumed a specific velocity difference between the central wavelengths of each peak of the emission line. The theoretical line strength of each ORL is calculated as the product of the ionic abundance and the emissivity of the respective ORL. For the emissivity, we adopted electron temperatures and electron densities for \ion{O}{ii} and \ion{N}{ii} based on the conditions of the `entire nebula', i.e., $\log(T_\mathrm{e})$ (\ion{O}{ii} ORLs) $ = 3.03 $, $\log(N_\mathrm{e})$ (\ion{O}{ii} ORLs) $ = 4.44 $, $\log(T_\mathrm{e})$ (\ion{N}{ii} ORLs) $ = 3.25 $, and $\log(N_\mathrm{e})$ (\ion{N}{ii} ORLs) $ = 3.92 $. Consequently, the ionic abundances of \ion{O}{ii} and \ion{N}{ii} were determined using the observed fluxes of \ion{O}{ii} $\lambda 4649.13$ and \ion{N}{ii} $\lambda 5679.56$, with the emissivities under above conditions.

The comparison shows that the theoretical spectra, generated using these electron temperatures and densities from our fitting results, match the observed spectra with remarkable accuracy. This close agreement validates our assumptions and the derived physical conditions of the nebula, demonstrating the robustness of our approach.

\begin{figure}
\centering 
\includegraphics[width=8.25cm]{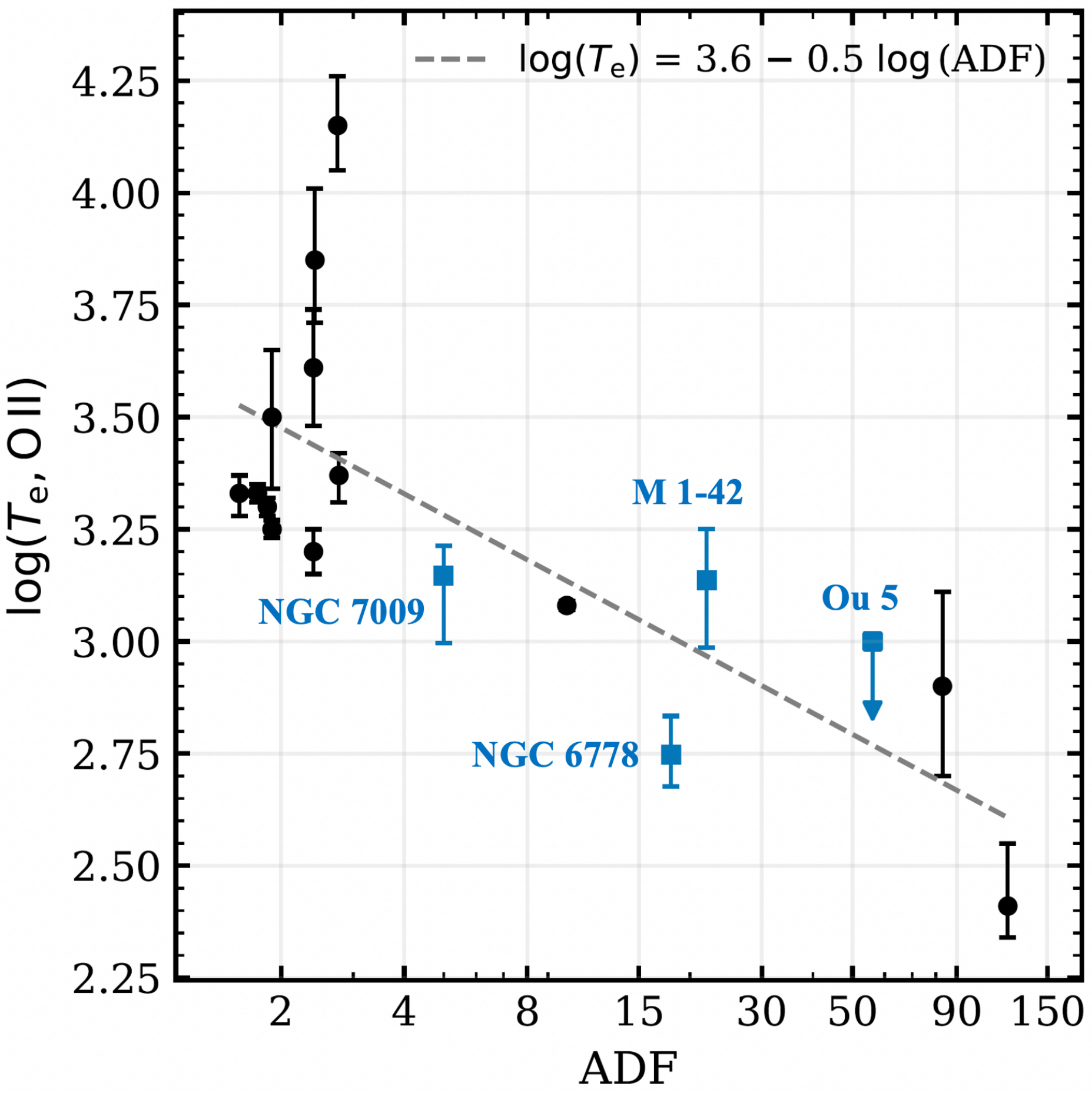}
\caption{Electron temperature derived with the O\,{\sc ii} recombination lines detected in PNe as a function of ADF in logarithm scale.  Black-filled circles are the sample of PNe from Table\,\ref{DIAGNOSTICS}, whose $\log{T_{\rm e}}$(\ion{O}{ii}) values were obtained through MCMC sampling.  Blue squares are the four PNe (NGC\,7009, NGC\,6778, M\,1-42, and Ou\,5; see text for the references) whose temperatures were derived from the observed O\,{\sc ii} $\lambda$4649/$\lambda$4089 line ratio, using the O\,{\sc ii} effective recombination coefficients calculated by \citet[][in Case\,B]{OIIcoe}; the blue downward arrow on Ou\,5 indicates the O\,{\sc ii} temperature of PN is an upper limit.  ADF values were adopted from literature (see Table\,\ref{DIAGNOSTICS} and description in the text).  The grey dashed line is a linear fit to the whole sample (see the legend on top of the figure).} 
\label{fig5}
\end{figure}

\subsubsection{The O\,{\sc ii} Temperature versus ADF} 

Among the sample with the \ion{O}{ii} and \ion{N}{ii} recombination-line temperatures derived using the MCMC method (see Table\,\ref{DIAGNOSTICS}), there seems to be a tendency wherein the PNe with extreme ADF values also have very low \ion{O}{ii} temperatures, $T_{\rm e}$(\ion{O}{ii}).  This trend is visually demonstrated in Figure\,\ref{fig5}, where we present the PNe from Table\,\ref{DIAGNOSTICS}, whose electron temperatures were derived through MCMC sampling.  To augment our sample of PNe and enhance the statistical significance of this trend, we added four additional high-ADF PNe from the literature, whose temperatures were obtained from the observed O\,{\sc ii} $\lambda$4649/$\lambda$4089 line ratio (retrieved from the literature), using the effective recombination coefficients for nebular O\,{\sc ii} lines calculated by \citet[][in Case~B]{OIIcoe}.  These newly added four PNe are:  NGC\,7009 with $T_{\rm e}$(O\,{\sc ii})=1400$\pm$200\,K and ADF$\approx$5 \citep{2013MNRAS.429.2791F}, M\,1-42 with $T_{\rm e}$(O\,{\sc ii})=1370$^{+570}_{-400}$\,K and ADF$\approx$22 (Huang et al. 2025, in prep.), NGC\,6778 with $T_{\rm e}$(O\,{\sc ii})=560$\pm$100\,K and ADF$\approx$18 \citep{2016MNRAS.455.3263J}, and Ou\,5 with $T_{\rm e}$(O\,{\sc ii})$<$1000\,K and ADF$\approx$56 \citep{2015ApJ...803...99C}. 

In Figure\,\ref{fig5}, an anti-correlation between $\log{T_{\rm e}}$(O\,{\sc ii}) and ADF in logarithm seems to be discernible; this relation is corroborated by a linear fit to the whole sample, which yields $\log{T_{\rm e}}$(O\,{\sc ii}) = 3.6 $-$ 0.5$\times\log{\rm ADF}$.  Figure\,\ref{fig5} indicates that the nebulae with larger abundance discrepancies tend to exhibit lower ORL temperatures; in particular, the PNe with extreme ADFs ($>$50), such as Abell\,46 and Hf\,2-2, have very low O\,{\sc ii} temperatures ($<1000$\,K).  The moderate-ADF nebulae (ADF$\sim$4--30), although very limited in sample size, display intermediate O\,{\sc ii} temperatures of $\sim$1000--2000\,K, while the objects with low ADFs ($<$3) are characterized by a broader range in the O\,{\sc ii} temperature, mostly from $\sim$1600\,K to 4000\,K, with the highest temperature reaching $\sim$14000\,K.

The anti-correlation exhibited by the sample in Figure\,\ref{fig5} is in line with the two-component nebular model of PNe with chemical inhomogeneity \citep{2000MNRAS.312..585L,2006MNRAS.368.1959L}.  The systematically low $T_{\rm e}$(O\,{\sc ii}) in high-ADF PNe indicates that the O\,{\sc ii} ORLs emission predominantly originates from a cool plasma, where the heavy elements (such as C, N, O, and Ne) are highly enriched, leading to very high cooling efficiency, effectively locking the electron temperature to a very low level ($\lesssim$1000\,K).  Consequently, the ADF generally reflects the abundance contrast between the cold, metal-rich component (as traced by heavy-element ORLs) and the warm ($\sim$10$^{4}$\,K), diffuse ionized gas (as mainly traced by CELs) with normal metallicity (i.e.\ near solar).  However, as demonstrated by \citet{2020MNRAS.497.3363G}, the ADF and the real abundance contrast between the two components may not always correlate in a straightforward way, especially in low-ADF PNe, where temperature and density structures can produce additional degeneracies (which is not within the scope of discussion in our paper); this might be the reason for the large spread in $\log{T_{\rm e}}$(O\,{\sc ii}) in the low-ADF PNe.  Nevertheless, in high-ADF PNe, the ADF tends to increase as the O\,{\sc ii} temperature decreases, more or less consistent with the presence of a strongly metal-enriched cold component.

\subsubsection{Error Analysis}

We need to be careful about the posterior distributions and errors for each parameter obtained from MCMC sampling. In our likelihood function, there are actually three implicit assumptions: (1) the predicted line fluxes $I_{\mathrm{pred}}(\lambda_l)$ are assumed to be accurate and free of error; (2) the observed flux errors are assumed to be appropriate; and (3) the assumption is made that the recombination lines originating from the same parent ions all come from the same region (with the same electron temperature and electron density as well as ionic abundance).

For the first assumption, this is because the process of calculating the effective recombination coefficients is difficult to give a reasonable error. These ORLs cannot be generated in the laboratory, so we cannot obtain accurate line fluxes at accurate electron temperatures and densities, resulting in an error in the model value that is dependent on the accuracy of the computational model, which in turn is difficult to quantify. In this case, we tentatively consider the model error to be small, estimated to be within 10 per cent \citep{2000MNRAS.312..585L}, and not to have a significant impact on the results.

The second assumption relies more on the quality of the data and the method of calculating the flux errors in the literature. In these samples we give, we calculated the theoretical line intensities of these ORLs under the optimal parameters obtained by the MCMC sampling and compared them with the observed line fluxes. Consequently, we found that in the vast majority of samples, the theoretical line intensities match well with the observed line fluxes, only a few of the theoretical line intensities deviate from the observed line fluxes by beyond 2 sigma. However, for the \ion{O}{ii} ORLs in the sample IC\,4776, nearly a quarter of the observed line fluxes deviate from the theoretical values by more than 2 sigma. if these data with excessively deviating observed line fluxes are removed, the results obtained will favor another set of solutions: $\log T_{\rm{e}}=3.67\pm0.04$, $\log N_{\rm{e}}=4.03_{-0.09}^{+0.12}$, and $12+\log(\mathrm{O^{2+}/H^+})=8.80\pm0.01$. Therefore, the accuracy of the observed line fluxes and the flux errors will directly affect the posterior distributions of MCMC sampling, especially at some points where the errors of the observed line fluxes are underestimated, which will have a more significant effect due to their larger weights.

The third assumption is the basis that we can use all the ORLs of the same parent ions for MCMC sampling. The real situation is more likely that these ORLs come from a larger continuous region where there are fluctuations in electron temperature and fluctuations in electron density. Thus, the observed line flux for an individual ORL in the one-dimensional spectra is the result of integrating the flux at each point over this continuous region.

\section{Summary and Conclusions}

This is the second paper on the PyEMILI code, focusing mainly on the test runs of the code on the emission lines detected in the deep, medium- and high-resolution spectra of PNe, \ion{H}{ii} regions, and HH objects from the literature of the last two decades.

In the present study, we expanded the AtLL database, which has already been incorporated into PyEMILI, by cross-matching with the Kurucz Line Lists.  A total of 6154 atomic transitions were added, addressing gaps in some observed emission lines of elements C, N, and O. Subsequently, utilizing PyEMILI, we performed line identifications for the emission line lists from 34 sample spectra, including 28 PNe and 5 \ion{H}{ii} regions and HH objects.  The results were compared and analyzed against manual IDs. Remarkably, PyEMILI demonstrated an agreement rate exceeding 90\% for these emission-line objects, affirming its efficacy and applicability. Additionally, our analysis using PyEMILI revealed some candidate IDs with high-rated (`A' or `B') given by PyEMILI that are deemed more reliable than manual IDs. Some of the high-rated candidate IDs can also serve as a reasonable explanation for previously unidentified lines in the literature. The amendments and supplements made to the line identifications serve as a robust research base for future spectroscopic analysis. 

In addition to this, PyEMILI now has a new module: based on the MCMC method, the identified \ion{O}{ii} and \ion{N}{ii} ORLs are used to further analyze their electron temperatures, electron densities, and ionic abundances. This is the first attempt to use the MCMC method and all ORLs originating from the same parent ion in the spectra for plasma diagnostics. Using all the ORLs for the fit can better constrain these parameters.

Our results show that for \ion{O}{ii} ORLs, we can obtain more accurate parameters that are basically consistent with the literature.  A key finding based on our analysis is a trend of anti-correlation between $\log{T_{\rm e}}$(O\,{\sc ii}) and the ADF in logarithm (Figure\,\ref{fig5}).  PNe with extreme ADFs ($>50$, such as Abell\,46 and Hf\,2-2), tend to have very low \ion{O}{ii} temperatures ($<$1000\,K).  PNe with moderate ADF values (ADF$\sim$4--30) seem to exhibit $T_{\rm e}$(O\,{\sc ii})$\sim$1000--2000\,K, while the low-ADF objects (ADF$<3$) show systematically higher O\,{\sc ii} temperatures, but with a large spread.  This trend is in line with a widely accepted conjecture that in high-ADF PNe, the bulk of the O\,{\sc ii} ORL emission probably originates from a cold ($\lesssim$1000\,K), metal-enriched plasma component, generally consistent with the two-component model of PNe with chemical inhomogeneity. 

For the \ion{N}{ii} ORLs, due to the weaker observed line fluxes of the \ion{N}{ii} ORLs compared to the \ion{O}{ii} ORLs, and the limitation of the spectral resolution, very few papers have results in analyzing the electron temperatures and densities of the \ion{N}{ii} ORLs. This, coupled with the fact that in low-excitation PNe, the fluorescence mechanism dominates the flux contribution to the transitions in the s, p, and most d states of the \ion{N}{ii}, results in a large number of unavailable \ion{N}{ii} ORLs, which makes it even more difficult to obtain well-confined electron temperatures and densities. We finally give six samples with \ion{N}{ii} ORLs plasma diagnostics that all have the electron temperatures that fit well into the electron temperature relation  $T_\mathrm{e}(\mathrm{ORLs})\leq T_\mathrm{e}(\mathrm{CELs})$\citep{2013MNRAS.428.3443M}. Further analyses will require high-resolution spectroscopic observations and more accurate spectral line measurements.

\begin{acknowledgments}
We thank the anonymous referee, whose insightful comments and constructive suggestions significantly improved this article.  Z.T.\ thanks Furen Deng and Xinlin Zhao for useful discussions and suggestions. 

J.L.\ acknowledges support from the New Cornerstone Science Foundation through the New Cornerstone Investigator Program.  J.G.-R. acknowledges financial support from grant PID-2022136653NA-I00 (DOI:10.13039/501100011033) funded by the Ministerio de Ciencia, Innovaci\'{o}n y Universidades (MCIU/AEI) and by ERDF ``A way of making Europe'' of the European Union.  
This work was also supported by the National Key R\&D Program of China (Grant No. 2023YFA1607902), and China Manned Space Program with grant No.\ CMS-CSST-2025-A14. 
X.F.\ acknowledges support from the Youth Talent Program (2021) from the Chinese Academy of Sciences (CAS, Beijing) and the ``Tianchi Talents'' Program (2023) of the Xinjiang Autonomous Region, P.~R.\ China. 
\end{acknowledgments}

\section*{Data Availability}

Tables\,2 and 5 are published in their entirety online in machine-readable format.  The tables presented in the Appendix -- the final emission-line tables (Tables\,A.1--A.34) and the formatted complete output of PyEMILI runs (Tables\,B.1--B.34) for all objects in the test sample -- are publicly available on Zenodo via:\dataset[10.5281/zenodo.17540949]{https://doi.org/10.5281/zenodo.17540949} \citep{tu_2025_17540949}.  Downloading and installation of the PyEMILI code are described in Paper~I (Section\,5 therein).


\appendix

\counterwithin{figure}{section}

\counterwithin{table}{section}

\section{Emission-line Table of IDs Assigned by PyEMILI}

Table\,\ref{Lines_NGC3918} shows the emission lines in PN NGC\,3918 in the test sample, with the most probable IDs assigned by PyEMILI.  The emission-line lists with PyEMILI's identification for all the objects in our test sample are presented in Tables\,A.1--A.34, which are publicly available on Zenodo \citep{tu_2025_17540949}.

\begin{deluxetable}{cclc}[!htp]
\label{Lines_NGC3918}
\tablecaption{Emisison-line List of NGC\,3918 with the Most Probable IDs Assigned by PyEMILI (A Section Is Presented Here for Demonstration)} 
\tablehead{\colhead{$\lambda_{\mathrm{obs}}$} & \colhead{$I_\lambda/I_{\mathrm{H\beta}}$} & \colhead{ID$^{a}$} & \colhead{Note} \\ 
\colhead{(\AA)} & \colhead{} & \colhead{} & \colhead{} } 
\startdata
3109.13 & 6.48E-03 & [\ion{Ar}{iii}] 3109.18 & \\
3115.66 & 2.19E-03 & \ion{O}{iii} 3115.674 & \\
3118.66 & 1.38E-03 & [\ion{Cl}{iv}] 3118.61 & \\
3121.65 & 3.25E-02 & \ion{O}{iii} 3121.633 & \\
3132.87 & 9.08E-01 & \ion{O}{iii} 3132.794 & \\
3187.74 & 3.25E-02 & \ion{He}{i} 3187.7436 & \\
3218.19 & 1.10E-03 & \ion{Ne}{ii} 3218.193 & \\
3230.12 & 7.51E-04 & [\ion{Fe}{ii}] 3230.1686, \ion{Ne}{ii} 3230.069 & * \\
3241.63 & 9.50E-03 & [\ion{Na}{iv}] 3241.63 & \\
3260.89 & 2.86E-03 & \ion{O}{iii} 3260.857 & \\
3265.33 & 1.90E-03 & \ion{O}{iii} 3265.329 & \\
3280.03 & 3.99E-04 & [\ion{Fe}{ii}] 3279.9221 & \\
3284.51 & 2.41E-04 & \ion{O}{iii} 3284.45 & \\
3287.56 & 1.01E-04 & \ion{O}{iii} 3287.647, \ion{O}{ii} 3287.471 & \\
3299.45 & 3.79E-04 & \ion{O}{iii} 3299.385 & \\
3312.35 & 9.83E-02 & \ion{O}{iii} 3312.329 & \\
3323.84 & 3.06E-04 & \ion{S}{iii} 3323.984, \ion{Ne}{ii} 3323.734 & \\
3328.86 & 3.19E-04 & \ion{Ne}{ii} 3328.693 & * \\
3334.87 & 1.20E-03 & [\ion{Fe}{iii}] 3334.954, \ion{Ne}{ii} 3334.836 & \\
\multicolumn{4}{l}{......} \\
3571.36 & 5.31E-04 & \ion{Ne}{ii}]  3571.231 & ? \\
\multicolumn{4}{l}{......} \\
\enddata
\tablecomments{
The asterisk ``*'' means the manual ID (given in the literature) of this line is not among the most probable ID(s) assigned by PyEMILI.  The question mark ``?'' indicates that the line was unidentified in the literature \citep[in this table, it is][]{2015MNRAS.452.2606G}, but is unambiguously identified by PyEMILI.\\
\smallskip\\
$^{a}$ The candidate ID with the ``A'' ranking assigned by PyEMILI.\\ 
\smallskip\\
(This table in its entirety is publicly available on Zenodo.)}
\end{deluxetable}

\section{The Formatted Complete Output of PyEMILI Runs}

The formatted complete output of PyEMILI runs on the emission lines of the objects in the test sample are presented in Tables\,B.1--B.34, which are publicly available on Zenodo \citep{tu_2025_17540949}.  A sample of PyEMILI output showing the ranked candidate IDs for an emission line in NGC\,3918 is presented in Table\,\ref{output_NGC3918} for the purpose of demonstration.

\begin{deluxetable}{rllcrllrcrlcll}[!htp]
\label{output_NGC3918}
\tabletypesize{\scriptsize}
\tablecaption{A Sample of PyEMILI Output Showing the Top 14 Candidate IDs of an Emission Line Observed at 3109.13\,{\AA} in PN NGC\,3918}
\tablehead{\colhead{(1)} & \colhead{(2)} & \colhead{(3)} & \colhead{(4)} & \colhead{(5)} & \colhead{(6)} & \colhead{(7)} & \colhead{(8)} & \colhead{(9)} & \colhead{(10)} & \colhead{(11)} & \colhead{(12)} & \colhead{(13)} & \colhead{(14)} 
} 
\startdata
+3109.115 & 3109.18   & [\ion{Ar}{iii}] & 3P\text{--}1S      & 4.04E-03  & 0/0 & 1  & A & -6.20820 & & &3s$^2$.3p$^4$\text{--}3s$^2$.3p$^4$ & 3.0 & 1.0 \\
+3109.133 & 3109.005  & \ion{Cr}{ii}    & c$^4$F\text{--}u$^4$D$^o$  & 4.46E-08  & 0/0 & 6  & B & 12.41221 & &&3d$^3$.4s$^2$\text{--}3d$^3$(4F).4s.4p.(3P$^o$) & 10.0& 8.0 \\
+3109.115 & 3109.14   & \ion{N}{iii}    & 4P$^o$\text{--}4P  & 4.63E-08  & 5/1 & 6  & B & -2.35136 & 3118.769& -11.9 & 2s.2p.(3P$^o$).4s\text{--}2p$^2$(3P).3s & 4.0& 6.0 \\
3109.133 & 3109.371  & \ion{V}{ii}     & b$^1$D\text{--}y$^1$D$^o$  & 4.59E-08  & 0/0 & 7  & C & -22.8774 &&& 3d$^4$\text{--}3d$^3$(a$^2$D).4p & 5.0 &5.0 \\
+3109.133 & 3109.254  & \ion{Mn}{ii}    & z$^3$G$^o$\text{--}e$^3$G  & 3.32E-07  & 1/0 & 7  & C & -11.5972 &&& 3d$^5$(4G).4p\text{--}3d$^5$(4G).5s & 11.0 &11.0 \\
+3109.133 & 3109.135  & \ion{S}{ii}     & $^2$F\text{--}1[2]$^o$     & 1.65E-07  & 2/0 & 7  & C & -0.12330 &&& 3s$^2$.3p$^2$(1D).3d\text{--}3s$^2$.3p$^2$(3P).5f & 6.0& 6.0 \\
3109.133 & 3108.812  & \ion{Mn}{ii}    & z$^3$G$^o$\text{--}e$^3$G  & 4.25E-07  & 1/0 & 8  & D & 31.02457 &&& 3d$^5$(4G).4p\text{--}3d$^5$(4G).5s & 9.0 &9.0 \\
3109.133 & 3108.764  & \ion{Si}{ii}    & $^2$P$^o$\text{--}D[3]     & 2.52E-07  & 0/0 & 8  & D & 35.65391 &&& 3s$^2$.9p\text{--}3s.3p.(3P$^o$).4f.D & 4.0 &6.0 \\
3109.115 & 3108.725  & \ion{Ne}{ii}    & $^2$P\text{--}$^2$D$^o$    & 1.93E-07  & 0/0 & 8  & D & 37.66918 &&& 2s$^2$.2p$^4$(3P).4s\text{--}2s$^2$.2p$^4$(3P).6p & 2.0& 4.0 \\
3109.133 & 3109.47   & [\ion{Cr}{iii}] & 5D\text{--}a$^1$D          & 4.46E-07  & 4/0 & 9  &   & -32.4215 &&& 3d$^4$\text{--}3d$^4$ & 1.0& 5.0 \\
3109.133 & 3108.67   & [\ion{V}{iii}]  & $^2$G\text{--}b$^4$F       & 9.37E-07  & 3/0 & 9  &   & 44.72012 &&& 3d$^3$\text{--}3d$^2$(3F).4s & 10.0& 8.0 \\
3109.158 & 3108.651  & \ion{Cr}{ii}]   & b$^4$G\text{--}z$^2$G$^o$  & 2.77E-07  & 5/1 & 9  &   & 48.90243 & 3118.135 &53.0 & 3d$^4$(3G).4s\text{--}3d$^4$(3H).4p & 8.0& 10.0 \\
3109.115 & 3109.538  & \ion{Na}{ii}    & $^2$P$_{3/2}$\text{--}$^2$P$_{5/2}$ & 8.29E-08 & 2/0 & 9 & & -40.7224 &&& 2s$^2$.2p$^5$(2P$_{3/2}$).3d\text{--}2s$^2$.2p$^5$(2P$_{3/2}$).5f & 5.0& 5.0 \\
3109.158 & 3108.622  & [\ion{Cr}{ii}]  & a$^6$D\text{--}b$^2$S      & 5.81E-08  & 0/0 & 9  &   & 51.69962 &&& 3d$^4$(5D).4s\text{--}3d$^5$ & 6.0 &2.0 \\
\enddata
\tablecomments{The observed line is 3109.13 \AA, with a flux of 6.48E-03 relative to $\mathrm{H\beta}$. \\
For a detailed description of the contents of the form, please refer to Paper I or the user manual. \\
\smallskip\\
(This table in its entirety is publicly available on Zenodo.)}
\end{deluxetable}

\section{Corner Plots Created Using MCMC Sampling}

\begin{figure}[htp!]
\centering
\includegraphics[width=1\textwidth]{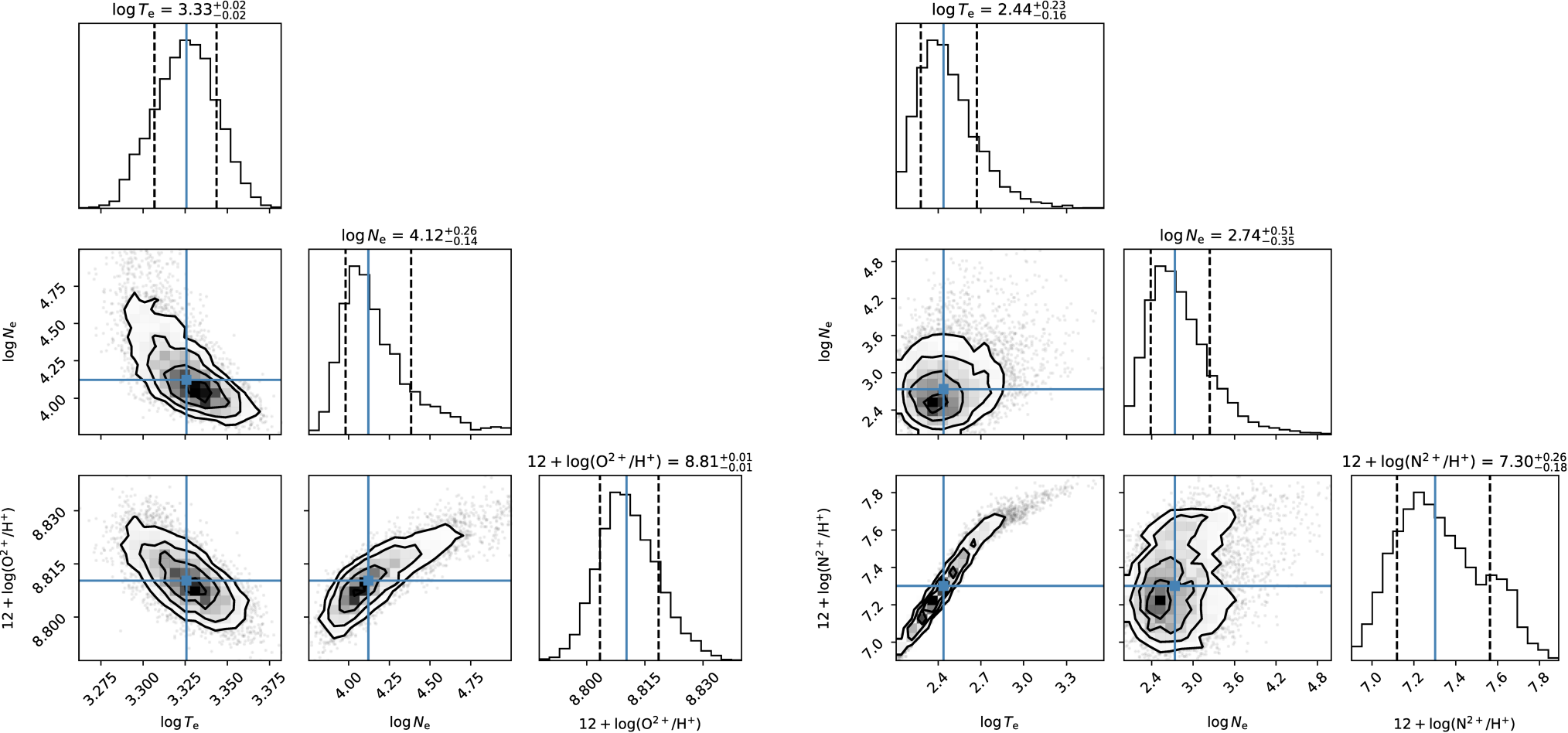}
\caption{\emph{Left}: The marginalized posterior distributions of electron temperature ($\log{T_{\rm e}}$), electron density ($\log{N_{\rm e}}$) and ionic abundance O$^{2+}$/H$^{+}$ optimized using the \ion{O}{ii} ORLs for IC\,4776.  \emph{Right}: same as the left panel but using the \ion{N}{ii} ORLs.  The vertical dashed black lines mark the 16th and 84th quantiles of the 1D distributions, while the blue lines indicate the 50th quantiles of each parameter in the 1D and 2D distributions.} 
\label{MCMC_IC4776}
\end{figure}

\begin{figure}[htp!]
\centering
\includegraphics[width=1\textwidth]{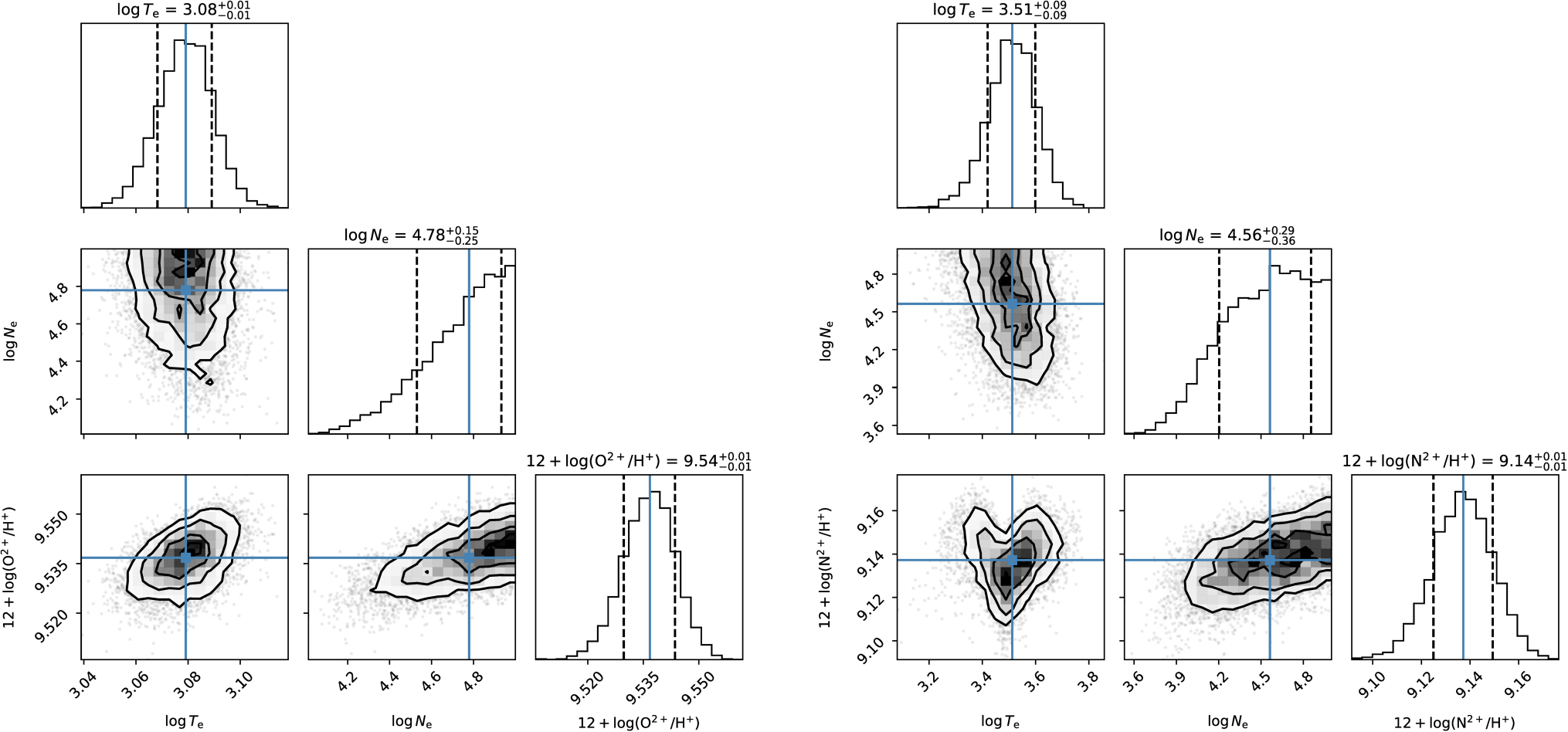}
\caption{Same as Figure\,\ref{MCMC_IC4776} but for the spectrum extracted along the minor axis of the main nebula of NGC\,6153 (see description in \citealt{2000MNRAS.312..585L}).}
\label{MCMC_NGC6153_minor}
\end{figure}

\begin{figure}[htp!]
\centering
\includegraphics[width=1\textwidth]{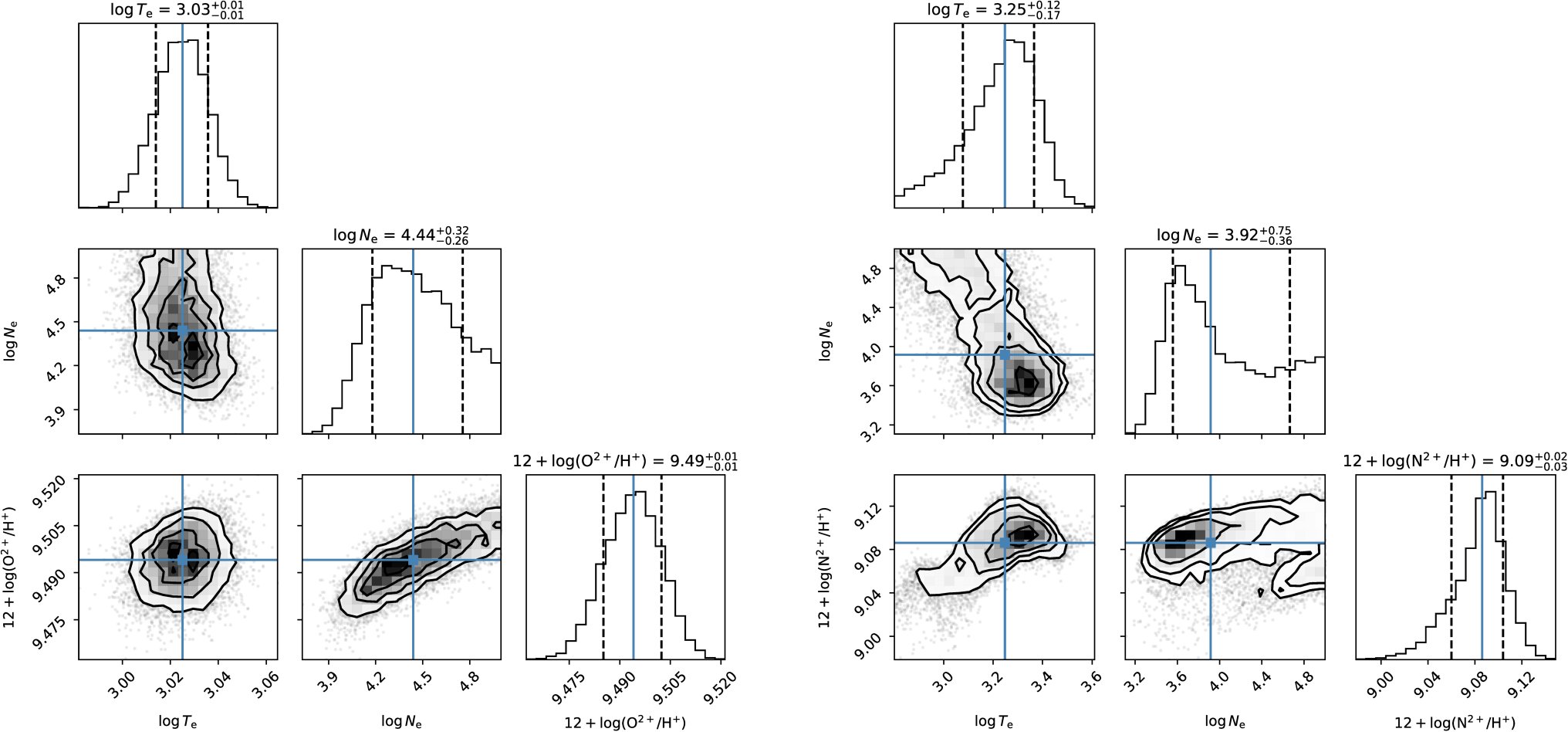}
\caption{Same as Figure\,\ref{MCMC_IC4776} but for the spectrum scanned across the entire nebula of NGC\,6153 (see description in \citealt{2000MNRAS.312..585L}).}
\label{MCMC_NGC6153_entire}
\end{figure}

\begin{figure}[htp!]
\centering
\includegraphics[width=1\textwidth]{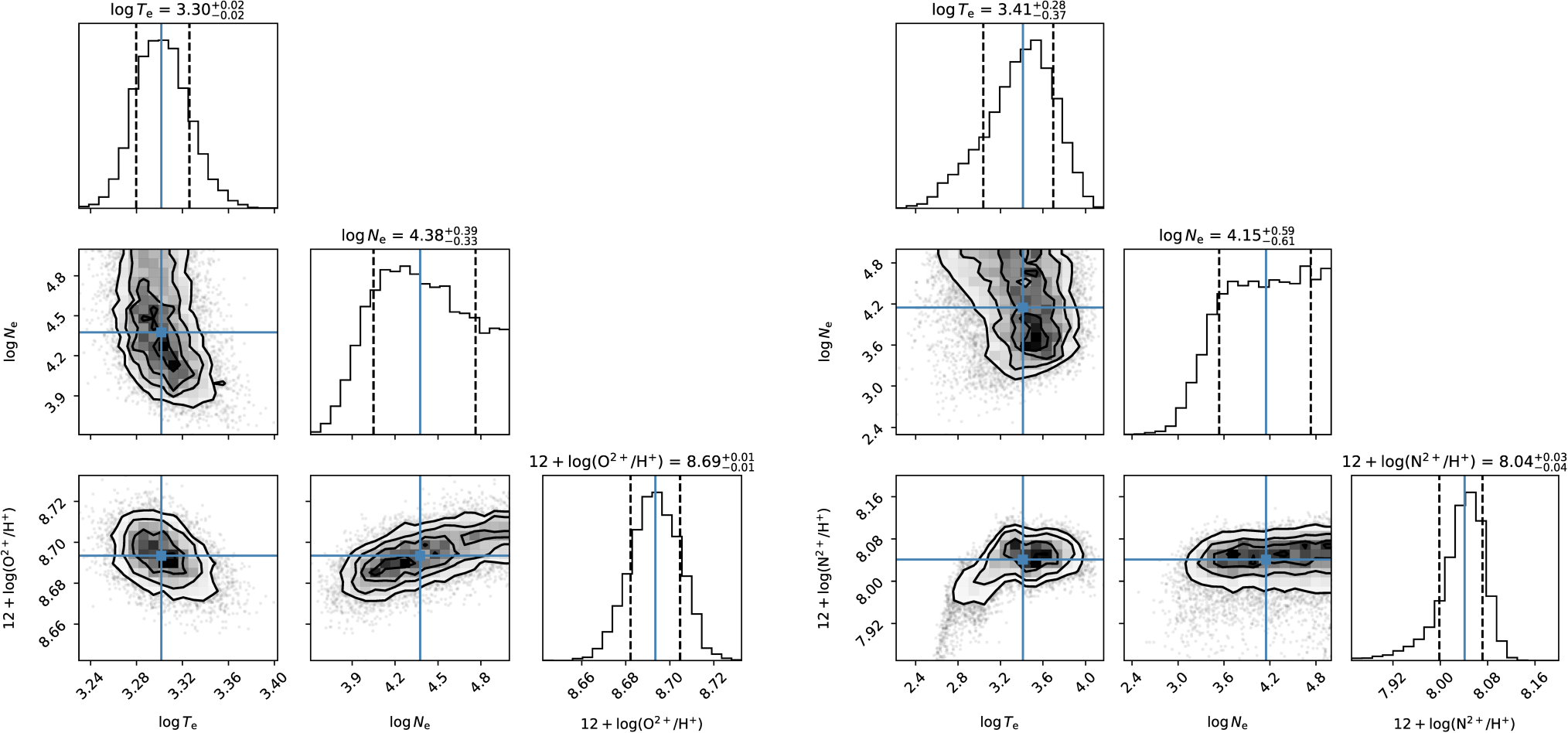}
\caption{Same as Figure\,\ref{MCMC_IC4776} but for the NGC\,3918.}
\label{MCMC_NGC3918}
\end{figure}

\begin{figure}[htp!]
\centering
\includegraphics[width=1\textwidth]{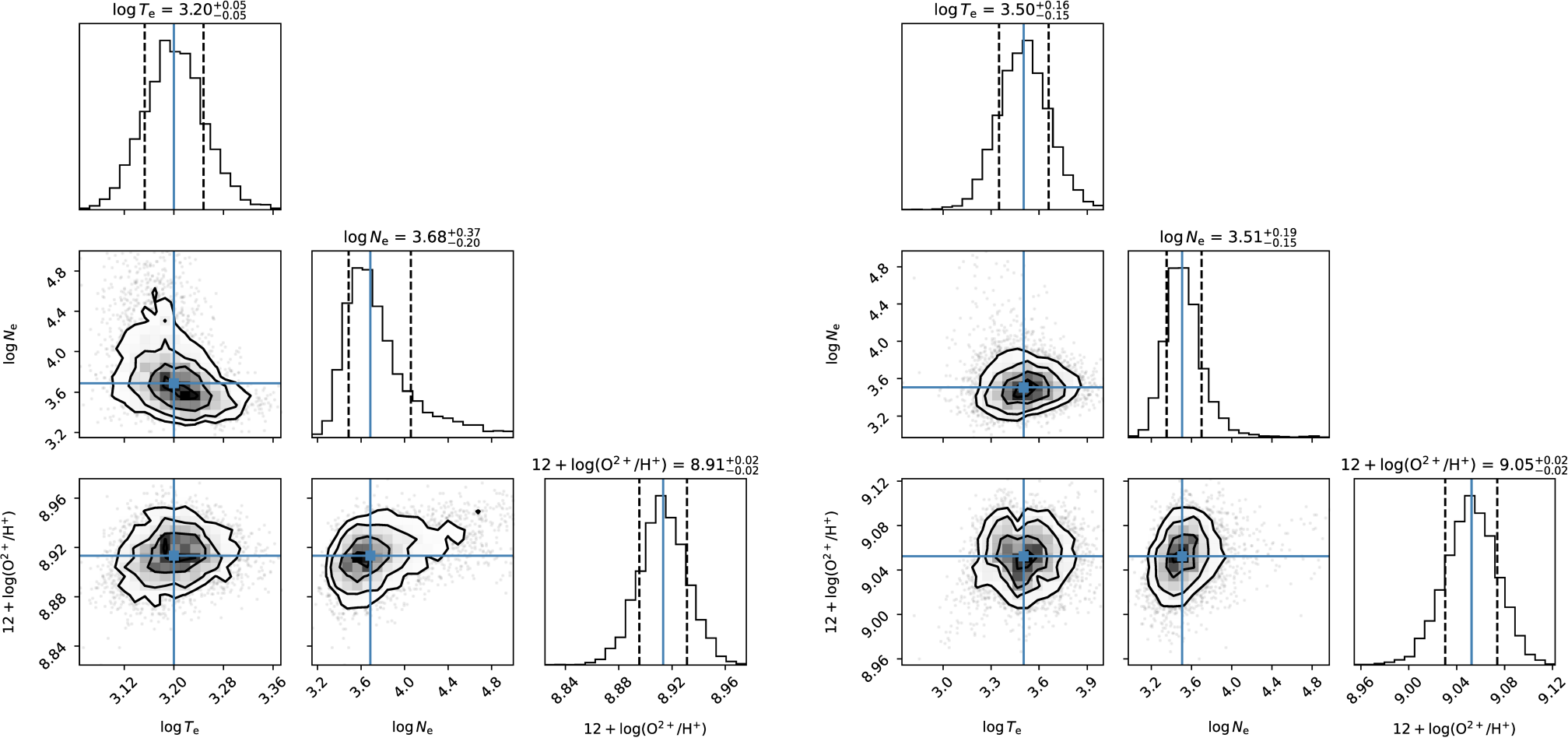}
\caption{Same as the left panel in Figure\,\ref{MCMC_IC4776} but for  M\,1-30 (\emph{left}) and PC\,14 (\emph{right}).} 
\label{MCMC_M1-30_PC14}
\end{figure}

\begin{figure}[htp!]
\centering
\includegraphics[width=1\textwidth]{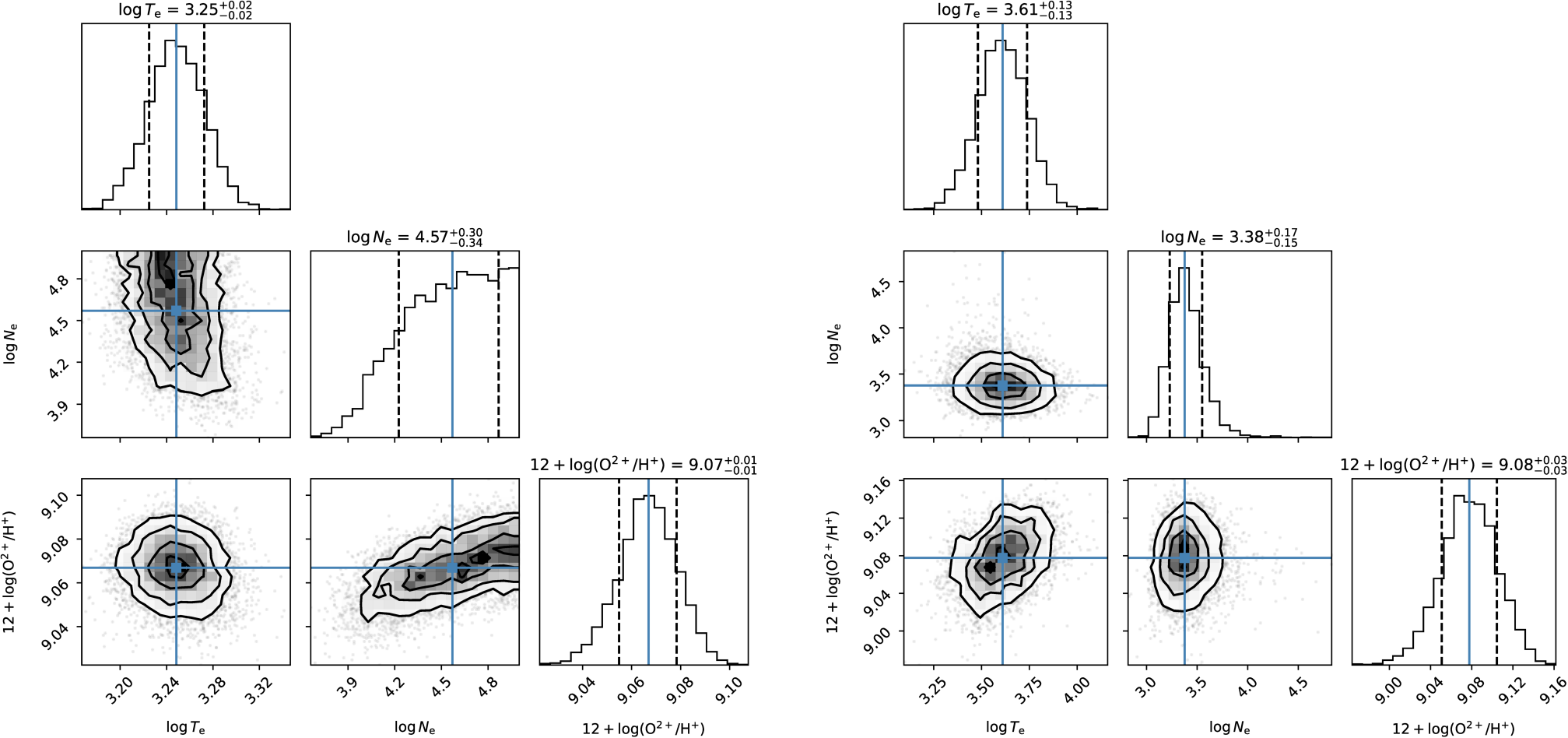}
\caption{Same as the left panel in Figure\,\ref{MCMC_IC4776} but for He\,2-86 (\emph{left}) and H\,1-50 (\emph{right}).} 
\label{MCMC_He2-86_H1-50}
\end{figure}

\begin{figure}[htp!]
\centering
\includegraphics[width=1\textwidth]{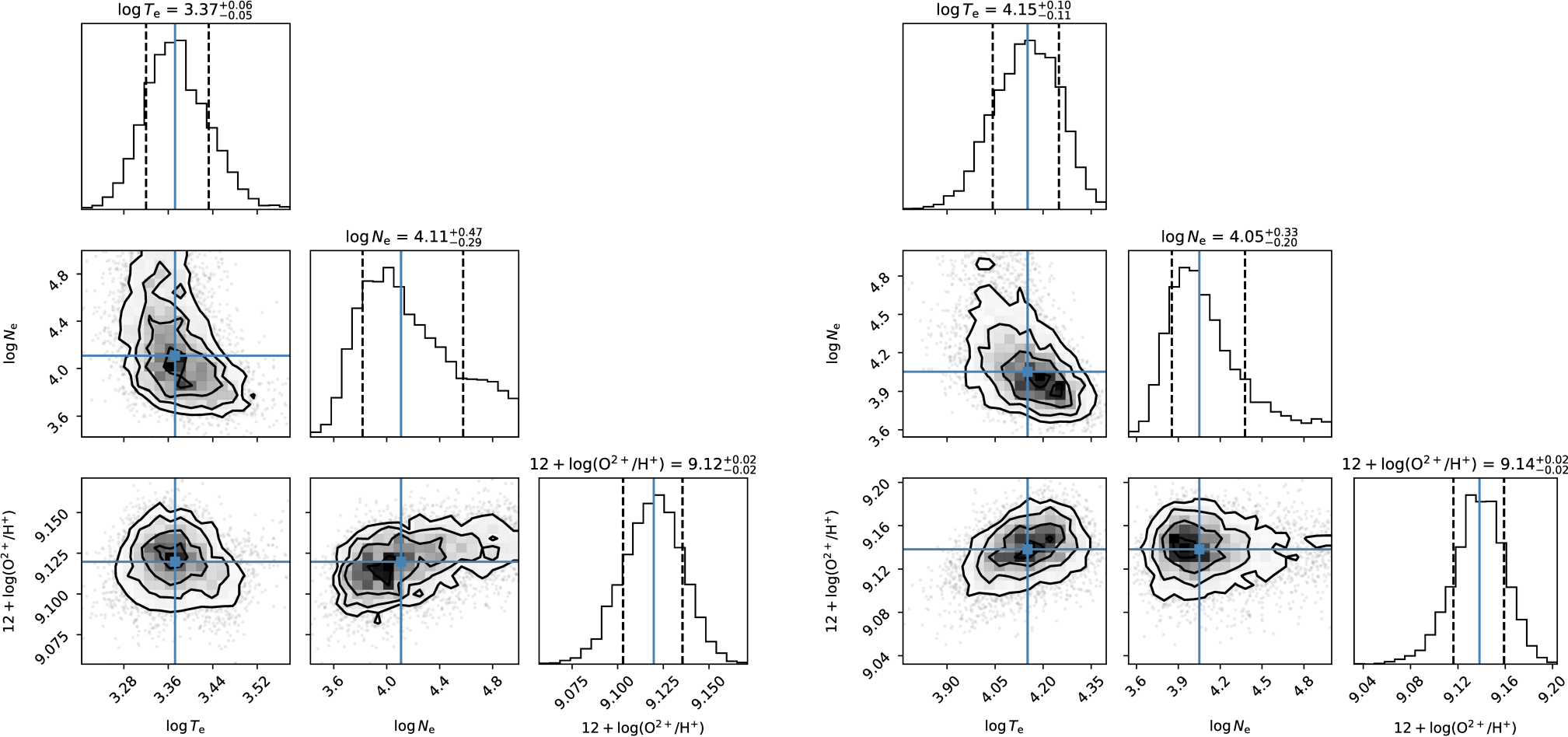}
\caption{Same as the left panel in Figure\,\ref{MCMC_IC4776} but for M\,1-33 (\emph{left}) and H\,1-60 (\emph{right}).} 
\label{MCMC_M1-33_M1-60}
\end{figure}

\begin{figure}[htp!]
\centering
\includegraphics[width=1\textwidth]{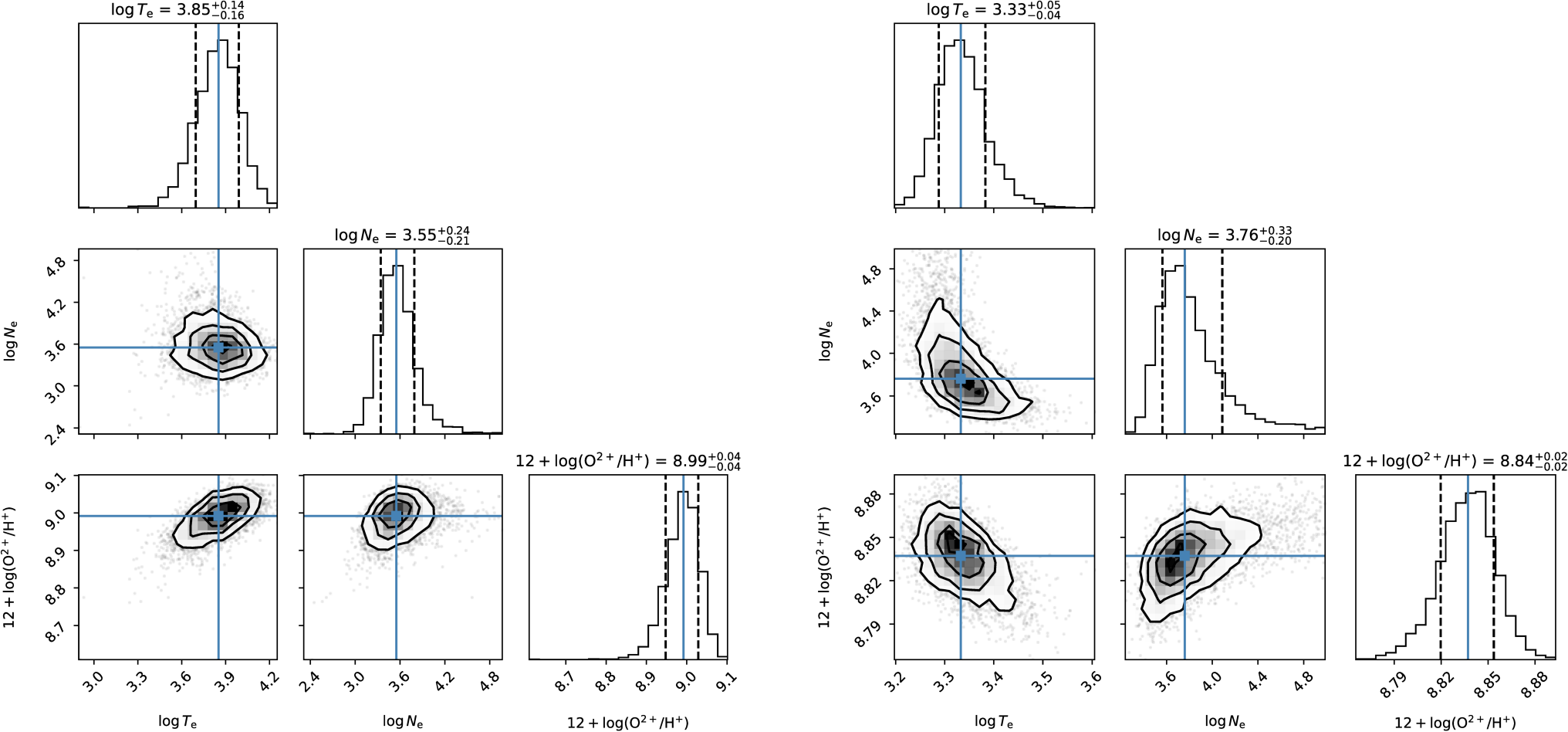}
\caption{Same as the left panel in Figure\,\ref{MCMC_IC4776} but for M\,2-31 (\emph{left}) and NGC\,5315 (\emph{right}).} 
\label{MCMC_M2-31_NGC5315}
\end{figure}

\begin{figure}[htp!]
\centering
\includegraphics[width=1\textwidth]{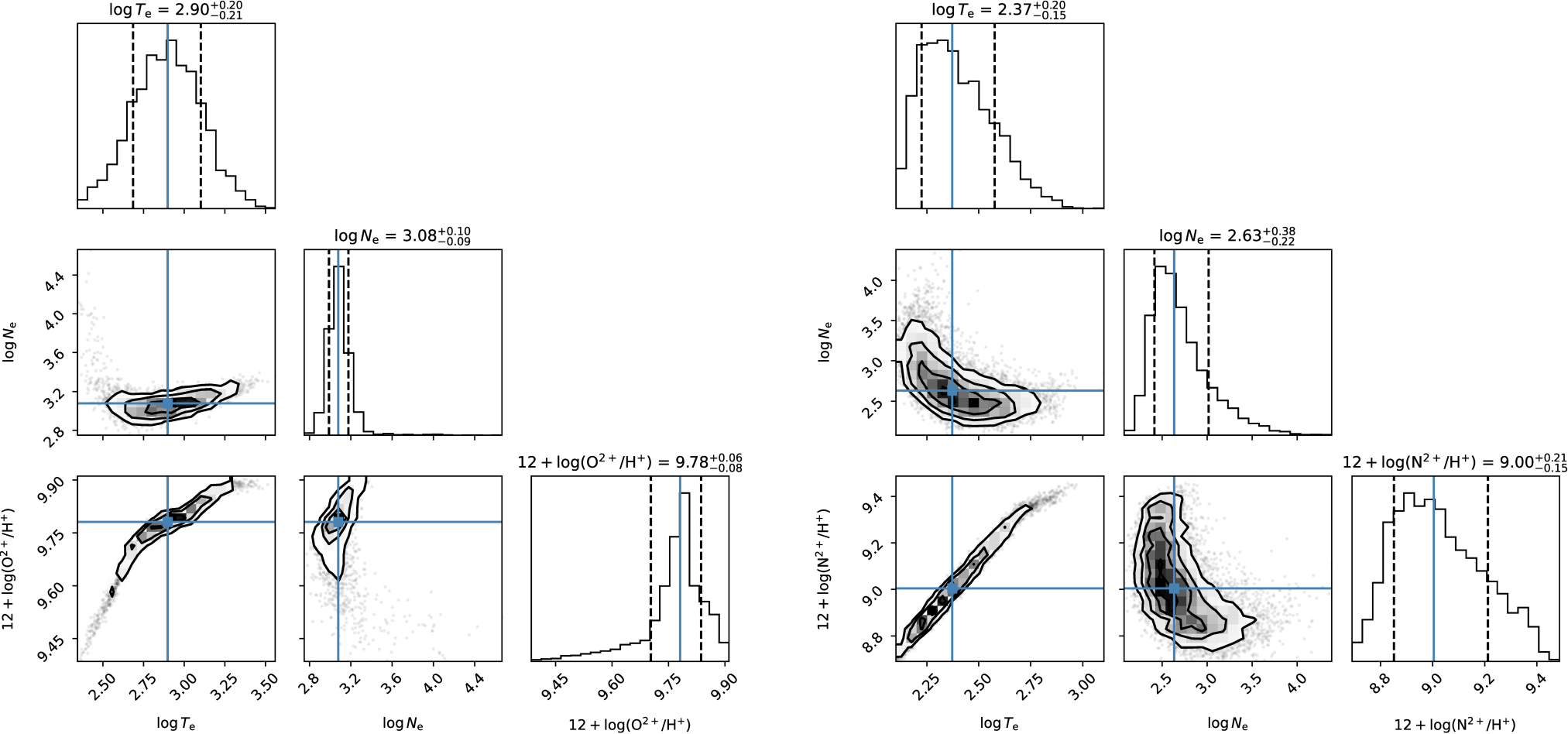}
\caption{Same as Figure\,\ref{MCMC_IC4776} but for the spectrum of Hf\,2-2 obtained with a slit width of 2\arcsec\ (see description in \citealt{2006MNRAS.368.1959L}).}
\label{MCMC_Hf2-2_2arcsec}
\end{figure}

\begin{figure}[htp!]
\centering
\includegraphics[width=1\textwidth]{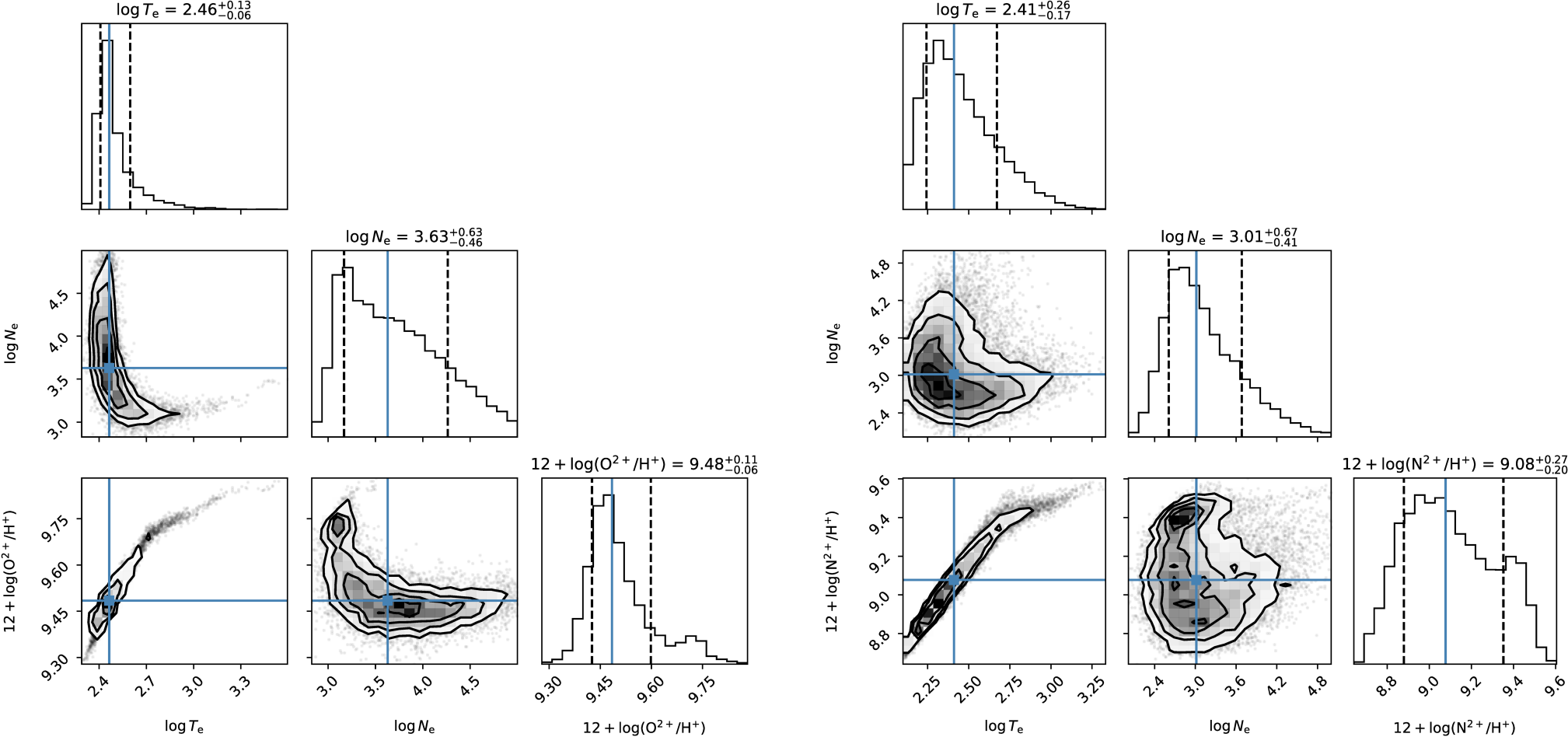}
\caption{Same as Figure\,\ref{MCMC_IC4776} but for the spectrum of Hf\,2-2 obtained with a slit width of 8\arcsec\ (see description in \citealt{2006MNRAS.368.1959L}).}
\label{MCMC_Hf2-2_8arcsec}
\end{figure}

\begin{figure}[htp!]
\begin{center}
\includegraphics[width=1\textwidth]{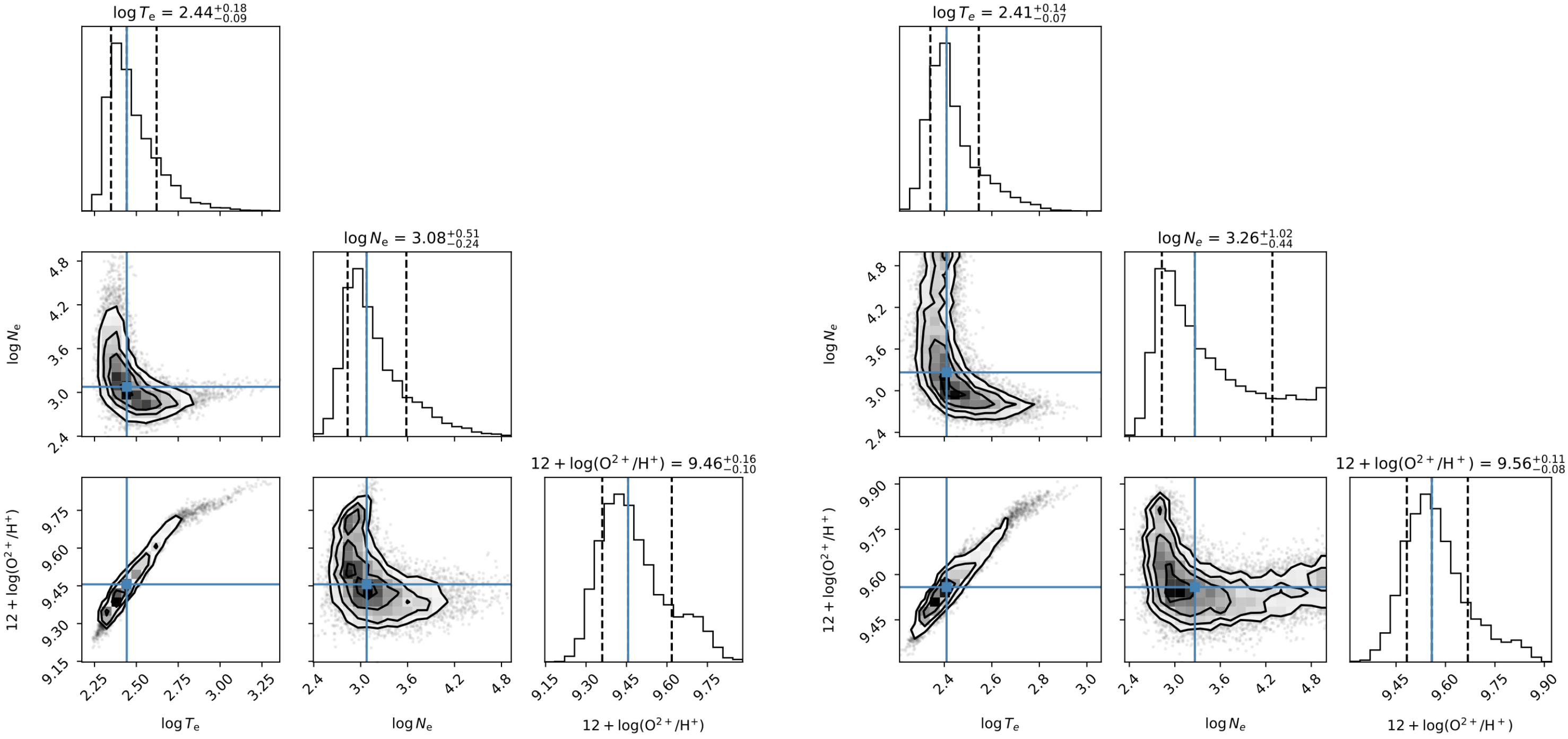}
\caption{Same as the left panel of Figure\,\ref{MCMC_IC4776} but for the spectrum of Hf\,2-2 obtained with a slit width of 4\arcsec\ (\emph{left}) and Abell 46 (\emph{right}).}
\label{MCMC_Hf2-2_4arcsec}
\end{center}
\end{figure}

The distribution of nebular parameters ($T_{\rm e}$, $N_{\rm e}$, and ionic abundances O$^{2+}$/H$^{+}$ and/or N$^{2+}$/H$^{+}$) of the PNe in our test sample are presented in Figures\,\ref{MCMC_IC4776}--\ref{MCMC_Hf2-2_4arcsec}, where corner plots created using MCMC sampling are shown. 

\clearpage

\bibliography{references}{}

\bibliographystyle{aasjournal}


\end{document}